%% file: main.tex
\documentclass[
               twocolumn, 
               astrosymbooll,
               paper]{aastex631}

\newcommand{\eg}{\emph{e.g.}}
\newcommand{\ie}{\emph{i.e.}}

\newcommand{\SNR}{\ensuremath{\mathrm{\emph{{S/N}}}}}
\newcommand{\Nlit}{165}

\newcommand{\elasticc}{ELAsTiCC}
\newcommand{\plasticc}{PLAsTiCC}

\newcommand{\Hy}{{H}}
\newcommand{\Halpha}{{H$\alpha$}}
\newcommand{\He}{{He}}
\newcommand{\Ca}{[\ion{Ca}{2}]}
\newcommand{\Si}{{Si}}

\newcommand{\synCo}{$^{56}$\rm{Co}}
\newcommand{\synNi}{$^{56}$\rm{Ni}}

\newcommand{\allbands}{\ensuremath{{w2,~m2,~w1,~U,~u,~B,~g,~V,~r,~R,~i,~I,~J,~H,~K_s}}}
\newcommand{\ubvri}{\protect\hbox{$U\!BV\!RI$} }
\newcommand{\bvri}{\Bband, \Vband, \Rband, and \Iband}
\newcommand{\bthoughi}{\Bband, \gband, \Vband, \rband, \Rband, \iband, and \Iband}

\newcommand{\ugri}{\uband, \gband, \rband, and \iband} 
\newcommand{\gri}{\gband, \rband, and \iband} 
\newcommand{\sloanugriprimed}{\protect\hbox{$u'g'r'i'$} }

\newcommand{\opticalbands}{\ensuremath{{U/u,~B,~g,~V,~R/r,~I/i}}}
\newcommand{\opticalbandsdetailed}{\ensuremath{{U, u,~B,~g,~V,~R,~r,~I,~i}}}
\newcommand{\NIRbands}{\ensuremath{{J,~H,~K_s}}}
\newcommand{\swiftbands}{\ensuremath{{w2,~m2,~w1}}}
\newcommand{\Uband}{\ensuremath{U}}
\newcommand{\uband}{\ensuremath{u}}
\newcommand{\Bband}{\ensuremath{B}}
\newcommand{\gband}{\ensuremath{g}}
\newcommand{\Vband}{\ensuremath{V}}
\newcommand{\Rband}{\ensuremath{R}}
\newcommand{\rband}{\ensuremath{r}}
\newcommand{\Iband}{\ensuremath{I}}
\newcommand{\iband}{\ensuremath{i}}
\newcommand{\zband}{\ensuremath{z}}
\newcommand{\Jband}{\ensuremath{J}}
\newcommand{\Hband}{\ensuremath{H}}
\newcommand{\Kband}{\ensuremath{K_s}}

\newcommand{\alltypes}{IIb,~Ib,~Ic,~Ic-bl,~and~Ibn} 
\newcommand{\blIc}{SN~Ic-bl}
\newcommand{\Ibc}{SN~Ibc}
\newcommand{\Ia}{SN~Ia}
\newcommand{\Ib}{SN~Ib}
\newcommand{\IIb}{SN~IIb}
\newcommand{\Ic}{SN~Ic}
\newcommand{\Ibn}{SN~Ibn}
\newcommand{\SESN}{SESN}
\newcommand{\blIcs}{SNe~Ic-bl}
\newcommand{\Ibcs}{SNe~Ibc}
\newcommand{\Ias}{SNe~Ia}
\newcommand{\Ibs}{SNe~Ib}
\newcommand{\IIbs}{SNe~IIb}
\newcommand{\Ics}{SNe~Ic}
\newcommand{\Ibns}{SNe~Ibn}

\newcommand{\maxep}{\ensuremath{\mathrm{JD}_\mathrm{max}}}
\newcommand{\Vmax}{{$\mathrm{JD}_\mathrm{Vmax}$}}

\newcommand\myfontsize{\fontsize{7pt}{8.4pt}\selectfont}

\defcitealias{drout11}{D11}
\defcitealias{taddia15}{T15}
\defcitealias{stritzinger2018carnegie1}{CSP}
\defcitealias{modjaz14}{M14}
\defcitealias{bianco14}{B14}
\defcitealias{ivezic2019lsst}{LSST}
\defcitealias{guillochon16}{OSNC}
\defcitealias{hosseinzadeh2017type}{H17}
\defcitealias{ho2023photometric}{Ho23}

\usepackage{hyperref}
\usepackage[toc,page]{appendix}
\usepackage{rotating}
\usepackage{booktabs} 
\usepackage{graphicx}
\usepackage[normalem]{ulem}
\usepackage{multirow}
\usepackage{url}

\begin{document}

\title{Multi-filter UV to NIR Data-driven Light Curve Templates for Stripped Envelope Supernovae}


\author[0000-0002-1910-7065]{
Somayeh~Khakpash}
\affiliation{Rutgers University, Department of Physics \& Astronomy, 136 Frelinghuysen Rd, Piscataway, NJ 08854, USA}
\affiliation{University of Delaware
Department of Physics and Astronomy
217 Sharp Lab
Newark, DE 19716 USA}
\affiliation{University of Delaware
Data Science Institute}

\author[0000-0003-1953-8727]{Federica~B.~Bianco}
\affiliation{University of Delaware
Department of Physics and Astronomy
217 Sharp Lab
Newark, DE 19716 USA}
\affiliation{University of Delaware
Joseph R. Biden, Jr. School of Public Policy and Administration, 
184 Academy St, Newark, DE 19716 USA}
\affiliation{University of Delaware
Data Science Institute}
\affiliation{Vera C. Rubin Observatory, Tucson, AZ 85719, USA}

\author[0000-0001-7132-0333]{Maryam~Modjaz}\affiliation{University of Virginia, Department of Astronomy, 530 McCormick Road
Charlottesville, VA 22904}

\author[0000-0001-7559-7890]{Willow~F.~Fortino}\affiliation{University of Delaware
Department of Physics and Astronomy
217 Sharp Lab
Newark, DE 19716 USA}
\affiliation{University of Delaware
Data Science Institute}

\author[0000-0003-4906-8447]{Alexander Gagliano}
\affiliation{The NSF AI Institute for Artificial Intelligence and Fundamental Interactions}
\affiliation{Center for Astrophysics \textbar{} Harvard \& Smithsonian, 60 Garden Street, Cambridge, MA 02138-1516, USA}
\affiliation{MIT Laboratory For Nuclear Science, 77 Massachusetts Ave., Cambridge, MA 02139, US}

\author[0000-0003-2037-4619]{
Conor Larison}
\affiliation{Rutgers University, Department of Physics \& Astronomy, 136 Frelinghuysen Rd, Piscataway, NJ 08854, USA}

\author[0000-0001-9227-8349]{Tyler~A.~Pritchard}\affiliation{NASA’s Goddard Space Flight Center, Greenbelt, MD 20771
USA}

\begin{abstract}
While the spectroscopic classification scheme for Stripped envelope supernovae (\SESN e) is clear, and we know that they originate from massive stars that lost some or all their envelopes of Hydrogen and Helium, the photometric evolution of classes within this family is not fully characterized. Photometric surveys, like the Vera C. Rubin Legacy Survey of Space and Time, will discover tens of thousands of transients each night and spectroscopic follow-up will be limited, prompting the need for photometric classification and inference based solely on photometry. We have generated 54 data-driven photometric templates for \SESN e of subtypes \alltypes\ in \opticalbands, \NIRbands, and Swift \swiftbands\ bands using Gaussian Processes and a multi-survey dataset composed of all well-sampled open-access light curves (\Nlit\ \SESN e, 29531 data points) from the Open Supernova Catalog. We use our new templates to assess the photometric diversity of \SESN e by comparing final per-band subtype templates with each other and with individual, unusual and prototypical SESNe. We find that \Ibns\ and Ic-bl exhibit a distinctly faster rise and decline compared to other subtypes. We also evaluate the behavior of \SESN e in the \plasticc\ and \elasticc\ simulations of LSST light curves highlighting differences that can bias photometric classification models trained on the simulated light curves. Finally, we investigate in detail the behavior of fast-evolving \SESN e (including SNe Ibn) and the implications of the frequently observed presence of two peaks in their light curves.

\end{abstract}

\section{Introduction}
\label{sec:intro_sec}

Stripped envelope supernovae (\SESN e) include a diverse ensemble of
explosive transients that arise from the core collapse of massive
stars that have been stripped of various amounts of their outer layers before
explosion. Their distinctive observational signature is the absence (or weakness)
of hydrogen (\Hy) in their maximum light spectra, and this family of
transients includes SNe type Ib and IIb, SNe type Ic and Ic-bl, which
also lack signatures of helium (\He), and their subtypes \citep{clocchiatti1997light, filippenko97, gal2016observational, modjaz2019new} including the ``transitional'' type Ibn \citep{pastorello15}. While intrinsically nearly as
common as Type Ia SNe \citep{shivvers16} they are less luminous, and,
unlike \Ias, \SESN e  do not show phenomenological relationships that can be exploited for standardization; thus, they are not an obvious cosmological tool (although some of the subtypes, especially
\blIcs\ with GRB, have been claimed to be promising standardizable
candles, see \citealt{cano14}, and references therein).  For these reasons
they are less observed and less well-studied than \Ias, and since
they are relatively less common, less observed than SNe~II.

In 2014, our groups presented what is still today one of the largest homogeneous photometric
sample of \SESN e, \citet{bianco14} -- \citetalias{bianco14} hereafter --  complemented by currently one of the largest spectroscopic
samples, \citet{modjaz14} -- \citetalias{modjaz14} hereafter. Both of these works are based on the CfA SN Survey\footnote{\url{https://www.cfa.harvard.edu/supernova/}}, comprising 64 photometric targets, and 73 spectroscopic targets, of which 54 are in both samples
and we began addressing the observational photometric properties
of the \SESN e population. Our work was followed by the data releases and analysis of \citet{stritzinger2018carnegie1} -- \citetalias{stritzinger2018carnegie1} hereafter -- and \citet{stritzinger2018carnegie2} -- based on the Carnegie Observatory data (CSP), and more recently the Palomar Transient Factory (PTF) and intermediate PTF (iPTF) data that supported studies including \citet{vincenzi2019spectrophotometric}, \citet{schulze2021palomar}, and \citet{barbarino2021type}, although a formal data release for these \SESN e is still to come.
Currently, fewer than 150 objects comprise a heterogeneous sample of well-studied \SESN e released in different publications (\citealt{drout11} -- \citetalias{drout11} hereafter --, \citealt{cano13}, \citetalias{modjaz14},
\citetalias{bianco14} ,
\citealt{lyman2016bolometric, prentice2016bolometric, modjaz14, liu16, stritzinger2018carnegie1}, \citetalias{ stritzinger2018carnegie1}, \citealt{ sollerman2021maximum, ho2023photometric} -- hereafter \citetalias{ho2023photometric}).

Our work is motivated by two distinct goals. {In the era of large photometric surveys, like the Vera C. Rubin Legacy Survey of Space and Time, hereafter LSST \citep{ivezic2019lsst}, increasing emphasis has to be placed on photometric classification and characterization of transients and variable phenomena, as spectroscopic resources will only enable follow up of a small fraction of observed transients \citep{najita2016maximizing}. With this motivation, photometric classifiers of astrophysical transients, even from sparse light curves, have seen rapid developments (see for example \citealt[][ and references therein]{Qu22, hlovzek2020results}). More recent work is emerging approaching classification from a more holistic perspective, for example, using photometry jointly with galaxy host information \cite{gagliano2023first}. Yet, primarily, the focus and drive for these works have been to ensure the purity of cosmological \Ia\ samples in high-$z$ synoptic surveys,
 rather than understanding the properties and diversity of \SESN e and setting the stage for \SESN e physics and stellar evolution studies in an astronomical era with
an unprecedented wealth of photometry, but when spectra of the photometric transients will be rare. As an example, some of the most successful supernova photometric classifiers, like \citet{boone2019avocado} or \citet{qu2021photometric}, to name just two, achieve $>80\%$ accuracy on the classification of \Ias, but for even for \SESN e as a collective class, they obtain modest classification power (52\% accuracy for \citealt{boone2019avocado} and as good as 78\% for \citealt{qu2021photometric} but only when a full light curve up to 50 days after discovery is used and accurate spectral information is available to measure the redshift). We note that both these studies are based on the \plasticc\ challenge dataset \citep{kessler2019models} and we will compare this simulated dataset, and its successor \elasticc, with our observations in \autoref{sec:plasticc_compare} to assess the accuracy of \Ibc\ simulated sample, and discuss implications for machine learning models trained on these data.}

Our ultimate goal is to assess the photometric diversity of \SESN e,
among and within subtypes, {setting the stage for studies aimed at relating the explosion taxonomy to diversity in
explosion properties and progenitor rates.  This inference requires
several steps to extract the underlying physics of the explosions
from the observed explosion properties. On the way to achieve our stated goal, in this paper, we will address the generation of {\it data-driven} templates for \SESN e subtypes, with a focus on addressing methodological challenges and designing best practices in the selection of the data on which the templates are based, and care in the algorithmic choices that minimize the introduction of biases in the template. }

As the authors strongly believe in the benefit of open science and open data, we chose to use the Open Supernova Catalog \citep[][hereafter \citetalias{guillochon16}]{guillochon16} as the backbone of our study; all SNe included in our work are sourced from this open library, or we contributed them this library, with a few instances in which the data from \citetalias{guillochon16} was augmented or corrected based on data available to the authors or published but not integrated in the catalog and for which we submitted pull requests though the \citetalias{guillochon16} GitHub project.\footnote{{\url{https://github.com/astrocatalogs/supernovae}}.} However, unfortunately, in the time span over which this paper was written, the \citetalias{guillochon16} stopped being regularly maintained, as it happens to a significant fraction of open software and data portals \citep{nowogrodzki2019support}. The interactive part of the \citetalias{guillochon16} broke down on March 8 2022 irrecoverably, and it is now only accessible through its API\footnote{\url{https://github.com/astrocatalogs/OACAPI}} and GitHub\footnote{\url{https://github.com/astrocatalogs}}. We will discuss the benefits of, and difficulties encountered in, using an open dataset, rather than a proprietary dataset owned by the authors or accessed through a colleague's network. 

Data-driven \emph{spectral} templates of \SESN, or \emph{mean spectra},
have been published in \citet{liu16} and \citet{modjaz14} and lead
to an improved understanding of the subtypes' typical characteristics,
and of the relation between subtypes. Additionally, photometric templates of \SESN e were produced by \citetalias{drout11}, \citet[][\citetalias{taddia15} hereafter]{taddia15}, and \citet{vincenzi2019spectrophotometric} and \Ibn\ templates by \citet[][ \citetalias{hosseinzadeh2017type} hereafter]{hosseinzadeh2017type} that we will discuss and compare to ours in \autoref{sec:compare_Ibc} and \autoref{sec:compare_GP}. 

We present here data-driven light curve templates for all
bands from ultraviolet (UV) to near-infrared (NIR) separately, which will allow us to correct for incomplete
observations when generating bolometric light curves in future work and allow us to probe the characteristics of the photometric evolutions of \SESN e.  After discussing relevant literature and our motivations (\autoref{sec:motiv}) and the data that supports our work (\autoref{sec:data}), we present two sets of templates we develop. The \Ibc\ templates are rolling median templates made for each band using all SNe in our sample, introduced and discussed in \autoref{sec:IBCtemplates}. These are also used to support the creation of templates created using Gaussian Processes (GP) for each \SESN e subtype separately in each band (\autoref{sec:gptempaltes}). We compare our templates with templates for \SESN\ produced by other authors, individual objects that we found or are thought to be peculiar and prototypical, and synthetic samples used in the literature for model development (\autoref{sec:compare_Ibc} and \autoref{sec:compare_GP}), and we examine in detail the photometric evolution of rapid evolving \SESN e (\autoref{sec:Ibn_compare} and \autoref{sec:compare_ZTF}). Our conclusions are summarized in \autoref{sec:sum_concl}. Our work is reproducible and we make available to the reader the templates produced with our curated photometric sample as well as the methods and code we developed to generate and update templates as observations of new SNe become available.\footnote{\url{https://github.com/fedhere/GPSNtempl}}.



\section{Supernova Templates: Motivation and Status of the Field}
\label{sec:motiv}

In this work, we generated two sets of templates for stripped-envelope supernovae:
\begin{itemize} 
\item a set of \Ibc\ templates that describe the behavior of the \SESN e as a single family (although we know that there are differences in the subtype phenomenology!), one for each photometric band in \opticalbandsdetailed, \NIRbands, and Swift UV \swiftbands. Using all subtypes together allows the sample to be large enough to successfully generate templates in all but the UV bands. We discuss these templates first in  \autoref{sec:IBCtemplates}.

\item individual subtype templates for SN type \alltypes\ in individual bands. Constructing these templates requires a Bayesian approach (GP) and templates can only be produced in some bands. However, these templates enable the assessment of similarities and differences in the photometric evolution of different SESN subclasses. These will be referred to as final templates or GP templates and will be discussed in more detail in \autoref{sec:gptempaltes}.
\end{itemize}

Light curve templates are instrumental in most inference one may envision doing on and with SNe. In our case, our first goal is to understand diversity, with the final aim to relate the diversity of explosion to different progenitors, and to be able to identify the presence of separate classes {\it vs} a continuum in the range of observed properties.

Spectral templates of stripped SNe have been published in \citet{liu16} and \citet{modjaz14}, leading to an improved understanding of the subtypes' typical characteristics, and of the relationship between subtypes. Photometric templates are often derived from a single well-observed SN \citep[\eg][]{2013MNRAS.434.1098C} or combined from small samples of objects from a single survey \citep[\eg][]{drout11}. More recently \citet{2019MNRAS.489.5802V} produced {\it single-object} spectro-photometric templates for 67 core-collapse (CC) SNe, including 37 \SESN e. 

The time behavior of SN light curves has been modeled from physical principles by many authors; just a few examples are the seminal work of \citealt{arnett82}, and more recent work including  \citealt{bernstein12},  \citealt{piro15}, and \citealt{2022A&A...667A..92O},. The simplest, most general data-driven functional form is provided by \citet{vacca96} and \citet{vacca97}, which models the light curves of SNe with an exponential rise, a Gaussian peak, and a linear decay (and a secondary peak if needed, as, for example, in red optical and NIR bands for \Ias  ).
For standard SN light curves, this parameterization can be very successful. 
However, it may fail to model subtle peculiarities of a SN, and, generally, we expect the above model to capture most of the variation in thermonuclear \Ias, but other types of SNe, such as our \SESN e, show more diversity. Therefore, we focus on a \emph{non-parametric data-driven model} in this paper.

Most models from which one can obtain explosion parameters rely on
bolometric flux information (e.g. \citealt{arnett82}). However, we
observe our SNe in individual photometric bands, occasionally
obtaining observations throughout a large portion of the photometric
spectrum, from UV through IR, only exceptionally even pushing into the
X-ray and Radio, but more often in just a much narrower
portion of the spectrum: optical, or optical and NIR. 
In this work, we create single-band templates for \SESN e in as many bands
as the data allows, with a comprehensive dataset that aspirationally includes
{\it all} SESN photometric data available in the literature (\autoref{sec:data}). Producing
bolometric light curves from these data and templates also requires that one corrects for dust extinction in both our and the host galaxies, which remains, in spite of recent attempts to find empirical relationship \citep{stritzinger2018carnegie2}, an unresolved issue for transients that, like SESN, are not standardizable (\ie, for which we do not know the un-extincted behavior). Uncertain host galaxy extinction
remains the largest source of uncertainty in the derivation of
absolute parameters from observed explosions. The creation
of light curve templates in each photometric band separately does not require extinction correction. 
In each band, we
only use the relative magnitude, so that neither a color correction
nor the absolute luminosity are needed (see also \autoref{sec:data}).
In other words, we create photometric magnitude templates by combining individual SNe relative photometry. Our photometric templates, then, are designed to have the same peak brightness in each band (nominally 0) to describe the {\it shape} of the light curve for each subtype, and its evolution. Therefore, we will work with relative magnitude throughout the paper and we do not deal with the diversity in the absolute magnitude of the \SESN e.

These templates pave the way to address a core open question about SESNe: what is the relation between subtypes, and are we indeed seeing distinct classes, arising from quantitatively distinct progenitors and/or explosion mechanisms, or are we in the presence of a continuum of observational properties that indicate a gap-less evolution between progenitor properties and similar explosion mechanisms: in other words, is the progenitor mass the only difference between stars that explode as \Ib, vs \Ic, or do environment and mechanisms for the ejection of the material prior to explosion differ? 

A word of caution is necessary: templates are statistical aggregates. In the presence of gappy or incomplete light curves, we can use templates to input missing data, but we run the risk of biasings ourselves to the aggregate behavior and impose homogeneity. So while templates can assist us in answering questions about progenitors, we must strive to
construct templates that respect the diversity within the sample. Using the largest possible dataset and paying exquisite attention to uncertainties in the data and in the interpolation assures we represent the distribution of our data as well as the dataset allows, and this is what we strived toward in this work.

\section{Data}
\label{sec:data}

A data-driven template requires a sufficient amount of data to capture
the details and the diversity of a phenomenon, and until recently, the
field of \SESN e~has suffered from a significant scarcity of
data.
However, the sample of observed \SESN e~bloomed in recent
years. Like any astronomical transient phenomenon, the dataset grew
rapidly with the advent of optical sky surveys, starting with SDSS
(although these \SESN e~are not always
``well-observed'').   Our study is enabled by the existence of the Open Supernova
Catalog~\citep[OSNC][]{guillochon16}. The OSNC was a platform, part of the Open Astronomy Catalogs GitHub organization, an effort to catalog and digitize \emph{all} SNe observed (and whose observations have been published) into a single consistent repository. It collected and digitized \emph{all} SN photometry \emph{and} spectroscopy. Sadly, the OSNC catalog is no longer maintained and the OSNC has been frozen as-is on April 8, 2022.\footnote{
\url{https://sne.space/}.} We regret the loss of this legacy. This study leverages the unique collection of SESNe in the OSNC which represents our collective (nearly) complete data knowledge of \SESN e as of the early 2020s.

\begin{figure}[b]
 \centerline{
 \includegraphics[width=1\columnwidth]{ 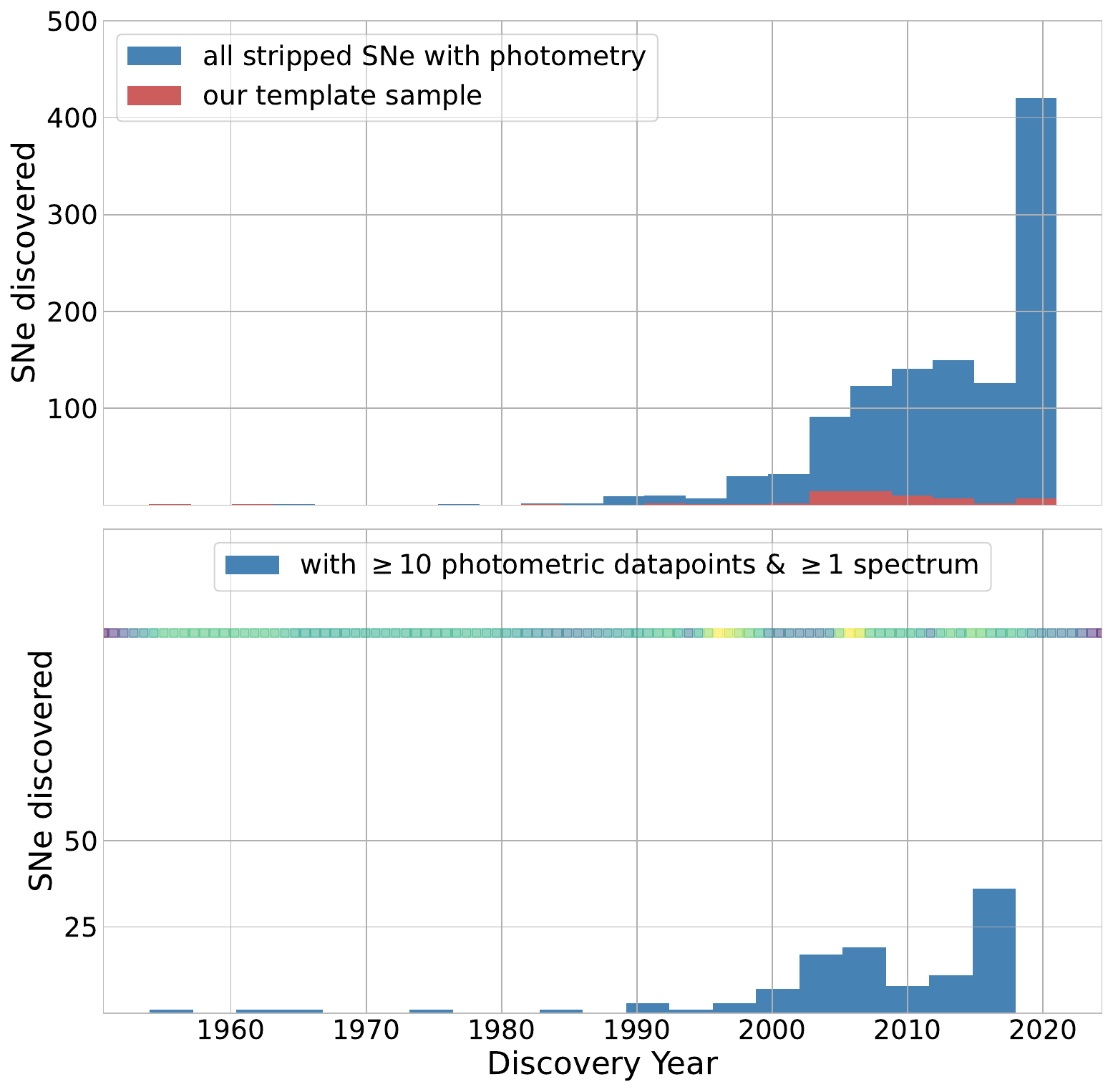}
 }
\caption{\SESN~discoveries since the first $\Hy$-poor SN, SN 1954A \citep{pietra55}. The top panel shows all \SESN e, including SNe~Ibc and peculiar subtypes (but excluding SLSNe) and the bottom panel shows only SNe for which $>10$ photometric data points have been published. The color band running at the top of the bottom panel represents the average number of photometric data points per SN, in log scale, and it runs from a minimum (purple) of 0 in years when no SNe were observed to a maximum (yellow) of 1838 data points in 1993, with the single best observed \SESN e: SN 1993J. The maximum number of discoveries with $>10$ photometric data points is reached in the five-year bin centered on 2004, but it is likely an artifact due to delayed publication: photometric observations often continue to appear in the literature in the decade following an object's discovery.}
\label{fig:SNdiscoveries}
\end{figure}

\autoref{fig:SNdiscoveries} shows the number
of \SESN e~discovered since the first hydrogen-poor SN Ib was
discovered in 1954 (SN~1954A, \citealt{pietra55}). 
The top panel
includes all SNe labeled as \emph{any} \SESN~subtype, including
general \Ibc\ classifications and
peculiar subtypes (\emph{e.g.}: Ib-Pec, Ibn, Ca-rich Ib, but no
He-poor super-luminous SNe - SLSN),\footnote{This selection  was initially
obtained by querying the Open SN Catalog with the constraint ``Ib, BL,
Ic'' used as input in the individual column search for the feature
``Type'' (1801 objects were retrieved with this query on August 27, 2021. No \SESN e data was added to the OSNC between this date and April 2022 when the catalog was frozen). The ``Ib/c'' classification is often used to indicate a lack of \Hy, or \Hy\ and \He , with a simultaneous lack of
\Si\ signatures in the spectrum, especially when a single
spectrum is available assuming that a moreconstraining identification requires observing the spectral evolution. It is
just slightly more constraining than the occasionally-seen classification
``non-Ia''. The community should move away from this classification,
since, as shown in \citet{liu16} the signatures of weak \Hy,
absent \Hy\ and absent \He, should be identifiable at all phases, allowing to separate subtypes of \SESN e with one spectrum (of sufficient quality).  SLSNe are
removed after downloading the data by searching for ``SLSN'' in the
Type column, as well as object type ``variable'', ``blazer'',
``microlensing'', and ``blue'', which are selected through these
search criteria due the ``Type'' string matching ``bl'' or
``ic''. Lastly, CasA is removed. CasA (SN1667A) is included in the
Open SN Catalog with a light echo spectrum, a single data point in
photometry, and with the discovery date set to the discovery of its radio
emission in 1948 \citep{ryle48}, but it should not be in the dataset
for the purpose of this statistics exploration. This leads to a dataset of 1194 \SESN e
with at least 1 photometric measurement.} while further into the
project we will limit our sample to well-identified and generally
non-peculiar subtypes. In the bottom panel we
show only SNe with at least 10 photometric data points and at least one
spectrum, as a first, crude cut to separate ``well-observed'' SNe
(308 \SESN e). The color bar in the bottom panel indicates the size of the best photometric sample of \SESN e discovered
per year.

We collected all photometry available in the \citetalias{guillochon16}
as described above, and had also contributed to OSNC by augmenting the catalog, which was an open-source catalog
and relied partially on the community data input. Two surveys contribute the bulk of the data that will be used in our work: CfA SN survey (\citetalias{bianco14}) and the Carnegie Supernova
Project (CSP~\citealt{stritzinger2018carnegie1,stritzinger2018carnegie2,stritzinger2018carnegie3}). 

The CfA SN survey contributed to the literature collection of
``well-observed'' \SESN e~(\ie, with multiple photometric bands, and
several spectra for each object to confirm classification and assess
evolution) by doubling the existing sample (\citetalias{modjaz14} and \citetalias{bianco14}), with 4,543
optical measurements on 61 SNe and 1,919 NIR measurements on 25 SNe.

The Carnegie Supernova
Project (\citetalias{stritzinger2018carnegie1}) released data for 34
supernovae, in optical and NIR bands
($u',~B,~V,~g',~r',~i',~Y,~H,~K_s$) with nearly 3,000 optical
data points and nearly 700 NIR data points. Of these, 10 SNe had no
previously published optical photometry.  Half of the \citetalias{stritzinger2018carnegie1} sample (18 out
of 34 SNe) already had published CfA photometry in \citetalias{bianco14}, and the
photometric measurements are generally in excellent agreement between
the two surveys, with generally comparable uncertainties. The
\citetalias{stritzinger2018carnegie1}  sample is a smaller dataset by number of objects compared to \citetalias{bianco14}
(34 objects compared to 62), although the size of the NIR samples is identical (25 SNe). Altogether, the surveys are
comparable in sampling, \citetalias{stritzinger2018carnegie1}  averaging 88 optical data points and 27 NIR
data points per SN, compared to 73 and 77 respectively for \citetalias{bianco14}.

\citetalias{drout11} published the first pioneering survey on \SESN e, including 25 objects
observed in \Vband\ and \Rband\ bands. However, as shown in \citetalias{bianco14}, this survey's
photometry can be contaminated by galaxy light due to \citetalias{drout11}'s extraction
method (aperture photometry, instead of template subtraction). While a
subset of the \citetalias{drout11} SNe is likely minimally contaminated, most of the
\citetalias{drout11} objects were also observed in the \citetalias{bianco14} and \citetalias{stritzinger2018carnegie1}  samples. Therefore,
without loss of representation, we can remove these photometric
data points from the sample we use to generate templates. 

\citet{taddia15} introduced a sample of 20 SESN in a study of early lightcurves behavior. 

Finally, the ZTF Bright Transient Survey, described in \citep{2020ApJ...904...35P} also observed and released light curves for SESN starting in 2018. Objects from this survey are included if they were integrated in the \citetalias{guillochon16}.
{Many additional recent publications include individual \SESN e and describe properties of the \SESN e as a family \citep[\eg][]{vincenzi2019spectrophotometric, schulze2021palomar} based on PTF, iPTF, and ZTF \citep{law2009palomar,kulkarni2013intermediate,bellm2018zwicky} data. However, these data are yet to be released as an ensemble. The individual SNe published in the aforementioned publications are typically available and in many cases have been ingested into the \citetalias{guillochon16}, in which cases they are included by default in our work.

In summary, the final photometric data sample used for our templates comprises data
from the large \SESN e observing programs by CfA (\citetalias{bianco14}) and \citetalias{stritzinger2018carnegie1},
individually published PTF/iPTF objects (\eg, iPTF13bvn, PTF11qcj, and PTF12hni), as well as
other single object campaigns that produced well-observed
 \SESN e (\eg, SN 1993J).   
{ Objects from PTF and iPTF also include:
iPTF15dld,
iPTF15dtg,
PTF10qts,
PTF11kmb,
PTF12gzk,
and
SN 2016hgs (iPTF16hgs) (See \autoref{tab:SESNessentials}).}

A collection of rapidly evolving \SESN e from the ZTF survey was recently published by \citetalias{ho2023photometric}. This is a sample selected specifically because of its fast-evolving photometric signature so it is inherently a biased sample. Therefore we do not include SESN type \Ib, \IIb, \Ic\ from this sample in the construction of our templates, rather we compare the photometric properties of this sample with our template (\autoref{sec:compare_ZTF}). We do, however, include \Ibn\ from \citetalias{ho2023photometric} in the construction of our \Ibn\ template as the expectation is that this subtype of SESNe is intrinsically rapidly evolving, so that the sample may be representative. These ZTF events are in fact included in the \citetalias{guillochon16} but do not appear to have spectra on the \citetalias{guillochon16} and therefore are automatically removed by our selection criteria (see \autoref{sec:photsel}).

Finally, in a few cases, we included \SESN e in our sample that had photometry missing from \citetalias{guillochon16} or
their photometry was published in non-standard formats. These objects include SN 2010as, OGLE16ekf\footnote{\url{http://ogle.astrouw.edu.pl/ogle4/transients/2017a/imagesSELECTED/OGLE16ekf.dat}}, iPTF15dld, PTF10qts, SN 1999dn, and SN 2002ji\footnote{\url{http://www.astrosurf.com/snweb/2002/02ji/02jiMeas.htm}} \citep{folatelli2014supernova, pian2017optical, walker2014optical, benetti2011type}. Also, DES16S1kt appears to have 1 photometric data point on our latest downloads from \citetalias{guillochon16} whereas we had photometric data for it from older versions on \citetalias{guillochon16} GitHub.

SN~2003lw has been removed from our sample because its light curves (obtained from \citealt{malesani09}) appeared to be extremely slow evolving changing (0.25 magnitudes between -15 and 70 days from \maxep) when our templates change by nearly 1.5 (between -20 and 71 days from \maxep) even though this SESN is considered in the literature to be a photometrically typical \blIc\ connected with GRB 032303 \citep{thomsen04}. We attempted to obtain photometric data from the authors but were unable and therefore we dropped this object from our sample and analysis.

Without claiming completeness, we believe we gathered the vast majority of
published photometric measurements of \SESN e~ up to April 2022, in the following
bands: 
\opticalbandsdetailed\ (optical), \NIRbands\ (NIR), and
\swiftbands\ (Swift UV). \footnote{We collected photometry based on photometric band names, without differentiating for example between the $r'$ Sloan and other \rband\ filter bandpasses and without applying filter conversions. This has the net effect of increasing the spread of photometry within a band, and thus the uncertainties in the templates, generally leading to more conservative conclusions wherever the impact of this choice is significant. Hereafter, we will use \ugri\ as generic names for filter bandpasses that in some cases may refer to the ``primed'' system \sloanugriprimed, or other variations. Similarly, most if not all of our \Kband\ band data is collected in the 2MASS \Kband\ filter (by the CfA or CSP surveys) but without loss of generality we will use the label $K$ in our figures and tables to refer to this band.} 

\subsection{Sample selection for template construction}\label{sec:photsel}

In the sample used to construct our templates, we include SNe with at least five photometric
data points in one band, with spectroscopic classification and at least one
published spectrum, and for which we are able to determine the epoch
of maximum \maxep. Note that
we refer to the $V$ band for the determination of the epoch of maximum
brightness, as in \citetalias{bianco14}.  Historically, for SNe~Ia the reference
band for maximum brightness is the $B$ band, where, especially after
$z$ correction, most objects are well sampled, and where the
sensitivity of photographic plates used prior to the advent of CCDs,
peaked. But for the fainter \SESN e~in our local sample, which do not
require any redshift correction (Section~\autoref{sec:selpreproc}), it
makes sense to choose the $V$ band as reference, as this filter is very
commonly used, and the SNe are brighter in $V$, thus most objects have
the best sampling in $V$, rather than $B$ band. However, the method described in \citetalias{bianco14} to measure \maxep\ can leverage photometry in bands other than $V$ by exploiting empirical relations observed in \SESN e between the evolution in different bands (\autoref{sec:selpreproc}).

In more detail, starting with the initial sample downloaded from \citetalias{guillochon16} (1194)\footnote{This selection is
obtained by querying the Open SN Catalog with the constraint ``Ib, BL,
Ic'' used as input in the individual column search for the feature
``Type''. The Ib/c classification is often used to indicate a lack of \Hy, or \Hy\ and \He , with a simultaneous lack of
Si signatures in the spectrum, especially in cases where a single
spectrum is available and a more constraining type identification is
difficult to obtain without observing the spectral evolution. It is
just more constraining than the occasionally-seen classification
non-Ia. The community should move away from this classification,
since, as shown in \citet{liu16} the signatures of weak \Hy,
absent \Hy\ and absent \He, should be identifiable at all phases.  SLSNe are
removed after downloading the data by searching for ``SLSN'' in the
Type column, as well as object type ``variable'', ``blazer'',
``microlensing'', and ``blue'', which are selected through these
search criteria due the ``Type'' string matching ``bl'' or
``ic''. Lastly, CasA is removed. CasA (SN1667A) is included in the
Open SN Catalog with a light echo spectrum, a single data point in
photometry, and with the discovery date set to the discovery of its radio
emission in 1948 \citep{ryle48}, but it should not be in the dataset
for the purpose of this statistics exploration (1194 \SESN e
with at least 1 photometric measurement as retrieved on August 27, 2021. No SESN data was added to the OSNC between this date and April 2022 when the catalog was frozen).}, we select all objects that satisfy the following selection criteria:

\begin{enumerate}
\item have at least one spectrum for classification, which leads to a classification that we judge reliable. That is: we find consensus in multiple literature sources about the classification or we  performed our own classification using the SNID~\citep{blondin07} classification code with the augmented \SESN\ SNID spectral template library published by the SNYU group \citep{liu14, modjaz2016spectral, liu16, liu17} to ensure the classification is correct (860 SNe);

\item have at least 5 photometric data points (upper limits not included) in at least one of the bands: \allbands\ (221 SNe.\footnote{Photometry from D11 is generally not included due to contamination from host galaxy and being superseded by \citetalias{bianco14}} We include only SN2004ge from D11.);

\item \maxep~ can be determined from the $V, B, ~r/R~$, or $i/I$ photometry following the re-sampling method described in \citetalias{bianco14} (128 SNe). 

\end{enumerate}

To these data, we add: 
\begin{itemize}
\item 7 SNe with photometry from other sources mentioned in \autoref{sec:data}, which we later added to the OSNC via pull requests (135 SNe);

\item 6 \Ibns\ from \citetalias{ho2023photometric} that were rejected in step 1, and for which we retrieve forced photometry from ZTF to extend the available time baseline (141 SNe, this sample is discussed in detail in \autoref{sec:compare_ZTF});

\item 7 SNe from \citet{sako14}. We did not inspect the spectra directly but used the SDSSII classification (148 SNe);

\item 17 SNe are in our list for which spectra were not originally available on the OSNC but were found in the literature or in other open repositories (\eg, \citealt{Yaron_2012}), which we later added to the OSNC via pull requests (a total of \Nlit\ SNe).

\end{itemize}

\subsubsection{Spectral classification}

Where the \citetalias{guillochon16} offered multiple classifications, we ran SNID with the largest database of SESN templates from \citet{liu14, modjaz2016spectral, liu16, liu17} to check the claimed \citetalias{guillochon16} classification. We ran SNID on their publicly available spectra closest to maximum light (and thus cannot comment on any type change that may have happened after maximum light). We revised the classifications for the following SNe (as reported also in \autoref{tab:SESNessentials}):
SN~1962L (changed from \Ic\ to \Ibc),
SN~1976B (removed from the sample since the noisy spectrum did not lead to a conclusive classification),
SN~1985F (for which the only spectrum is nebular and it does not allow us to specify a  type beyond a generic \Ibc\ classification),
SN~2005fk (confirmed as \blIc),
SN~2006el (confirmed as \IIb),
SN~2006fe (removed from the sample because all photometric data points are upper limits, the available spectra appear heavily contaminated by emission lines and the classification remains very uncertain),
SN~2006nx (changed from \Ic\ to \blIc),
SN~2007ms (changed from \Ic\ to \Ic/Ic-bl),
SN~2007nc (confirmed as \Ib),
SN~2007qx (removed from the sample since the spectral classification based on noisy spectra matches \Ib's, \Ic's, as well as \Ia's),
SN~2012cd (confirmed as \IIb),
SDSS-II~6520 (removed because of the poor quality of the spectrum),
SDSS-II~14475 (confirmed as \blIc),
OGLE-2013-SN-091 (confirmed as \Ic),
OGLE-2013-SN-134 (confirmed as \Ic),
ASASSN-14dq (which has only a spectrum at early times, and seems to show a plateau in the light curve, so we removed it from the sample as it may be a type SN II),
OGLE15xx (removed from our sample as we re-classified it as a SLSN Ic).



\subsubsection{\maxep\ determination}\label{sec:selpreproc}

\begin{figure*}
 \centering
 \begin{minipage}{0.45\linewidth}
\includegraphics[height=4inch]{ 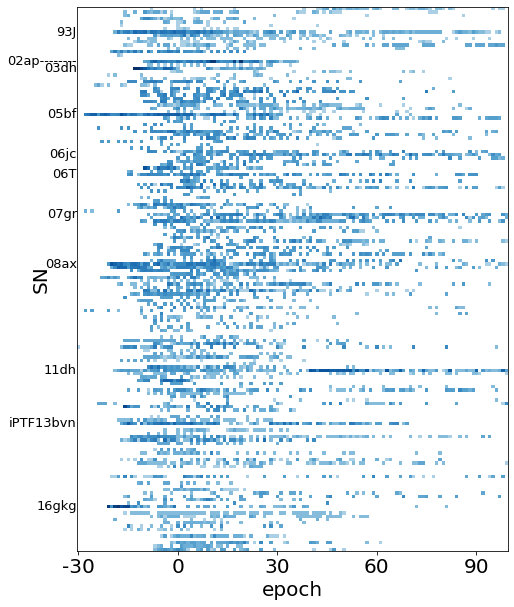}
\end{minipage}
\begin{minipage}{0.45\linewidth}
\includegraphics[height=4inch]{ 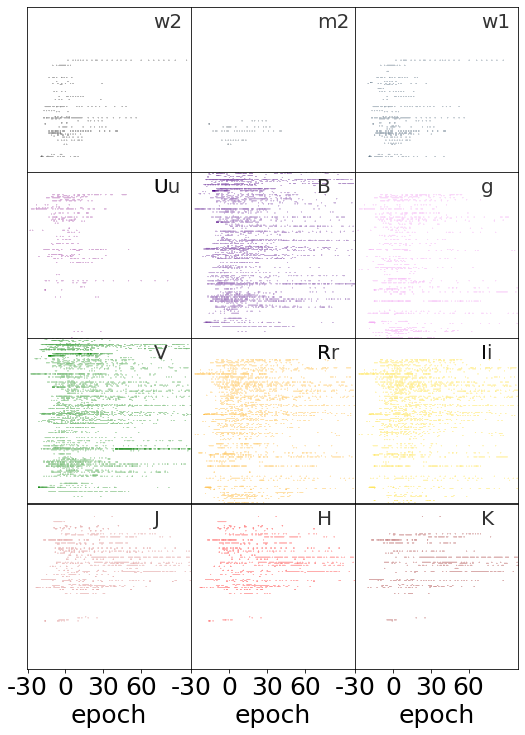}
\end{minipage}

\includegraphics[width=0.7\linewidth]{ 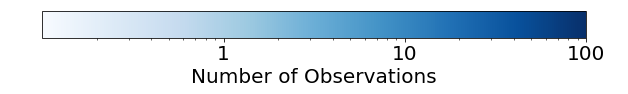}



\caption{\emph{Left}: Photometric coverage, in any band, for all 165 light curves in our sample, out to 100 days past peak. The name of SNe with more than 500 observations are explicitly indicated. As expected, most coverage is seen near peak, although a few light curves have consistently dense coverage throughout (\eg, SN 2007gr, SN~2011dh, SN~2006jc).
\emph{Right}: Photometric coverage for all 165 light curves in our sample in each of the considered photometric bands: \allbands, as labeled. $Uu$, $Rr$ and $Ii$ are plotted together. $B, V, Rr, Ii$ have the most coverage. The color bar reflects the mapping of color saturation to number for all subplots (illustratively shown in blue but consistent for all panels).}


 
\label{fig:coverage}
\end{figure*}

It is necessary to have a self-consistent definition and determination of the epoch of maximum to properly align the light curves when creating templates. Thus, in order to
retain an object in our sample we require knowledge of the epoch of
maximum brightness in \Vband\ band (\maxep), which we determine as
described in \citetalias{bianco14} and \autoref{sec:IBCtemplates}, either from the  \Vband\ band itself, or from $B,~R/r'$,
or $I/i'$ photometry. As an example, SN~2010S is removed from the sample because \maxep~ cannot be determined in any of these bands.
 
 We look for a primary peak in the light curve in  \Vband\, $R/r'$, $I/i'$ or  \Bband.
The epoch of maximum \maxep\ and the observed peak
brightness and their uncertainties are obtained through a Monte Carlo
method as described in \citetalias{bianco14}: a time region around maximum deemed by visual inspection suitable for fitting
second-degree polynomial is chosen. Data in this
region are re-sampled over the observed uncertainties and 
$n_\mathrm{edge} \leq3$ data points at each edge of the chosen region are
added with $n_\mathrm{edge}$ chosen by a stochastic draw for each realization.  This bootstrapping method allows the determination of the epoch of maximum and its confidence interval. 

For each band, if we do not have a direct measure of the \maxep\ from the light curve in that band we derive it from the peak location in other bands, using peak offsets measured in \citetalias{bianco14} (Table 1, as calculated from the
CfA data). To extend the procedure to bands not included in \citetalias{bianco14} ($g$ and UV bands \swiftbands), we fit the values in \citetalias{bianco14} table 1 as a function of the filter's effective wavelength ($\lambda_\mathrm{eff}$ in $\AA$)
with a second-degree polynomial,  which 
gives us the equation:
\begin{eqnarray}
\mathrm{Peak}_V - \mathrm{Peak}_{\lambda_\mathrm{eff}} [{\rm days}] = &-& 4.5_{-5.5}^{-3.4}\times10^{-8}\lambda_\mathrm{eff}^2\\
&+& 1.9^{2.2}_{1.6}\times10^{-3}\lambda_\mathrm{eff}\\
&-& 9.5^{-8.2}_{-11}~[{\rm days}].
\end{eqnarray}
Note that this procedure leads to an \emph{inter}polation for  \gband\ band, but an \emph{extra}polation in UV bands, with decreased reliability.
\autoref{tab:SESNessentials} contains the list of the \SESN e in our sample along with their subtypes, time of \maxep, error in \maxep, and redshift.

\begin{figure*}[ht!]
 \includegraphics[scale=0.4]{ 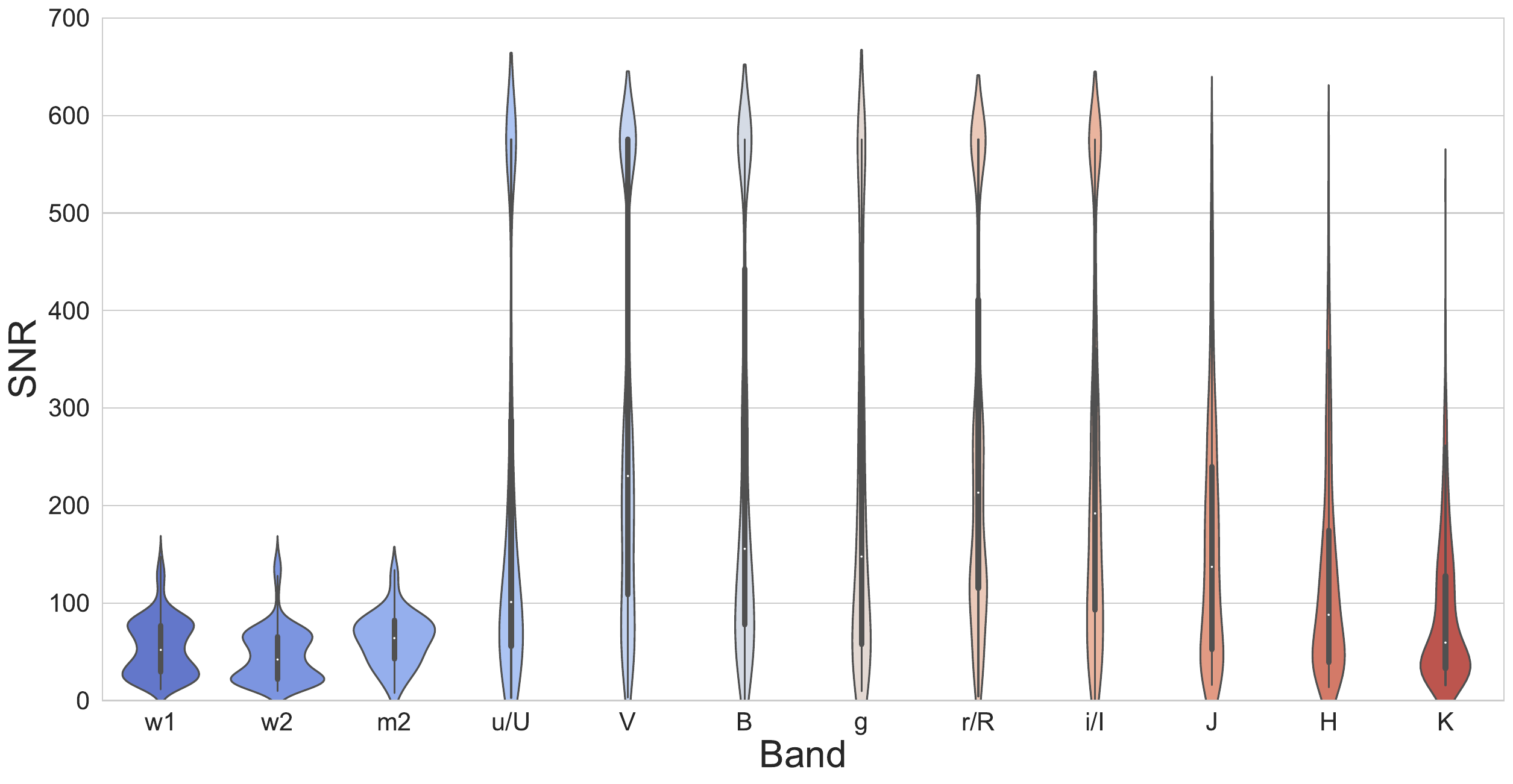}
\caption{Distribution of the \SNR\ of the photometry used in this analysis, different photometric bands shown along the $x$-axis. For each band, the \SNR\ distribution is shown as a violin plot where the distribution is smoothed via Kernel Density Estimation and the median, inter-quartile-range (IQR), and full extent of the distribution are shown by a white dot, thick gray bar, and thin gray line respectively. While our photometric templates will be generated separately for each band, for the purpose of this plot the \SNR\ measurements for $u'/U$, $r'/R$ and $i'/I$ are considered together. As expected, optical bands, and in particular  $V$, $B$, and $g$, have a higher median \SNR\ (101, 230, 156, 148, 213, and 192 for $u'/U$, $V$, $B$, $g$, $r'/R$, and $i'/I$ respectively) than NIR and UV bands (median \SNR\ = 88, 60, 52, 42, 64 for $H$ and $K_s$, $w1$, $w2$, $m2$, and median \SNR\ of 137 for $J$) with the upper IQRs stretching out to $\SNR>400$ for optical photometry. }
\label{fig:snr_plot}
\end{figure*}

We removed all upper limits, along with extremely late or extremely early phases and only included photometric measurements over a time range of -20 days $\leq$ \maxep $\leq$ 100 days. For example, SN~2008D and SN~2002ap had measurements later than 3 years after \maxep, and OGLE targets typically have upper limit photometry available for one hundred days before the SN explosion. 
In 
\autoref{app:appB} we summarize in three additional tables the photometric features of our sample including the number of photometry measurements for each SN, the first and last photometric epoch (in days from \maxep), and the mean and standard deviation of the photometric sample. 

When we create the individual templates for each subtype (\alltypes) we select from this sample the non-peculiar SNe of that sub-type and further reject light curves whose photometry does not lead to a good GP interpolation (as we will describe in \autoref{sec:gptempaltes}).

\subsubsection{Final SESN sample}

At the end of this process, we assembled a sample of \Nlit\ SNe for constructing \SESN\ templates that include all \SESN~subtypes \IIbs\ (34), \Ibs\ (38), \Ics\ (33), \blIcs\ (25), \Ibs\ (14), Ca-rich SNe Ib/Ic (3), peculiars (Ib-pec, Ic-pec, 5), Ib/c (14, including one \IIb/Ib, one \Ib/c-bl and one \Ic/c-bl). \autoref{fig:coverage} shows the coverage in any band, and in each band separately, for our sample, limited to $-30 \leq $\maxep$ \le 100$ days.

\begin{center}\begin{small}
\input{photsample}

\end{small}
\end{center}

\autoref{fig:snr_plot} shows the distribution of the signal-to-noise ratio (\SNR) of the photometry in our sample in different bands (where the \SNR\ is the ratio of flux to flux error as derived from the magnitudes). Some extremely high \SNR\ measurements are available, especially in the optical bands where the distribution of \SNR\ appears bimodal, with the median $\SNR _{u/U}\sim100$ in $u/U$ band, and median $\SNR _{V,~r/R}>200$ in $V$ and $r/R$ bands, and a secondary distribution peak  $500 \leq \SNR \leq600$ in all optical bands, while, as expected, we see that the NIR bands have lower \SNR\ compared to the optical bands with median  $50\leq \SNR _{\NIRbands}\leq100$ and lower yet in the UV bands (median $\SNR _{\swiftbands}<150$).
Our photometric sample has diverse sampling, diverse noise, and generally a smooth behavior, with greater diversity at early times, compared to later phases (after the \synNi\ decay starts dominating). Three examples of time series in our sample are shown in \autoref{fig:lcvs}.  

\begin{figure}
 \centerline{
 \includegraphics[width=0.63\columnwidth]{ 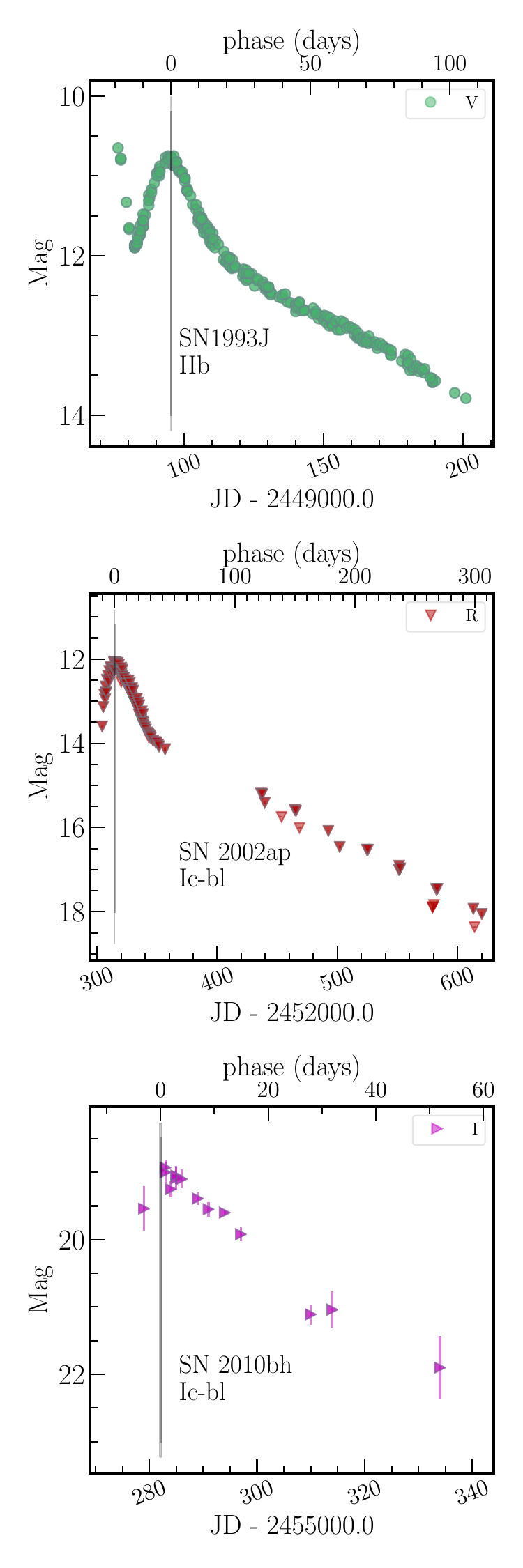}
 }
\caption{Examples of SN light curves, from heterogeneous photometric sources. 
For each SN the date of V band maximum \maxep\ is marked by a vertical black line, and its $1\sigma$ uncertainty is marked as a gray region (to enhance visibility, since the uncertainty is typically very small, the $1\sigma$ uncertainty region extends further above and below the \maxep\ line). SN~1993J $V$-band (top) is very well-sampled, possibly with underestimated uncertainties. SN~2002ap $R$-band (center) is well-sampled around the peak but it shows large gaps later. SN~2010bh $I$-band is a sparsely sampled light curve with uncertainties conveying the scatter properly (bottom). Many publications are dedicated to each of these supernovae: for SN~1993J the first articles appeared in 1993, \citealt{1993PASJ...45L..63O}, and the most recent at the time of writing is \citealt{2022MNRAS.509.3235Z}. For SN~2002ap  \citealt{2002MNRAS.332L..73G} and recent modeling work include \citealt{2017ApJ...842..125Z}. For SN~2010bh see  work including \citealt{2011ApJ...726...32F} and \citealt{2012ApJ...753...67B}.
}

\label{fig:lcvs}
\end{figure}

\subsection{Photometric corrections}

\begin{figure}[ht!]
 \centerline{
 \includegraphics[width=1\columnwidth]{ 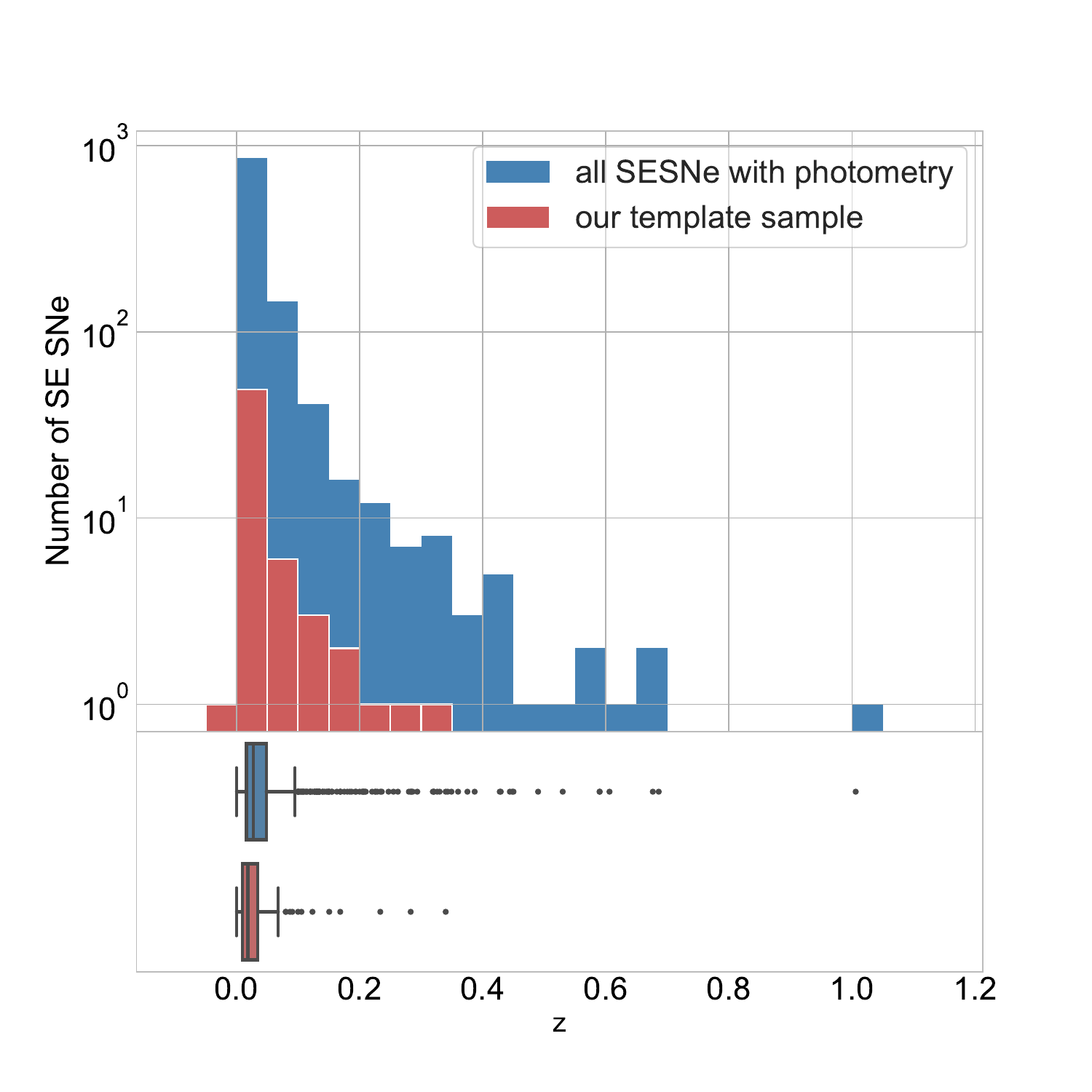}
 }
\caption{Redshift $z$ distribution for all \SESN e~in the literature, as collected by the Open SN Catalog (\citetalias{guillochon16}), and for the subsample we will use for template construction. Top: a histogram of the distributions in 10 bins. Bottom: box and whiskers plot of the distributions. A box and whiskers plot shows the median of the distribution as a vertical line, the 25-th percentiles as a colored box (\emph{i.e:} the inter-quartile range, IQR), and the bars, joined to each end of the box by a line, represent the minimum and maximum of the distribution \emph{excluding statistical outliers}, where outliers are defined as any point farther than 1.5$\times$IQR from the edges for the IQR. The ``outliers'' thus defined are shown as data points. None of the SNe in the full literature sample exceeding $z=0.4$ is in the selected ``well-observed'' sample which is used for template construction. At a  $z=0.4$ the spectra are shifted by the typical width of an optical photometry band, so that the $r'$ rest-band, for example, would be observed in the $i'$ filter. The template sample has a median of 0.017 and a 99.7 percentile (equivalent to a $3-\sigma$ in a Gaussian distribution) of 0.31. Notice that the low-$z$ edge of the distribution is below zero: SN~1993J and SN~2004gk, extremely near-by objects, have negative recession velocities ($z~ =~ -1.13 \times 10^{-4}$ and $z~ =~ -5.50 \times 10^{-4}$ respectively.)}
\label{fig:zdistrib}
\end{figure} 

We do not apply any $k-$correction to the data, because the vast majority of the
well-observed \SESN e are in the local Universe. We
show the redshift $z$ distribution for the complete \SESN e sample of
literature data from the \citetalias{guillochon16} and of
the ultimate selection of SN that we will use to construct templates
in \autoref{fig:zdistrib}, as a histogram in
10 bins in the top panel, and as a box-and-whiskers plot in the bottom panel. The
16-th, 50-th (median), and 84-th percentiles of the $z$ distribution
are 0.01, 0.02, 0.08 and 0.00, 0.02, 0.06 for the full sample and the
subsample selected for photometry (described in \autoref{sec:photsel}) respectively, requiring a median wavelength
shift $<=3\%$ in the SED for our sample.
There are no SNe in our template subsample that exceeds $z=0.4$, which
is the redshift value at which a spectrum is shifted by roughly a
full optical band. Only three SNe exceed $z=0.2$: SN~2012bz \citep{schulze14}, and
SN~2013cq \citep{melandri2014diversity}, both \blIcs\ connected with GRBs and first discovered in the
$\gamma$-rays, and SN~2005fk (\blIc). $k$-correction requires an understanding of the full spectral
evolution, including the effects of reddening, in order to interpolate
between bands. The reddening in particular is extremely uncertain
due to the intrinsic diversity of \SESN e and to the difficulties in
determining the reddening for objects that are not standardizable. So
in order to not introduce additional errors, we refrain from
applying $k$-corrections.

%
%

\section{Combined SN I\lowercase{bc} Templates}
\label{sec:IBCtemplates}

\begin{figure}
  \begin{center} \includegraphics[width=0.95\columnwidth]{ 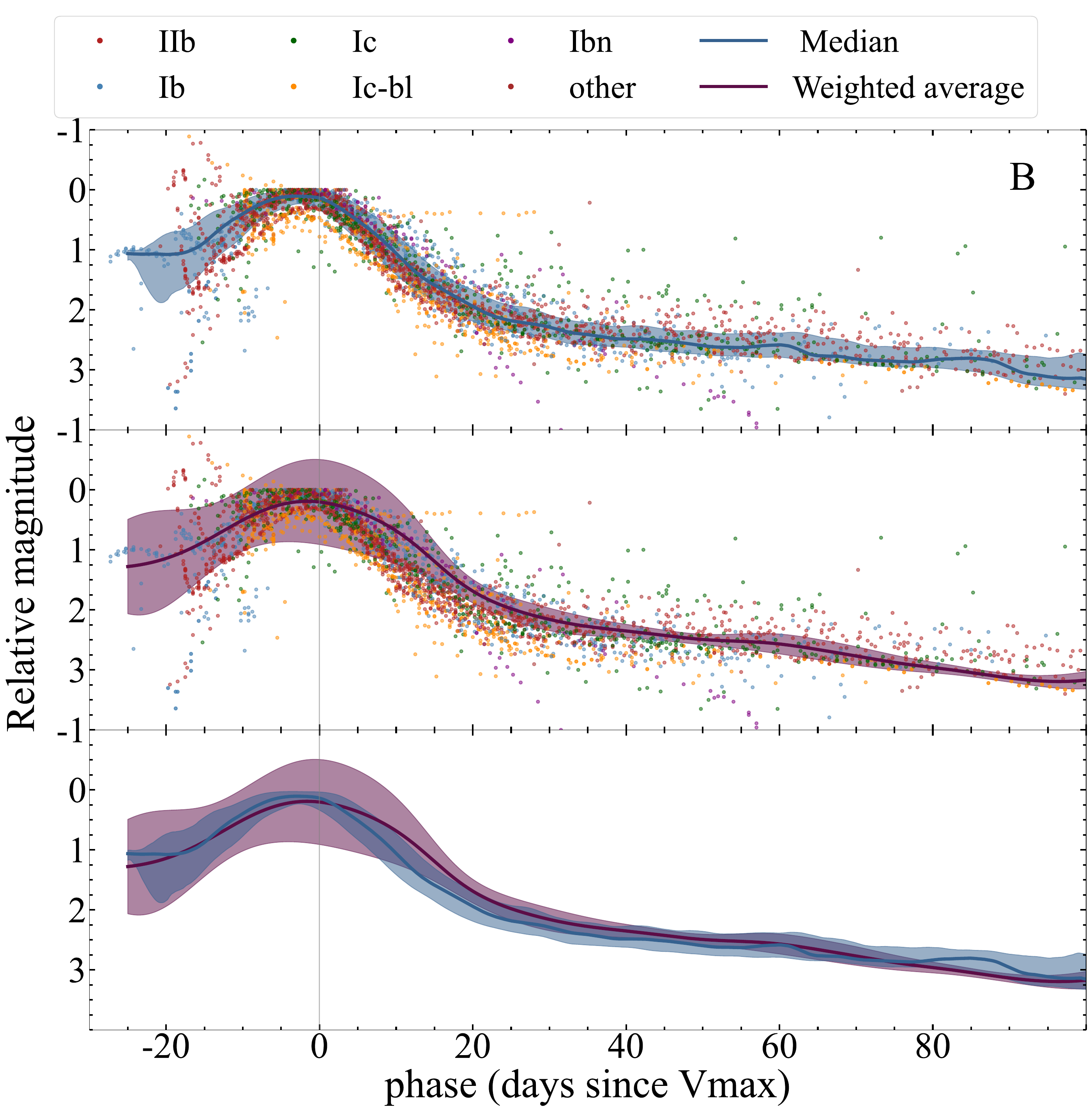} 
  \caption{{\it Top panel:} The B-band median template for all \SESN e generated as described in \autoref{sec:IBCtemplates} is shown as a blue line along with the IQR (shaded region). Datapoints for all light curves used in the generation of the template are shown, labeled and colored by \SESN\ subtype.
  {\it Middle panel:} Same as the top panel for the weighted-average templates and their weighted standard deviation uncertainty region (IQR). {\it Bottom panel:} The median template (blue line) and the weighted-average template (purple line) are shown together to aid comparison, along with their respective uncertainty as shaded red and grey areas respectively. The weighted-average template has larger uncertainties and is typically brighter due to a photometric bias introduced by the use of the photometric uncertainty to weigh the data points. We ultimately select the rolling median procedure to generate \Ibc\ templates.}  \label{fig:B_compare_temshownplates} \end{center}
\end{figure}

We begin by creating a preliminary template for \emph{all} \SESN e in each band, without differentiating by subtype. Following the literature, we call these templates 
``\Ibc\ templates''. We wish to emphasize that this is a diverse class of transients. If we can identify distinct photometric behaviors, lumping all of them in one class will have a negative impact on our ability to classify transients photometrically. This is further addressed in \autoref{sec:plasticc_compare}. However, we need to create \Ibc\ templates to represent the average time-dependent behavior of \SESN e in each of the bands to then subtract this template from the individual SN light curves and enable the GP fits (refer to \autoref{sec:gptempaltes}).

The construction of \Ibc\ templates as a mean or median of the photometric measurements is a relatively standard procedure, done first in ~\citet{drout11} and later in ~\citet{taddia2018carnegie} (both of these templates will be compared with ours in \autoref{sec:compare_Ibc}). We now can do this in optical and NIR bands where we have several light curves, amounting to many data points for phases between -20 and 100 days. 


First, we align the light curves in each band by their date of maximum
brightness, with care to ensure we are indeed looking at the maximum brightness
due to nucleosynthesis processes (rather than peaks that may occur due, for example,
to shock breakout or interaction with interstellar material).

\begin{figure*}
  \begin{center}
  \includegraphics[width=0.9\textwidth]{ 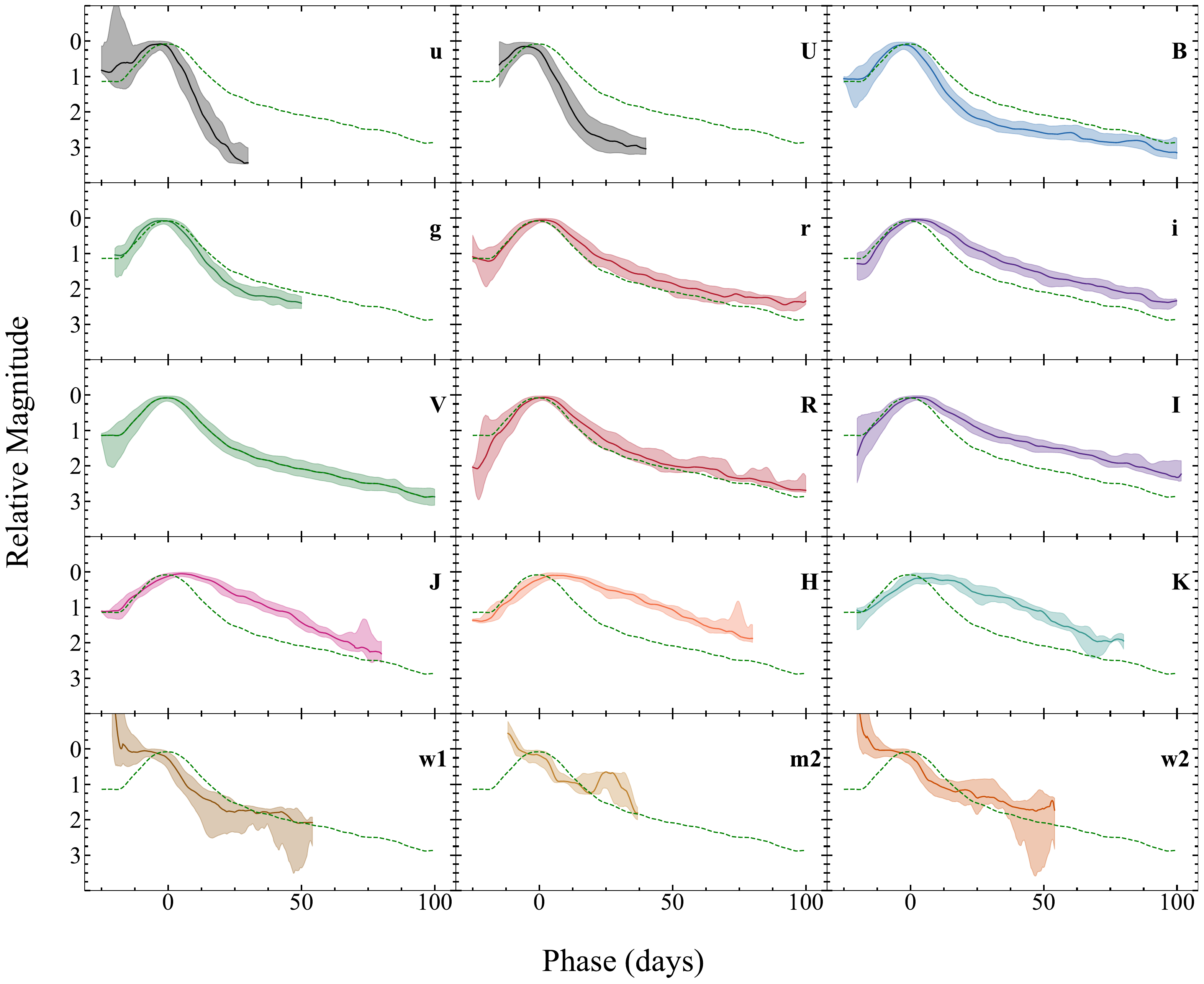} 
  \caption{The data-driven SN Ibc templates produced as a rolling median of all photometric data points available in each band as described in \autoref{sec:IBCtemplates}. The number of SNe used to construct each template is: \Uband: 39, \uband: 20, \Bband: 75, \Vband: 80, \Rband: 41, \rband: 57, \Iband: 27, \iband: 55, \Jband: 23, \Hband: 22, \Kband: 18, $w1$: 15, $w2$: 15, and $m2$: 9. The \Ibc\ median template and its IQR is shown in each band as a solid line and shaded region. For reference, the \Vband\ band template is plotted in each panel (as a green dashed line).}
  \label{fig:ubtcompare}
  \end{center}
\end{figure*}

Next, with the light curves now in phase space with peak brightness corresponding to phase = 0, we scale the observed magnitude by setting to 0
the magnitude of the brightest data point 
between phases -5 and 5 days. We remove light curves that have no data points in that range. 
Scaling the light curves' brightness to an observed data point, rather than to the maximum found by fitting a smooth interpolation to the data, adds some scatter in the templates' photometry but it prevents us from
making model-driven assumptions. The additional scatter generated by small misalignments, in fact, makes our estimate of the template
uncertainties conservative. However, in the presence of shock breakout or cooling envelope signatures, the brightest point near \maxep\ may not be associated with the \synNi-driven evolution of the SN, thus we reject these final adjustments where the visual inspection suggests the presence of such contamination, and scale those light curves to $m($\maxep$) =0$. The light curve of SN2013df (\IIb) in the $R$ band is an example of such contamination.

\begin{figure*}
  \begin{center} \includegraphics[width=0.8\textwidth]{ 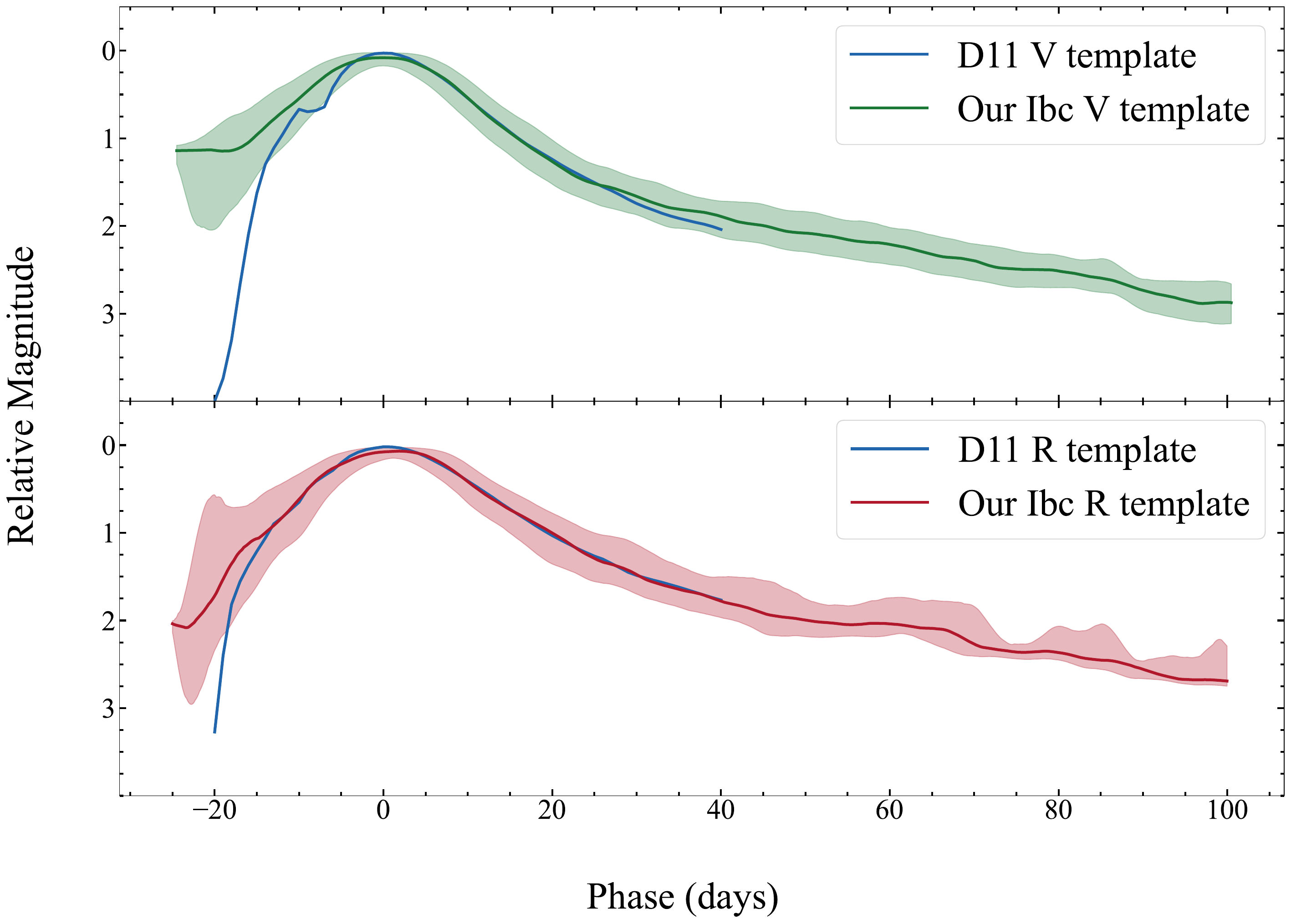} \caption{Comparison of our templates with the templates published in \citetalias{drout11}. Our \Ibc\ templates are plotted as a solid line in color and the \citetalias{drout11} templates are plotted as a solid blue line.  \citetalias{drout11}'s \SESN e templates are available for \Vband\ and \Rband\ band \url{https://www.cfa.harvard.edu/~mdrout/data/V_template.txt} and \url{https://www.cfa.harvard.edu/~mdrout/data/R_template.txt}, accessed on February 12th, 2021. \citetalias{drout11}'s photometry suffers from galaxy contamination (see \citetalias{bianco14}) but the templates are mostly based on well-sampled literature light curves and therefore the contamination is likely not significant. The post maximum light curve decay is consistent with what we measure, however, the light curve rise is far steeper for \citetalias{drout11}. Notice that the uncertainty on \citetalias{drout11}'s templates, plotted in their figures 2, 3, and 5 of \citetalias{drout11}, is not reproduced in our figure since it is not available online. Nonetheless, since this uncertainty simply represents the standard deviation of the sample and the early time the template behavior is based on 1-to-a few light curves, the uncertainty as plotted in \citetalias{drout11} goes to 0 in the earliest epochs and does not help reconcile the pre-maximum discrepancy between our template and \citetalias{drout11} templates.}  \label{fig:ubtcompareD11} \end{center}
\end{figure*}

\begin{figure*}
  \begin{center} \includegraphics[width=0.8\textwidth]{ 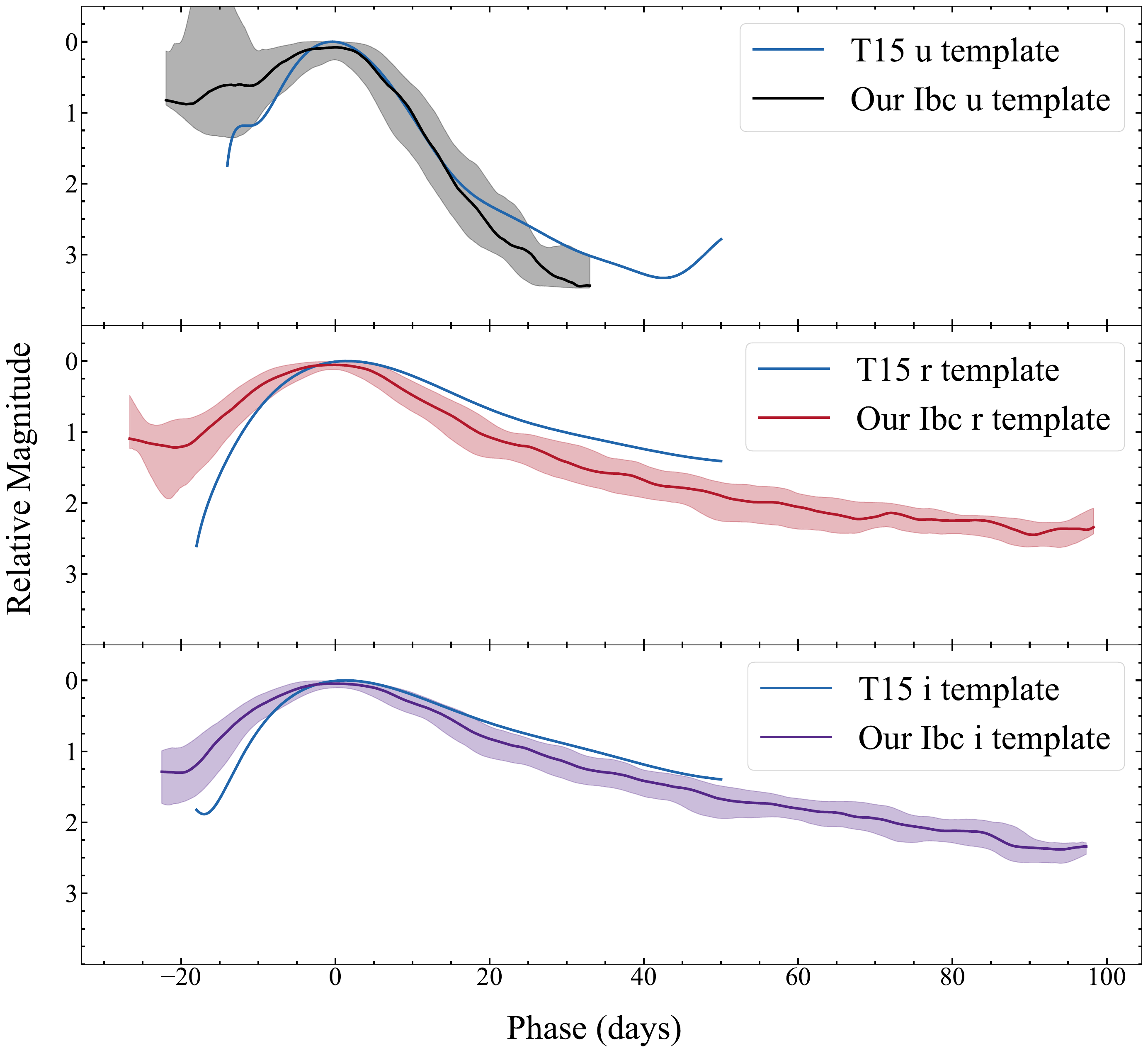} \caption{Comparison of our templates with the templates published in \citetalias{taddia15} for the common bands $u',~r',~i'$, accessed at Vizier catalog J/A+A/574/A60/template on February 12th, 2021. The rise portion of the templates, which was the focus of \citetalias{taddia15}, is steeper than our rise, and the decay is slightly shallower in all bands. In $u'$ and $i'$,~\citetalias{taddia15} templates seem to have weak signatures of shock-breakout, but they seem to be dimmer than the lower range of the uncertainties of our templates. } \label{fig:ubtcomparet15} \end{center}
\end{figure*}
We can now proceed to aggregate the light curves into
templates. We produced templates with two different methods: First, we create a smoothed, rolling \emph{weighted-average template}.  In order to
account for the heteroscedastic nature of the photometric measurement
uncertainties, especially when observing a diverse sample of SNe with
diverse brightness characteristics and with different instrumentation,
we use, as customary, averages weighted by the inverse of the squared
uncertainty of each measurement.

Second, compute a smoothed, rolling \emph{ median template}, with a 5-day rolling
window. This rolling median is noisy, getting noisier
at late and early times where fewer SNe are observed, with the IQR
artificially shrinking where the number of SNe in the window drops to only a
few. The rolling median template is then smoothed using the Savitzy-Golay filter \citep{press1990savitzky} (see details below).

\autoref{fig:B_compare_temshownplates} shows a comparison of the \emph{ median} and  \emph{ weighted-average} templates for band $B$. The weighted average template tends to be brighter than the median template. This effect is even more extreme in other bands. This is simply because of an observational bias: the brighter data points generally have larger weights (smaller uncertainties). While the SNe themselves have a range of brightness and are observed by systems with different accuracy, in the aggregate, the slow-evolving light curves have smaller uncertainty in the late measurements which causes them to dominate the late time average templates, introducing a bias. For this reason, we use the smoothed rolling median as \Ibc\ templates for the rest of the analysis.

Our final \Ibc\ templates are shown for each band in \autoref{fig:ubtcompare} 
 as a
gray line, with the IQR shown as a pale-filled region. The steps taken to generate these \Ibc\ templates are as follows:

\begin{itemize}
\item[1.] Phases are defined as an array between -20 and 100 with 24 points within each day for each hour.
\item[2.] At each hour, data points from all of the selected SNe between that hour and its next five days are selected.
\item[3.] The medians and IQRs are calculated within each 5-day window.
\item[4.] The median is then smoothed using a Savitzy-Golay filter within windows of 171 data points using a third-degree polynomial.
\end{itemize}

In some bands, towards the beginning and end of the light curves where data is sparse, the rolling median shows sudden jumps that could not be smoothed by our procedure. To avoid these undesirable features we restrict the range of phases over which we produce the templates to whatever is appropriate in that band. These phase boundaries are shown in \autoref{tab:boundaries} for each band. The data is insufficient to produce effective and accurate templates in the ultraviolet bands. The $w1, w2$, and $m2$ templates are included for completeness, but we caution the reader that these templates are produced from small datasets (see \autoref{fig:ubtcompare}).

\begin{table}
\begin{center}

\caption{Phase Range for Ibc templates in each band.}
\centering
\begin{tabular}{ |c|c|c|c| } 
 \hline
 Band & Phase Range & Band & Phase Range \\ 
  & (days) &  & (days) \\ 
  \hline
 u & -25,30 & &\\
 U & -15,40 & w2 & -20,50  \\
 B & -25,100 & m2 & -20,40\\
 g & -20,50 &  w1 & -20,50\\
 V & -25,100 &  &\\
 r & -25,100 &J & -25,100\\
 R & -25,100 &  H & -25,100 \\
 i & -20,100&K & -25,100\\
 I & -20,100 &&\\
 
 \hline
\end{tabular}
 \label{tab:boundaries}
\end{center}
\end{table}

\subsection{Comparison of the Ibc templates}
\label{sec:compare_Ibc}

\begin{figure*}
  \begin{center}
  \includegraphics[width=0.95\textwidth]{ 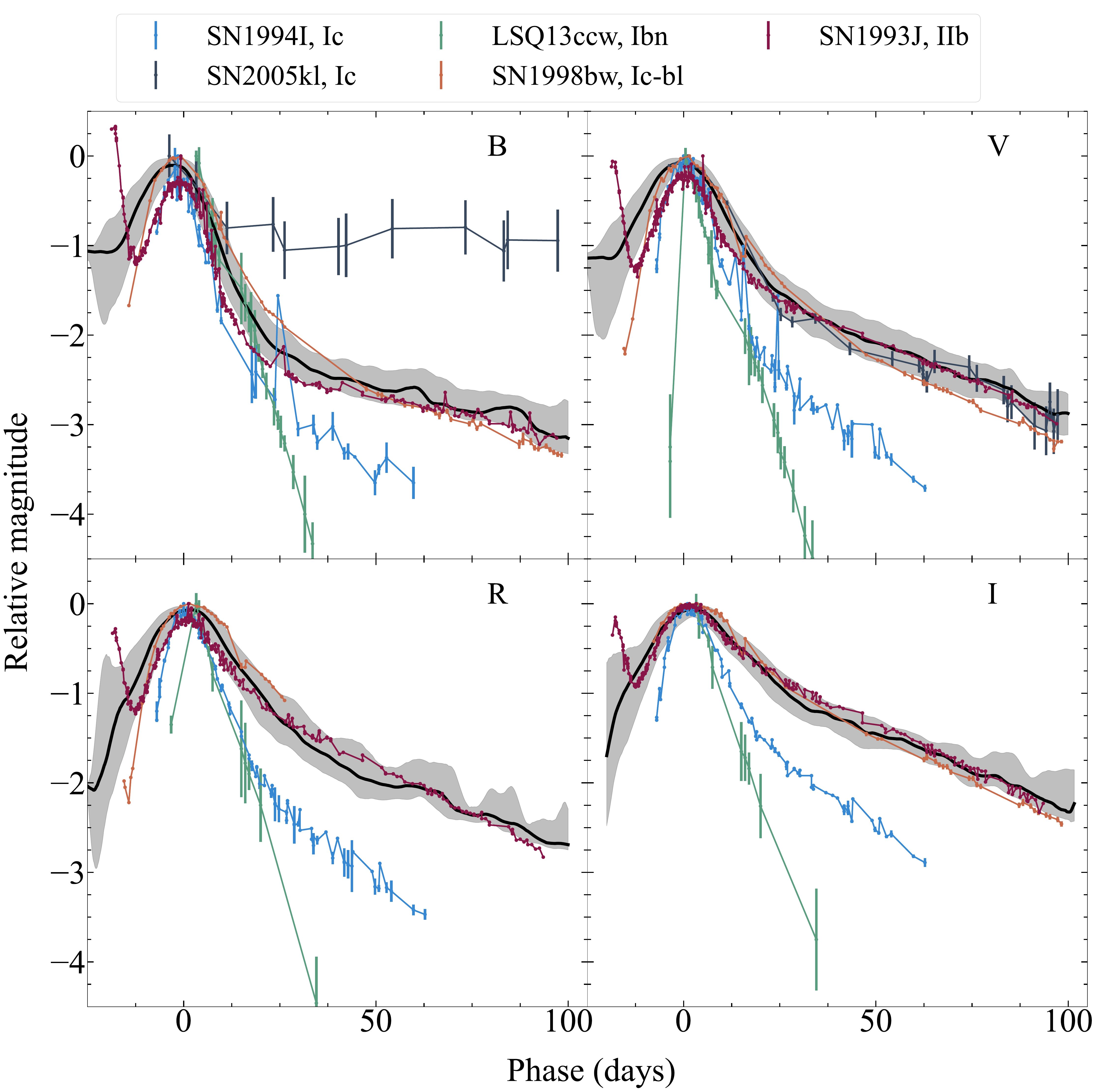}
  \caption{The \Ibc\ templates made of all photometric data points available for all \SESN e subtypes (black curves) in each of the bands \Bband, \Vband, \Rband, \Iband. These templates are plotted along with light curves of those SNe that have been claimed as prototypical of their class. For example, SN1994I and SN2005kl have been claimed to be prototypical \Ic, but clearly deviate from our template.}
  \label{fig:ubtcompareSNe}
  \end{center}
\end{figure*}

{In \autoref{fig:ubtcompare} we show
our \SESN\ Ibc templates for all the available photometric bands: \allbands. Although we
do calculate and plot the templates for UV Swift
\swiftbands\ bands, the sparsity and
low \SNR\ of the data in these bands only allow us to generate very noisy templates, which we show for completeness. The error bands represent the
epoch-by-epoch IQR, and the \Ibc\ template in \Vband\ band is plotted for reference, as a dashed green line in each panel. We can immediately notice that the light curves decline faster in bluer bands. }
We quantify the rise and decline rate of the templates with the parameters $\Delta m_{-10}$ and $\Delta m_{15}$. $\Delta m_{15}$ was first introduced by \citet{phillips1993absolute} to measure the evolutionary rate of \Ias\ (leading to their standardization).  This parameter measures the magnitude change from \maxep\ to ${\rm phase} =\maxep+15$~days, and $\Delta m_{-10}$, similarly, measures the rate of change between ${\rm phase} =\maxep-10$~days and \maxep. The measured $\Delta m$'s for our \Ibc\ templates are shown in \autoref{tab:del_m15} and \autoref{tab:del_m10}. The uncertainty on these quantities (also reported in \autoref{tab:del_m15} and \autoref{tab:del_m10}) is simply the $\Delta m$ measured at the edge of the template's IQR.


\input{  del_m_15_10_data2}


Next, we compare our \Ibc\ template with \Ibc\ templates previously released in the literature. The comparison with
~\citet{drout11} for \Vband\ and \Rband\ bands, shown in \autoref{fig:ubtcompareD11}. The light curve decline is consistent
in both bands with our \Ibc\ templates within the IQR. However, the early light curve
behavior, the hardest to model typically due to the rarity of SESN observations in the early phases of evolution, differs significantly, especially in the \Vband\ band. In addition,
the uncertainty in the \citetalias{drout11} templates artificially shrinks at early
epochs, due to sparse observations. Meanwhile, in our templates, the
intrinsic diversity in behavior at early times is properly represented
by the growing uncertainty. Early bumps in light curves are explained by shock-breakout
(for \blIcs) and shock cooling \citep[see for example][and discussion in \autoref{sec:compare_ZTF}]{filippenko93,ciabattari2011supernova,ciabattari2013supernova}. Since 2011 several \SESN~light curves
have been observed to show such features. These SNe contribute to lifting our
template at early times, especially in bluer bands. 

\citetalias{taddia15} also released Ibc templates, obtained as a 12th-degree polynomial fit to the $z<0.1$ \SESN e in the SDSS sample. In \autoref{fig:ubtcomparet15}, we show how they compare to our templates. \citetalias{taddia15}'s templates have a slightly shallower fall (slower decay) than our rolling-median templates, while the rise is significantly faster than we measure. 

Comparing our \Ibc\ templates generated with a larger, yet heterogeneous sample of \SESN e and a more statistically sophisticated method for data aggregation than previous work,  with \citetalias{drout11} and~\citetalias{taddia15}, we find that:

\begin{itemize}
    
\item The new \Ibc\ templates include a measure of the uncertainty, previously missing.

\item Adding more SNe to the sample allows us to extend the templates to earlier and later epochs compared to \citetalias{drout11} and \citetalias{taddia15}.

\item The new \Ibc\ templates are generated for additional bands.

\item  While the new templates do not dramatically change our understanding of the time evolution of \Ibc\ in any band for which templates previously existed (\Vband, \Rband, \uband, \rband, and \iband\ bands), differences arise with significance above the IQR at early and late epochs.

\end{itemize}



\subsubsection{{Photometrically prototypical and unusual \Ibc's}}
\label{sec:outlierUber}

Below we discuss a few \SESN e that stand out as prototypical and {\it a}-typical in comparison with our \Ibc\ templates in \Bband, \Vband, \Rband, and \Iband\ bands. These objects are shown in \autoref{fig:ubtcompareSNe}. 


SN~1994I, often considered a prototypical SN~Ic, has narrow light curves rising and declining by one magnitude around the peak in less than 10 days. We suggest that SN1994I should never be regarded as a typical SN Ic, since is neither a spectroscopically typical SN Ic (see \citetalias{modjaz14}), nor a photometrically typical SESN, as shown here.

LSQ13ccw stands out compared with the \Ibc\ template as it has a narrow light curve, with a fast rise and a fast decline.

SN~2005kl shows a plateau in its \Bband\ band light curve but shows normal behavior in \Vband.

We have also plotted the light curves of SN~1998bw and SN~1993J(\autoref{fig:ubtcompareSNe}), which are known as prototypical \blIcs\ and \IIbs\ respectively to show that these light curves are overall well fit by our \Ibc\ templates within the IQR and can indeed be considered photometrically prototypical \Ibcs.

\section{GP Templates}
\label{sec:gptempaltes}

\begin{figure*}
\begin{center}
\includegraphics[width=0.5\textwidth]{ 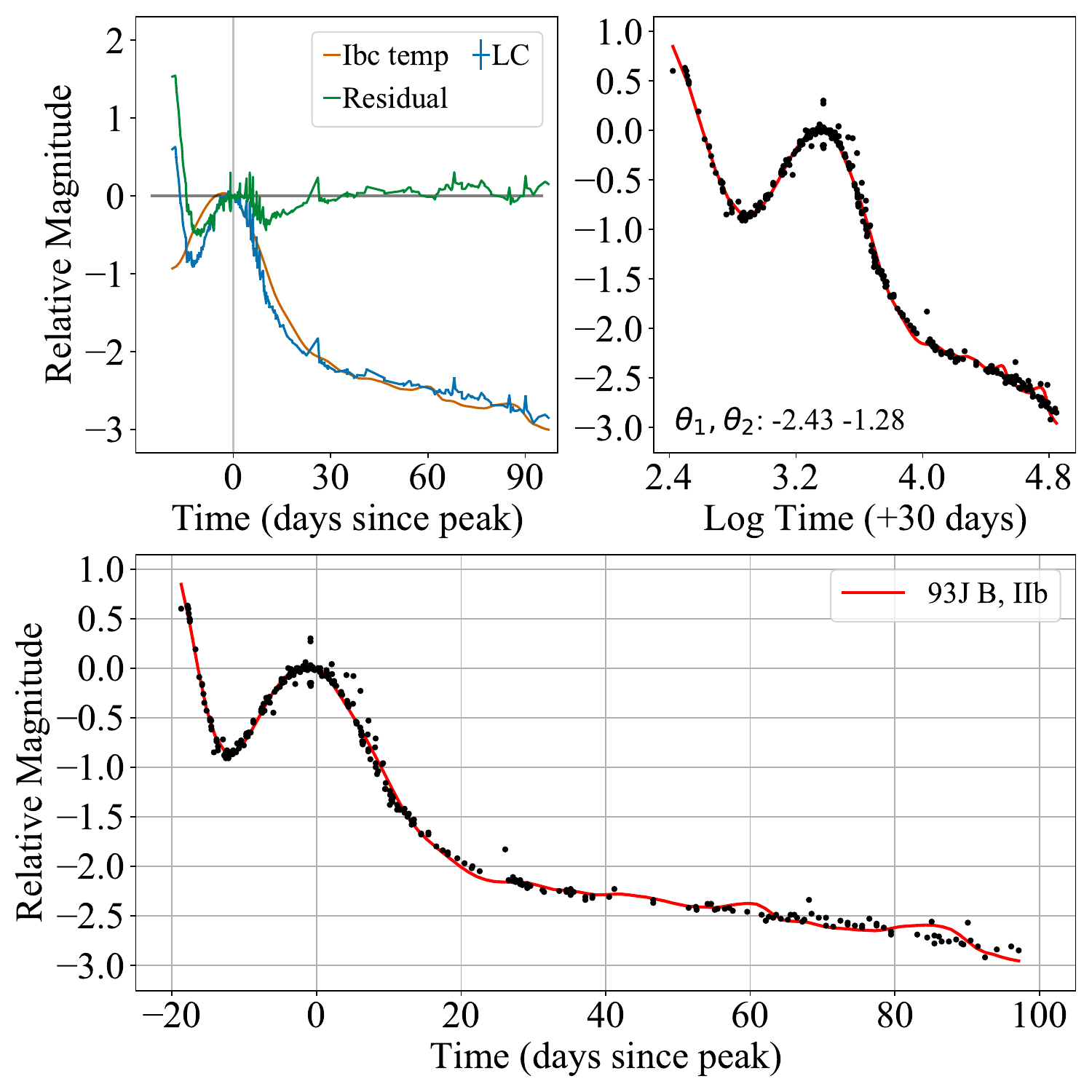}\includegraphics[width=0.5\textwidth]{ 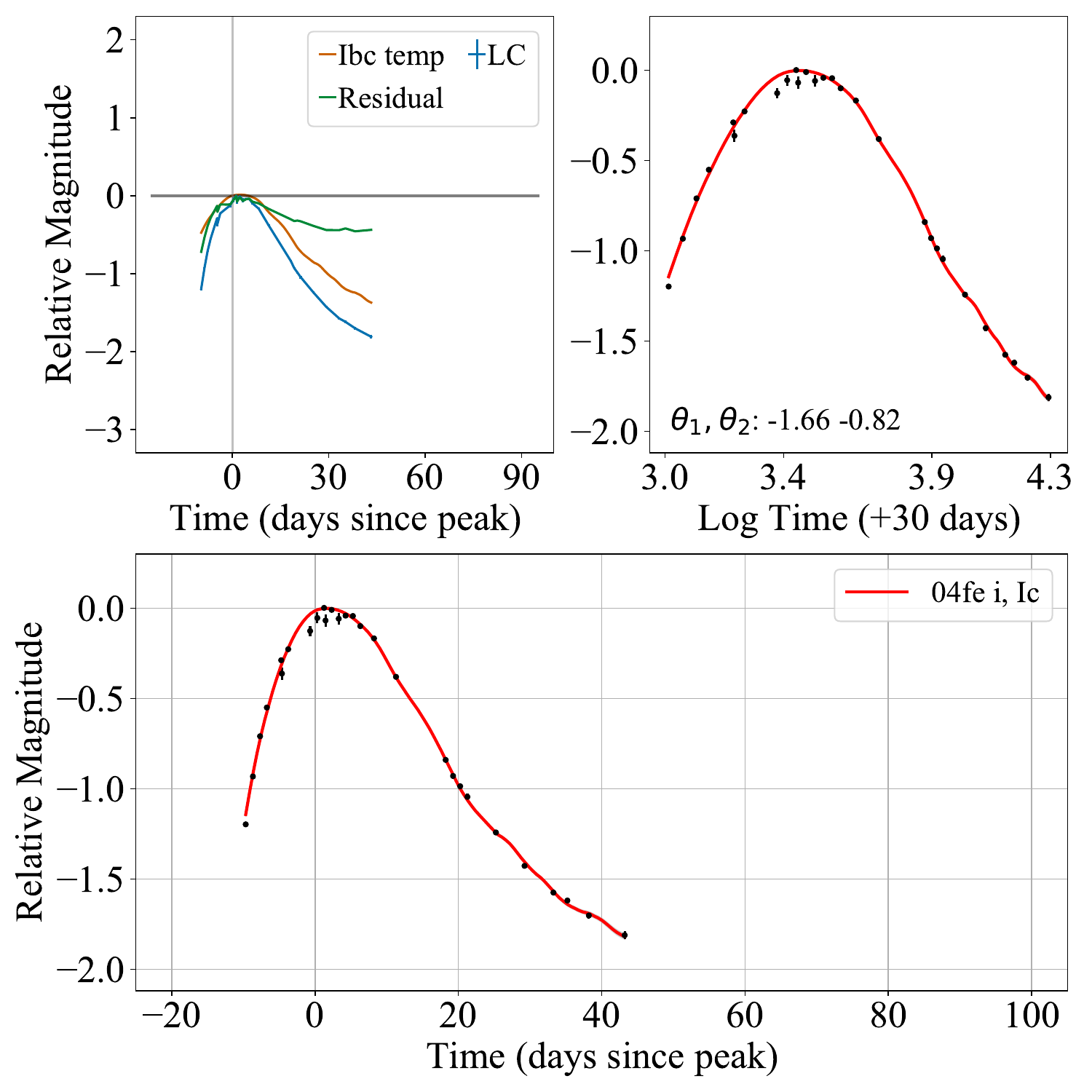}

\includegraphics[width=0.5\textwidth]{ 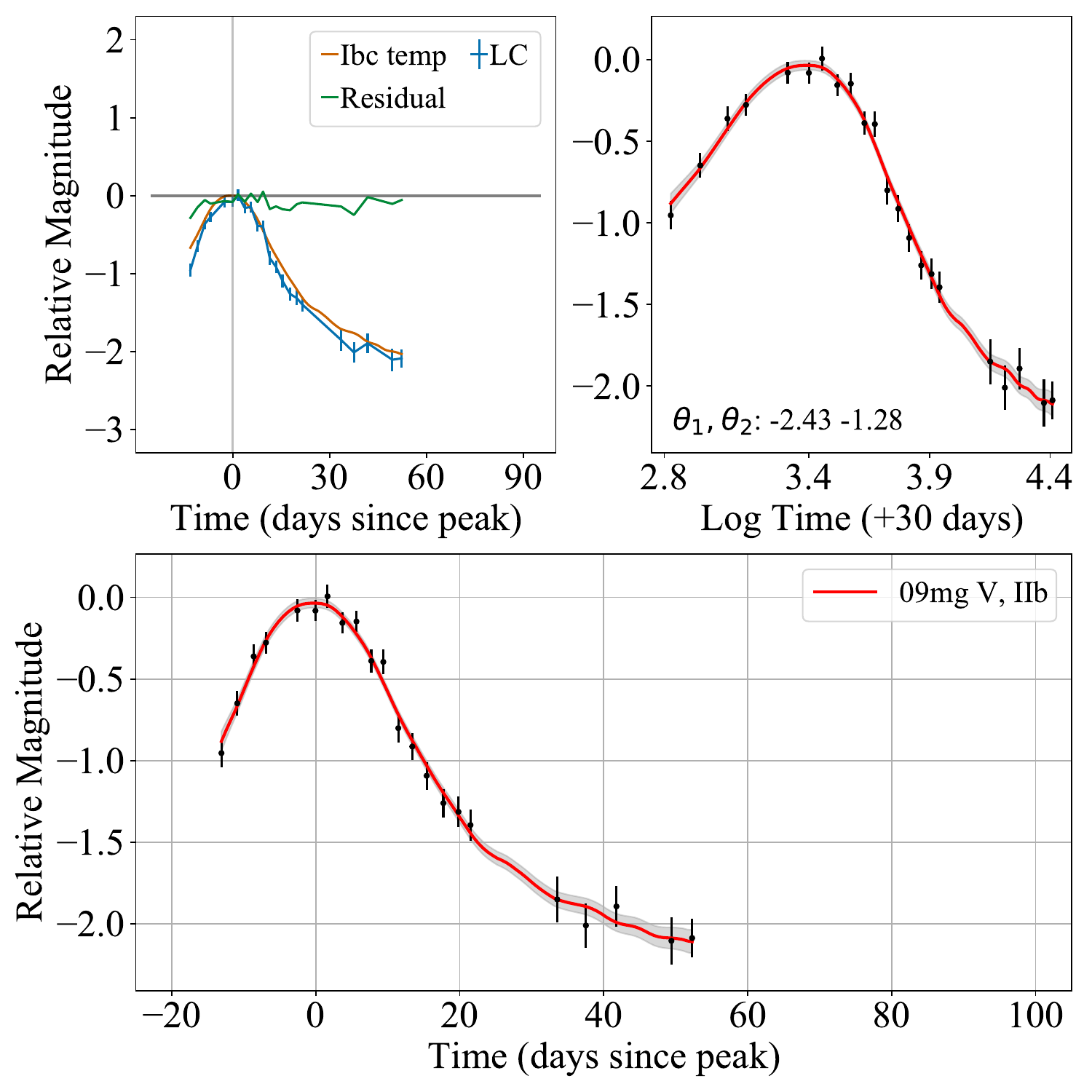}\includegraphics[width=0.5\textwidth]{ 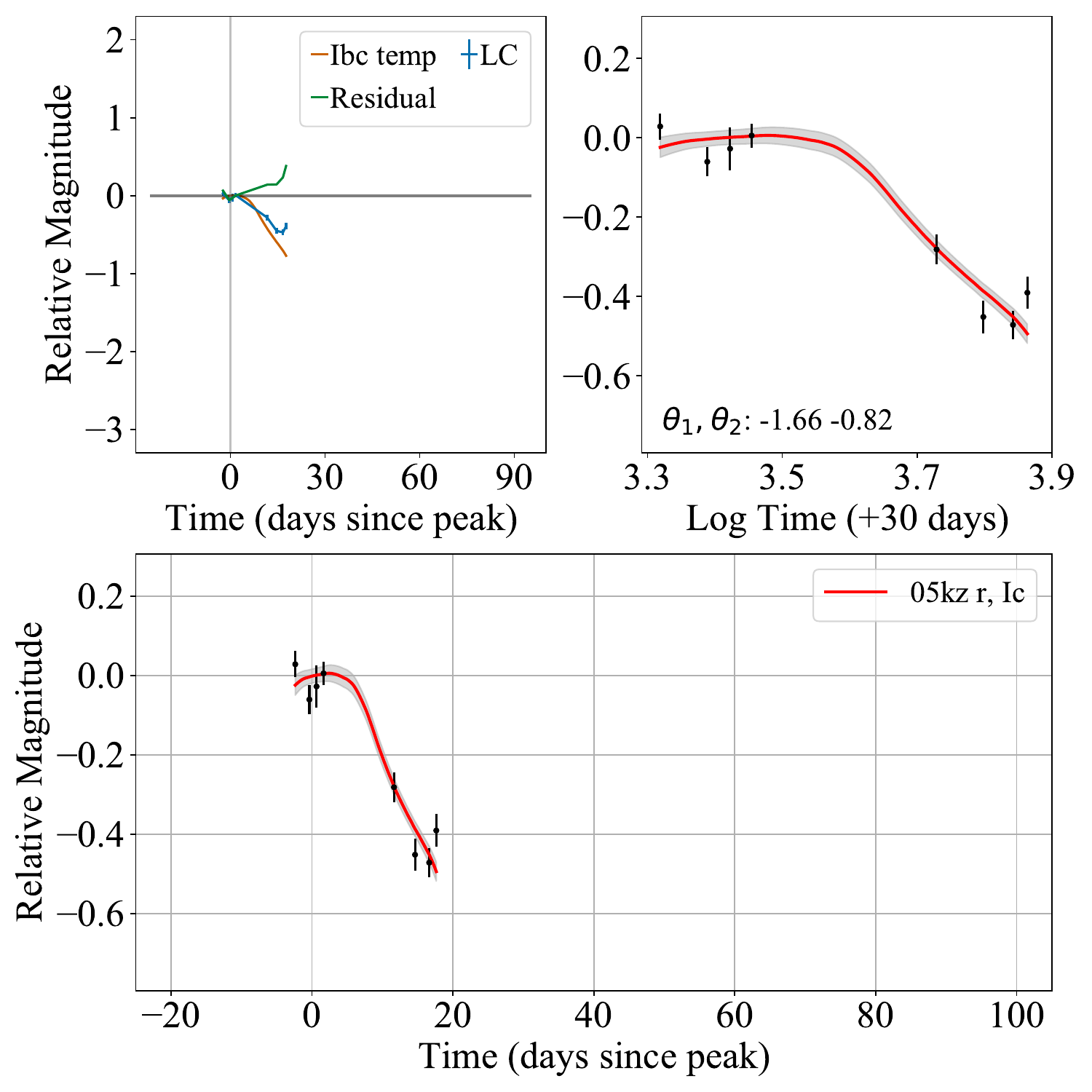}
\caption{{From top left and moving clockwise: SN1993J  \Bband\ band, SN2004fe \iband\ band, SN2009mg \Vband\ band, and SN2005kz \rband\ band. For each plot, the top left panel includes the time series, \Ibc\ template for that band, and the residual between the two, and the right panel shows the light curve along with its GP fit in the logarithmic time domain where the fit is performed. The bottom panel is the fit in natural time. The red line is the mean of the fit and the gray band (occasionally too thin to see) represents the uncertainty (point-by-point standard deviation of the GP realizations). To learn more about SN 2009mg see \citealt{2012MNRAS.424.1297O}.
}}\label{fig:goodfit}
\end{center}
\end{figure*}

We want to generate templates for each SN subtype, \alltypes, so we need a more sophisticated scheme than simply averaging over all SN photometry since the data per band per SN subtype is generally scarce and sparse. We want an interpolation procedure that:
\begin{enumerate}
\item{{\bf Captures the diversity in the SN sample:} Using a non-parametric model allows us to capture any peculiar behavior. We fit the light curves with Gaussian Processes (GP) using the {\tt george} {\tt Python} module.}
\item{{\bf Captures the early time variability, and the late time smoothness:}
Since variability is expected to be larger at early times, we fit the time in the logarithmic time domain.}
\item{{\bf Fits the observed data points :} 
{The goodness of fit is measured with a $\chi^2$ term in the objective function and is minimized to choose the hyperparameters of the GP.}}
\item{{\bf Maintain smoothness:} With the choice of a squared exponential kernel for the GP, which is infinitely differentiable, a smoothness requirement is implicitly enforced. However, non-discontinuous sharp features are still possible and not desirable, since we always observe SNe to be smooth on short time scales. An additional smoothness requirement is enforced by modifying the GP objective function adding a regularization term that minimizes the second derivative. }
\end{enumerate}

Despite being a rather old and established technique in many domains, first conceived by Daniel Krieg, a South African statistician in 1951 in the context of geospatial statistics \citep{Krige51}, only recently GP have gained increasing traction in astronomy as a tool for data-driven Bayesian interpolation and modeling (e.g.
\citealt{thornton2024extrabol},
\citealt{boone2019avocado}, \citealt{Ambikasaran14},  \citealt{McAllister17}, \citealt{pruzhinskaya2019anomaly}, \citealt{qu2021photometric}). For a review of GP applications in astronomy see \citet{aigrain2023gaussian}. 
GPs are ideal in the creation of SN templates and have been used in \citet{Reese15} to construct templates for \Ias\ and \citet{vincenzi2019spectrophotometric} to create single-object spectrophotometric templates of CC SNe. \citet{Reese15} used a training set of 1500 SNe Ia and 7000 CC SNe. However, the fraction of stripped SNe in the CC sample is very small (15 objects) thus the parameter choice is not optimized for these subtypes. \citet{vincenzi2019spectrophotometric} fitted SNe individually to produce object-level models. We will build on these results with one important difference: GP interpolation is generated in log-time (point 2 above).



\begin{figure}[!ht]
\begin{center}
\includegraphics[width=0.95\columnwidth]{ 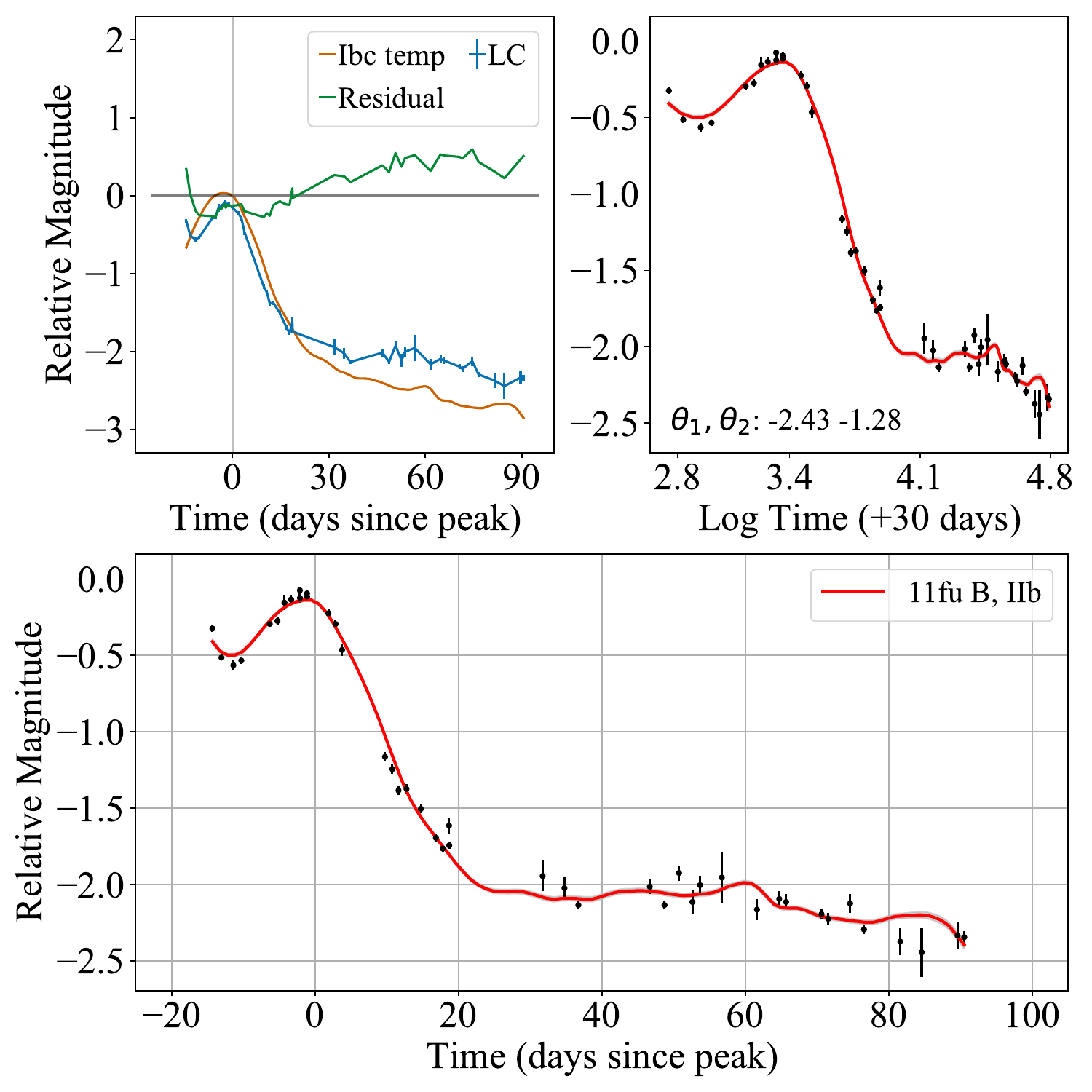}


\caption{{Panels and colors as in \autoref{fig:goodfit}, GP fit to Type IIb SN2011fu, \Bband. Our model captures well the small excesses at early times which is a known characteristic of many \IIbs\ \citep[see for example ][]{2018Natur.554..497B,  2017hsn..book..239A, 2022A&A...667A..92O}, associated with a shock cooling component.  However, 
 the late fit shows high confidence features on days time scales that are likely unphysical (see phase $\sim{60}$ days) To learn more about SN 2011fu see \citealt{2013MNRAS.431..308K} and \citealt{ 2015MNRAS.454...95M}.\label{fig:11fu_B}
}}
\end{center}
\end{figure}

\begin{figure}[h!]
\begin{center}
\includegraphics[width=0.95\columnwidth]{ 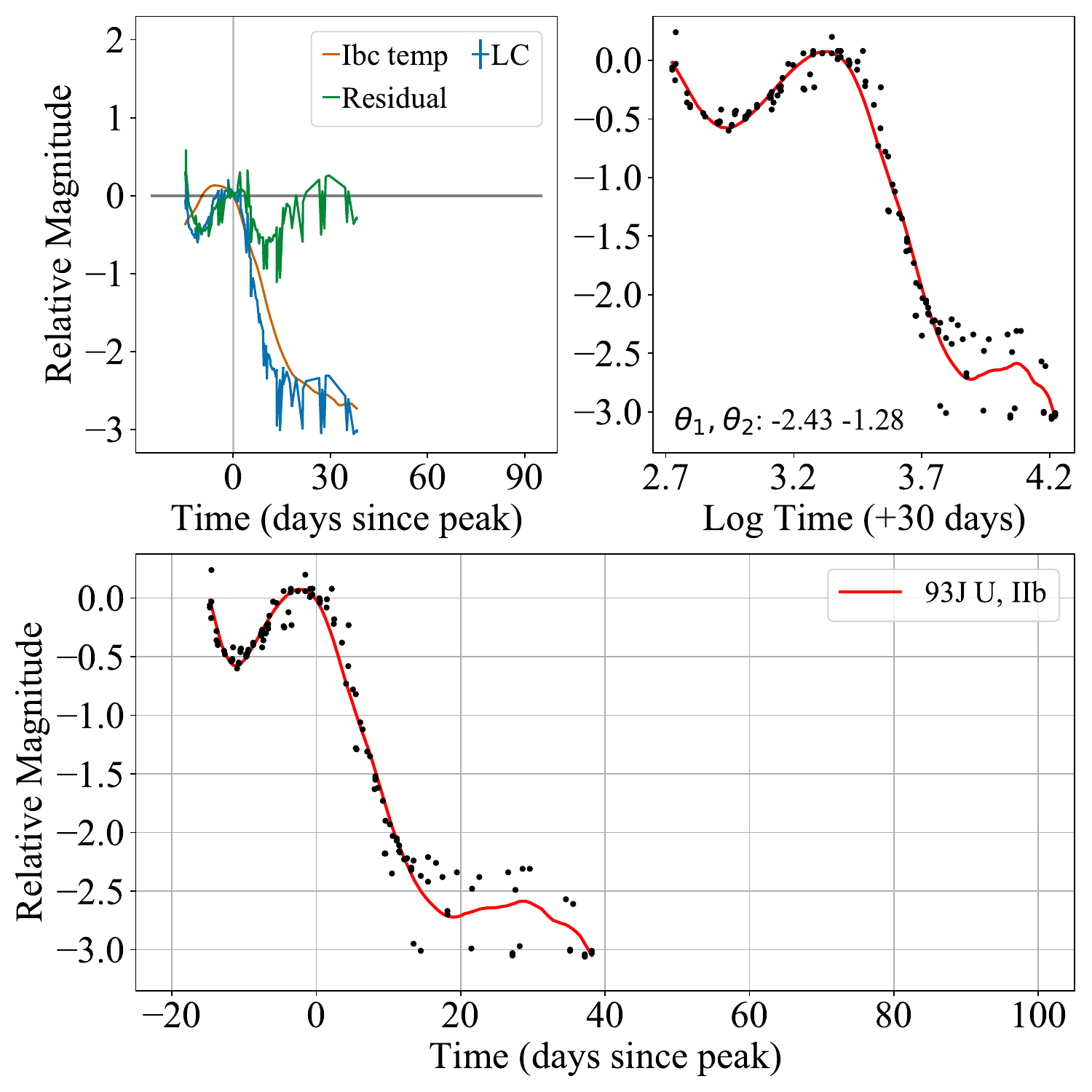}
\caption{{SN 1993J  \Uband: Color and panels are as in \autoref{fig:goodfit}. The small uncertainty in the data causes overfit with 
any choice of parameters that leads to the late time bump at $\sim 30$  (which is most apparent in the log fit top panel). Inflating the error bars by a factor $\sim10$ gives a more realistic fit at late times at the cost of losing details at early times (including the cooling envelope-related early excess). }}\label{fig:93J_U}
\end{center}
\end{figure}

Because at this stage we are interested in comparing the photometric behavior in different bands, we are applying GP only on the temporal axis, while recent work has applied it in the two-dimensional space of time and wavelength.

To improve the stability of the GP fit, we subtract the \Ibc\ templates (\autoref{sec:IBCtemplates}) in each band from each SN and fit the residuals. We choose a square exponential kernel ({\tt ExpSquaredKernel} in {\tt george})
\begin{equation}
k(d)=\theta_1^2  \exp(\frac{-d^2}{\theta_2^2}).
\end{equation}

To choose the best values of the model's hyperparameters, $\theta_1^2$ and $\theta_2^2$ and $s$, we minimize the objective function: $\chi^2(m(t|\theta_1^2,\theta_2^2)) ~+~ {\sum_t { \left| \frac{d^2 m(t|\theta_1^2,\theta_2^2)}{dt^2}  \right|}}^s,$ where $m$ is the GP-predicted magnitude and the second term is the second derivative of the predicted magnitude to a power of $s$ (points 3 and 4). $s$ determines the importance of the term that ensures a smoother fit in the objective function. Setting $s=1$ leads to good fits for most light curves except particularly fast-evolving light curves that have a narrow peak and fast magnitude variation in early phases. For fast-evolving SNe only (\autoref{sec:compare_ZTF})\footnote{sn2015U, sn2010et, sn2015ap, sn2019aajs, sn2018bcc, sn2019rii, sn2019deh, sn2019myn, LSQ13ccw, sn2010jr, sn2011dh, sn2006aj, ASASSN-14ms} we choose $s=0.25$.

The hyperparameters $\theta_1$ and $\theta_2$ are first optimized for each SN in the sample independently. A subset of light curves that shows a good fit for phases between $-20$ and 20 is identified by visual inspection (110 SNe). Within this subset, for each \SESN e subtype, we take the median of $\theta_1$ and $\theta_2$ as the chosen hyperparameter values. These values are reported in \autoref{tab:gp_params}. Next, the \SESN e of types \alltypes\ are fit again with their subtypes' hyperparameters. The fits are again visually inspected, and some light curves with bad fits are removed from the sample (read \autoref{sec:bad_fits}). The final selected fits will be used to create the GP templates as explained in \autoref{sec:subtypeTempaltes}.

\begin{table}
\begin{center}

\caption{Selected kernel parameters for each subtype}
\centering
\begin{tabular}{ |c|c|c| } 
 \hline
  & $\theta_1^2$ & $\theta_2^2$ \\ 
  \hline
 Ib & 5.95 & 1.02 \\
 IIb & 5.90 &  1.64\\
 Ic & 2.76 & 0.67\\
 Ic-bl & 1.39 &  1.99\\
 Ibn & 0.004 & 8.70\\
 
 \hline
\end{tabular}
 \label{tab:gp_params}
\end{center}
\end{table}

\subsection{Successful fits}
Generally, the fits capture well the time behavior of the single band light curves. \autoref{fig:goodfit} shows a few examples of good fits. For each SN we show in the top panel the original observations, the \Ibc\ templates, and the residuals between the two, which is what we actually fit with the GP procedure. On the right panel, we show the result of the GP in the logarithmic time domain.\footnote{the fit in log-time domain, the epochs are shifted by 30 days since no SN has data at -30 days from \Vmax.}
In the bottom panel, we show the final fit in natural time and the uncertainty bands generated by the GPs. In general, the uncertainty is very small for well-sampled light curves. Below are a few successful fits for SNe with extremely good sampling (SN 93J, \Iband\ band), with good sampling (05mf, \Rband), and more sparse and noisy data (05mf, \Hband). Features like small upturns of the light curves at early time are well fit by our method, as, for example, SN 2011fu in (\autoref{fig:11fu_B}).

\subsection{Bad fit and pathological cases}\label{sec:bad_fits}
We identify three kinds of undesirable behaviors in the GP fits obtained through the procedure outlined above. 

\begin{figure}[ht!]
\begin{center}
\includegraphics[width=1\columnwidth]{ 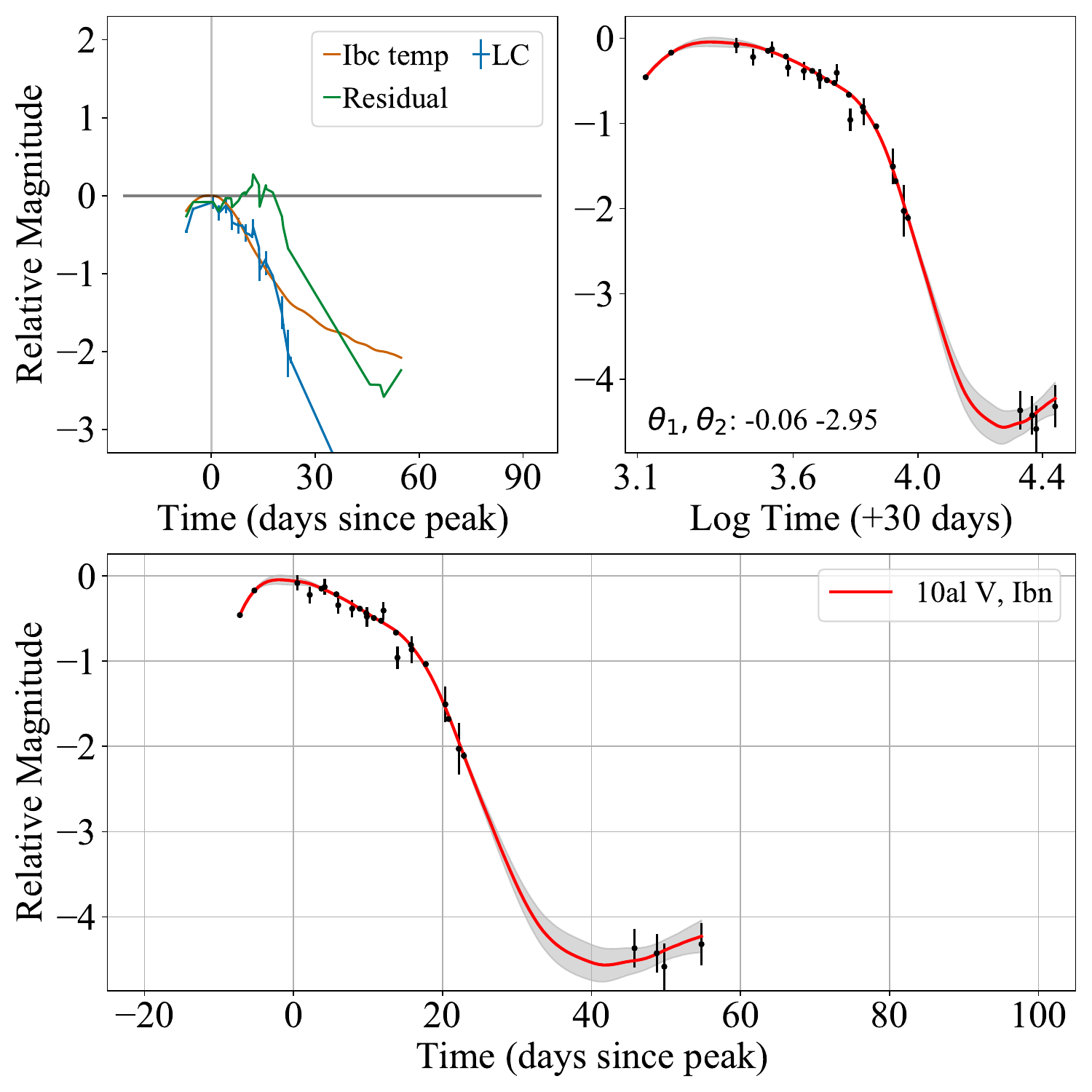}
\caption{SN 2010al \Vband: Color and panels are as in \autoref{fig:goodfit}. The light curve sampling ends before 60 days and the GP fit shows an upturn towards the end which is an artifact of the fit and not a feature of the supernova photometry.}
\label{fig:10al_V}
\end{center}
\end{figure}
\subsubsection{Late time behavior}
An upturn is seen in a few ($\sim 4$) light curves at the last few data points (\eg, in \autoref{fig:93J_U} and \autoref{fig:10al_V}). Generally, these epochs are outside of the range where we create our templates.

As in the 2010al \Vband\ example, the GPs propagate the light curve continuing the latest trend in the data. If there are missing data points, the GP extrapolation will generate a light curve with an increasing trend in brightness in later epochs. This is a problem with the data, not with the GPs, that cannot be solved in a non-parametric fashion, without imposing strict constraints on the sign of the derivative, but that in return may not allow us to characterize individual peculiarities in SN light curve behavior.

\subsubsection{Unphysical short-term variability}

Some of the fits exhibit unphysical features at the regions with no or very few data points. In these regions, the GPs tend to converge to the mean of the distribution which in this case is the \Ibc\ template, and the same features in the \Ibc\ templates are projected into the GP fits. For instance, in \autoref{fig:11fu_B},  there are only two data points around epoch 60 in the bottom panel where the fit exhibits some bump. If we look at the same interval for the \Ibc\ template (orange curve in the upper left panel), we see the same features.

\subsubsection{Bad fits}

In some cases, the GP fit overall is not acceptable. In some bands like ultraviolet bands, this is due to a lack of data, large uncertainties, and a noisy \Ibc\ template, and in other bands, due to having no data points between epochs $-5 \leq \maxep \leq 5$ days. These fits are removed by visual inspection and not included in the GP template creation. 56 light curves out of 710 total light curves were removed for these reasons.

\begin{figure*}

  \includegraphics[width=0.95\textwidth]{ 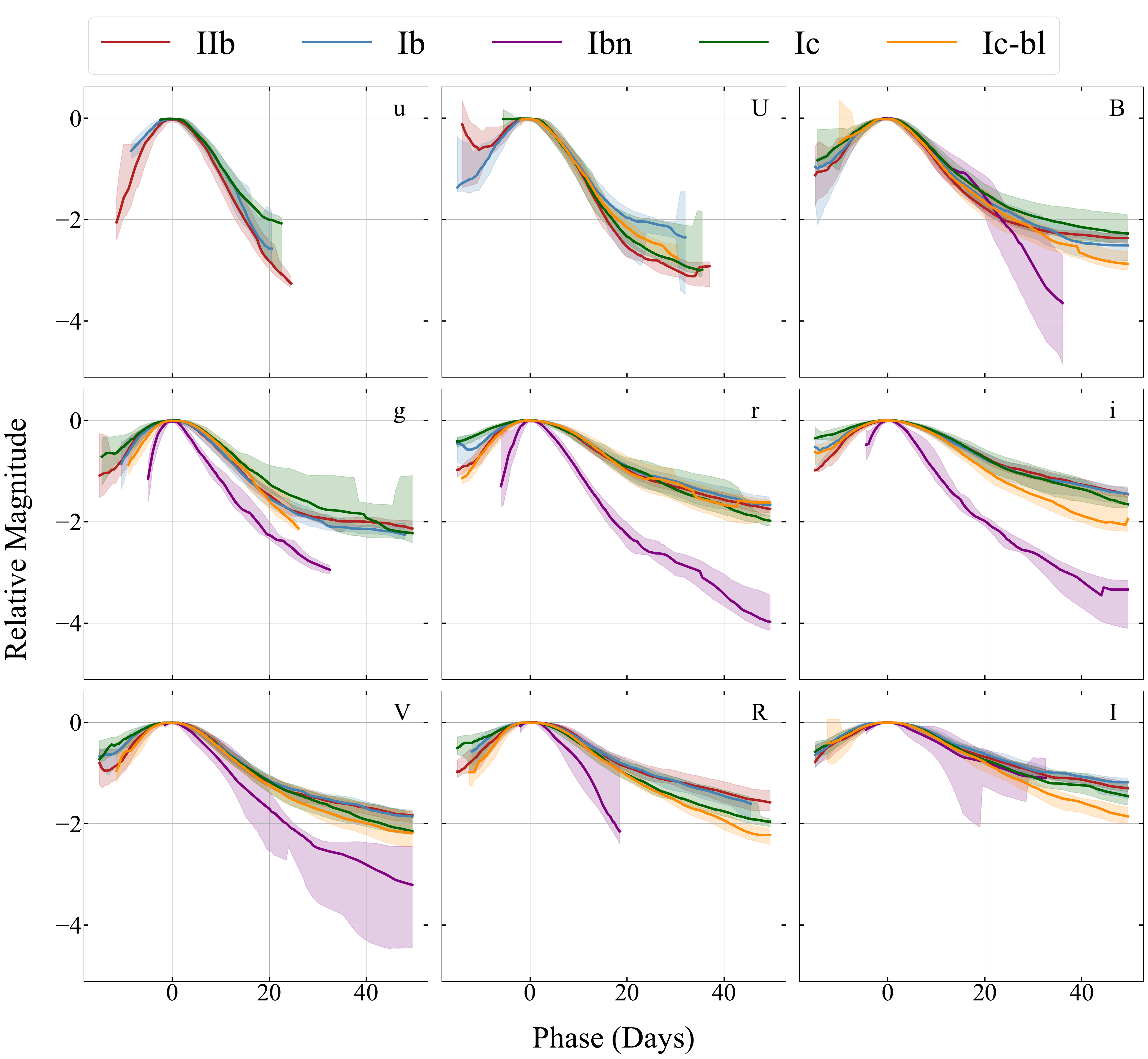} 
  \caption{ GP templates are shown in optical bands of $U,~u',~B,~V,~g',~R,~r',~I,~i'$ bands for each SESN subtype (as per legend at the top). Templates of different subtypes in each band are behaving mostly similarly within the uncertainties. The behavior of the templates in different bands is consistent with our expectation of the decline becoming shallower as we move to redder bands. 
  Sudden changes in the number of light curves used to generate the templates at some epochs may lead to unphysical behaviors such as the rapid increase in the uncertainty of the Ib $U$ band template near $\sim 30$ days or the uncertainty drop in the Ic $g$ template between $\sim35$ and $\sim50$  days from \Vmax. }\label{fig:gp_subtype_per_band} 
\end{figure*}

\begin{figure*}

  \begin{center} \includegraphics[width=0.95\textwidth]{ 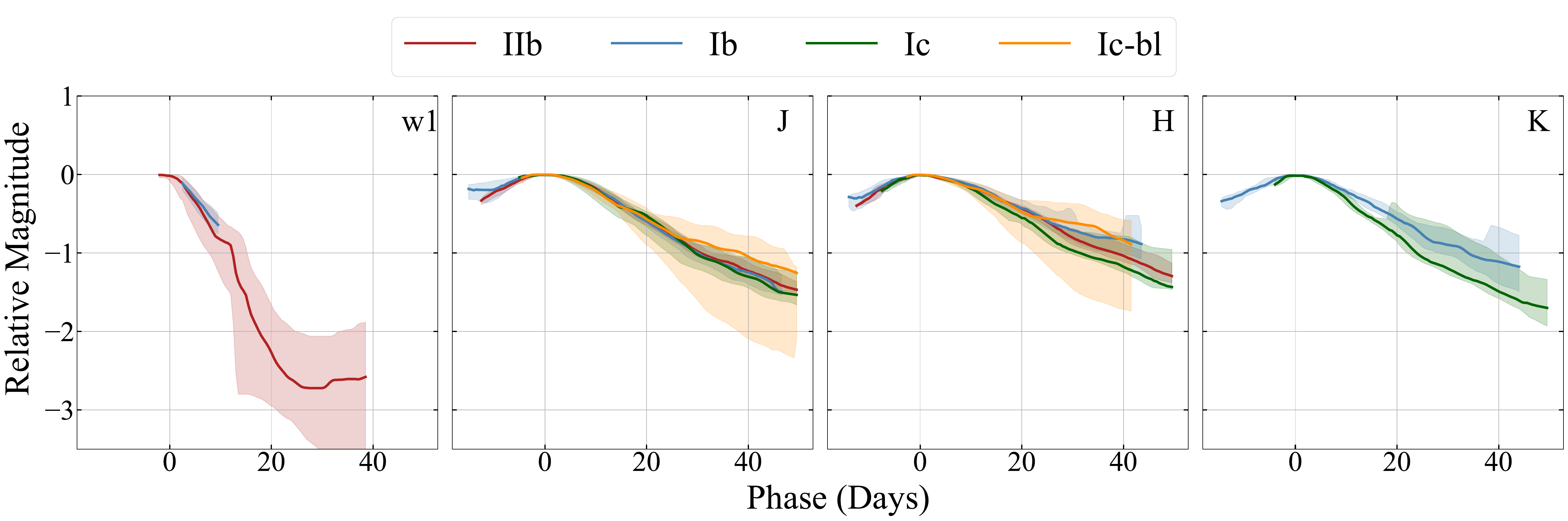} 
  \caption{GP templates for the $~w1,J,~H,~K_s$ bands. There are less than three light curves in $w2,~m2~$ bands and we are not able to release a template for these bands. }\label{fig:gp_subtype_per_band_jhk} \end{center}
\end{figure*}

\begin{figure*}[!t]
\includegraphics[width=0.5\textwidth]{ 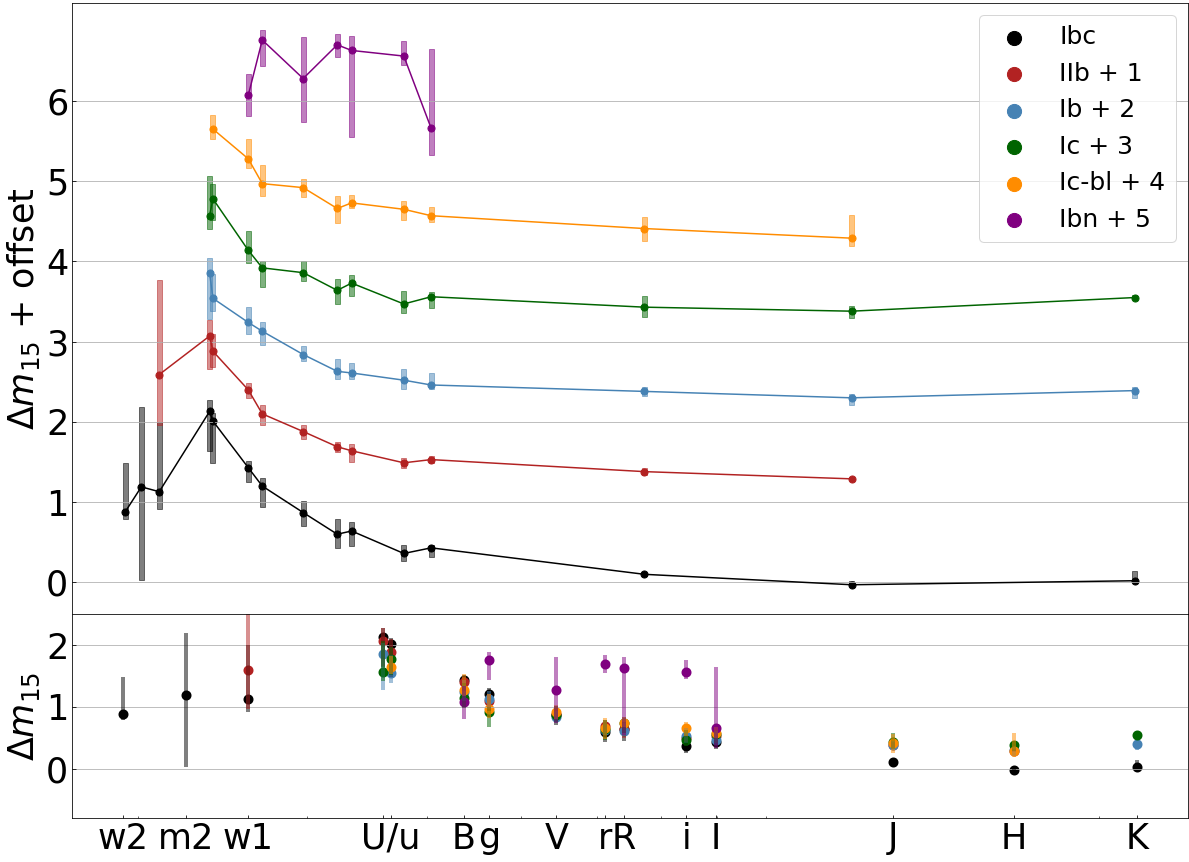}
\includegraphics[width=0.5\textwidth]{ 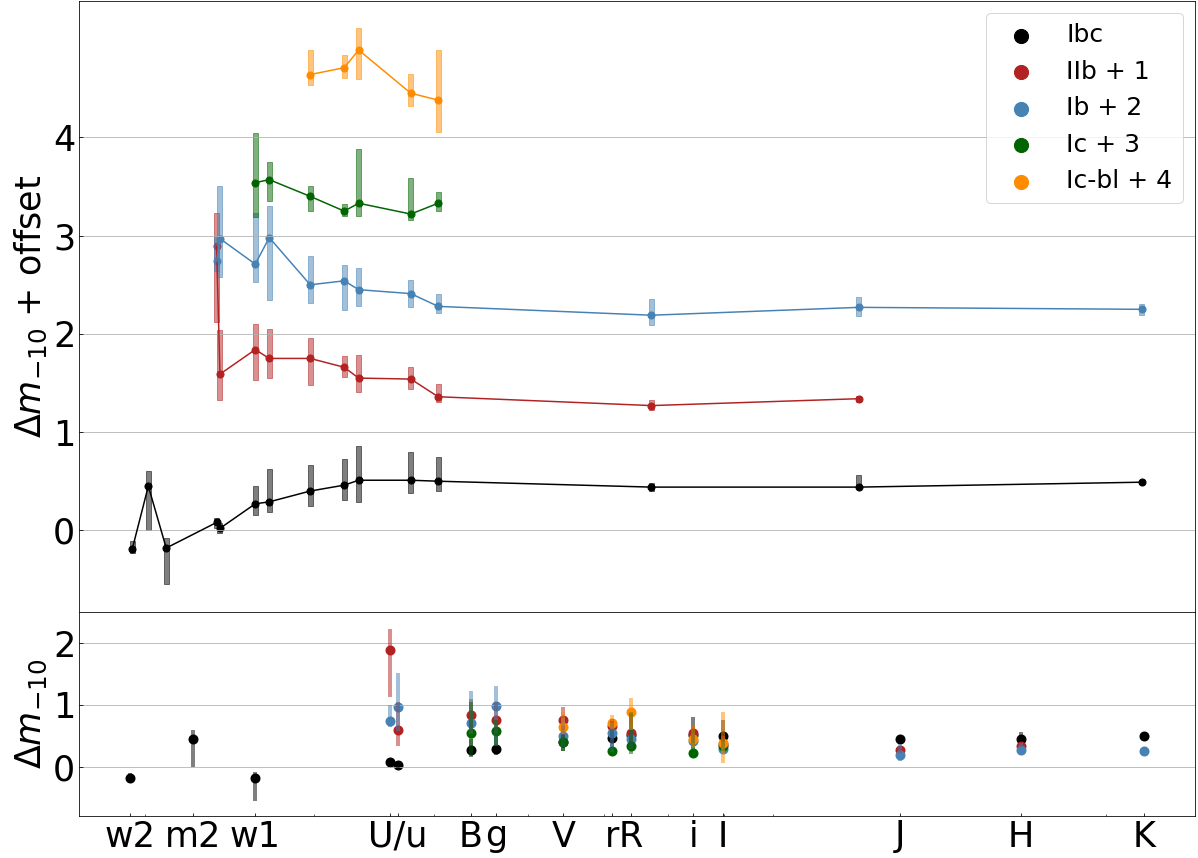}
\caption{$\Delta m_{15}$ (\emph{left}) and $\Delta m_{-10}$ (\emph{right}) for our \Ibc\ templates and each SESN subtype template. The location of the point on the $x$ axis is set to the effective wavelength of the filter (assuming the Swift UVOT system for \swiftbands, Johnson and SDSS filters for the \ubvri\ and \sloanugriprimed\ respectively, and 2MASS/PAIRITEL	for \NIRbands). In the top panel, the values are separated on the vertical axis by an arbitrary value (as per legend) for readability. On the bottom panel, the values are plotted in their natural space to make the overlap and separation of measurements more obvious to the eye. We see an overall slower evolution past peak as wavelengths get longer, both in the Ibc templates and in each individual subtype template. At the bluest wavelengths ($w2$, $m2$, $w1$) the photometric measurements are too sparse and too scarse to assess any wavelength-based dependence on the evolutionary time scales. In most photometric bands, the  $\Delta m_{15}$ and $\Delta m_{-10}$ are consistent for all subtypes within the uncertainties (IQR) except for the rapid post-peak evolution of SN Ib-n, discussed further in \autoref{sec:compare_ZTF} (for which measurements of $\Delta m_{15}$ only are available in selected bands). Occasionally, we see inconsistencies between the values measured for the \Ibc\ templates and for the individual subtype GP templates. For example, the \Ibc\ templates show a more rapid evolution at NIR wavelengths than each subtype's template. We attribute this inconsistency to the poor quality and sparsity of photometric measurements in NIR bands. Values plotted are reported in \autoref{tab:del_m15} and \autoref{tab:del_m10}. This figure is further discussed in \autoref{sec:compare_GP}.}
    \label{fig:dms}
\end{figure*}

In the remainder of this section, we aim to investigate the behavior of the GP templates. 
We first compare the behavior of different subtypes relative to each other (\autoref{sec:compare_GP}). Then, we compare the GP templates to a set of well-known simulated \Ibc\ light curves from the LSST photometric simulations \plasticc\ and \elasticc\ (\autoref{sec:plasticc_compare}). Finally, we compare the GP templates to individual \SESN e to identify subclasses and peculiar SNe (\autoref{GP_atypical} and \autoref{sec:compare_ZTF}).

\subsection{GP Templates for SESNe Subtypes}
\label{sec:subtypeTempaltes}

With the individual GP fits in hand, we create templates for each subtype in the bands that have at least three light curves as the rolling median of the light curve fits (and IQR describing the uncertainty). The median is calculated within an adaptive time window. We require a larger time window at late epochs, where fewer light curves are available: the window size is three days up to \maxep + 15 days,  four days in the $15 \leq \maxep \leq 20$ days range, five days in the $20 \leq \maxep \leq 27$ days range, six days in the $27 \leq \maxep \leq 35$ days range,  seven days $35 \leq \maxep \leq 45$ days range, and eight days thereafter. These numbers were optimized empirically.  \autoref{fig:gp_subtype_per_band} and \autoref{fig:gp_subtype_per_band_jhk} shows the GP templates of different subtypes in each band for epochs $-20 \leq \maxep \leq 40$ days along with their IQR between.\footnote{In some bands the templates are obtainable at longer epochs as well but for generalization purposes, we have shown all of them up to epoch 40.} 

 In general, there is significantly less data available for \SESN e in the UV and IR when compared to optical bands. The UV in particular provides a significant challenge as stripped-envelope supernovae tend to be less UV bright than other core-collapse supernovae \citep{Pritchard14}. We were only able to create GP templates for a single subtype in a single filter in UV frequencies: Type IIb in Swift $w1$ (the reddest UV filter) and for a fairly limited set in the IR (3/5 subtypes in \Jband, 1/5 in \Kband, and 2/5 in \Hband).

\subsection{Comparing behavior of different subtypes}\label{sec:compare_GP}

We have created separate templates for different subtypes of \SESN e including \alltypes. 

We present ${\Delta m}_{15}$ and ${\Delta m}_{-10}$ values of the GP templates in different bands in  \autoref{tab:del_m15} and \autoref{tab:del_m10} and \autoref{fig:dms}. 
We generally see a progressively slower decline for all subtypes in redder bands, except for SNe Ibn (to which we will return in \autoref{sec:compare_ZTF}). Otherwise, there are no statistically significant differences in evolutionary time scales near the peak. We warn the reader that this can be due to the limited size of the sample when we split the \SESN e by subtype. We note that the Ibc templates measure lower values of $\Delta m_{15}$ in the reddest bands (\NIRbands) and $\Delta m_{-10}$ in the reddest bands ($u$ and $U$) than the values measured for individual subtype templates available in those bands.
\citet{barbarino2021type} measured the ${\Delta m}_{15}$ and ${\Delta m}_{-10}$ of 44 spectroscopically normal SNe Ic in $R$ band and found the majority of them to have 
$0.2\leq {\Delta m}_{15}\leq0.7$ and $0.1\leq{\Delta m}_{-10}\leq 0.7$.
This result is consistent with the  ${\Delta m}_{-10} = 0.33_{-0.13}^{+0.55}$ we measure on our Ic templates. 
However, we measure ${\Delta m}_{15} = 0.73_{-0.16}^{+0.10}$, statistically inconsistent with the slower evolving values measured by \citet{barbarino2021type},.

We investigate to what extent the SESN subtypes are photometrically distinguishable by comparing pairs of templates. We only compare templates for subtypes and bands where there are at least three light curves. 
Most notably, we see that Ic-bl have a faster evolution than other types at late times. \autoref{fig:GP_compare_Ic_Ic_bl} shows a comparison of Ic and Ic-bl templates. Type Ic-bl SN are distinguishable from Ic type by the broad lines in their spectra, but they have not been known to show significantly different photometric behavior. Our templates show that these two subtypes are mostly similar within their IQR range, although in $R$, $I$, and $i$ bands, Ic-bl subtype seems to have a faster evolution starting at $\sim20$ days, and in $B$ in phases later than 40 days. In \autoref{fig:GP_compare_Ib_Ic_bl} and \autoref{fig:GP_compare_IIb_Ic_bl} we see that Ic-bl are faster evolving at late times in the $R$ and $I$ band when compared to SNe Ib and IIb as well, with an even more significant separation. This may be true in other bands redder than $V$, but photometry is generally poor for SNe Ic-bl in $i$ and NIR bands, leading to large uncertainties in the template. Additional pair-wise template figures are shown in \autoref{app:appA}.

Several examples of \IIb\ light curves with shock-cooling double-peaked signatures have been observed in the literature (\citealt{1994AJ....107.1022R, Arcavi_2011, 10.1093/mnras/stt162, 10.1093/mnras/stu1837,
10.1093/mnras/stw3082, 2021MNRAS.507.3125A}, and see review by \citealt{modjaz2019new}). However, we do not see a clear signature of the cooling envelope emission in our templates (\autoref{fig:GP_compare_IIb_Ic_bl} and \autoref{app:appB}). While in the \Uband\ \IIb\ templates we do see the evidence of two peaks in the median GP time series (solid red line), the uncertainties are large in the pre-peak phases, especially in the bluer wavelengths (\Uband, \Bband, \gband). The large uncertainties are due to the concurrent effects of paucity and low \SNR\ of early time data, and intrinsic diversity of \IIbs\ in these phases: the cooling envelope signature is not always present, and when it is it has different time scales, consistent with different progenitor sizes. We will return to double-peaked lightcurves in \autoref{sec:compare_ZTF}.

\begin{figure*}
    \centering
\includegraphics[width=0.9\textwidth]{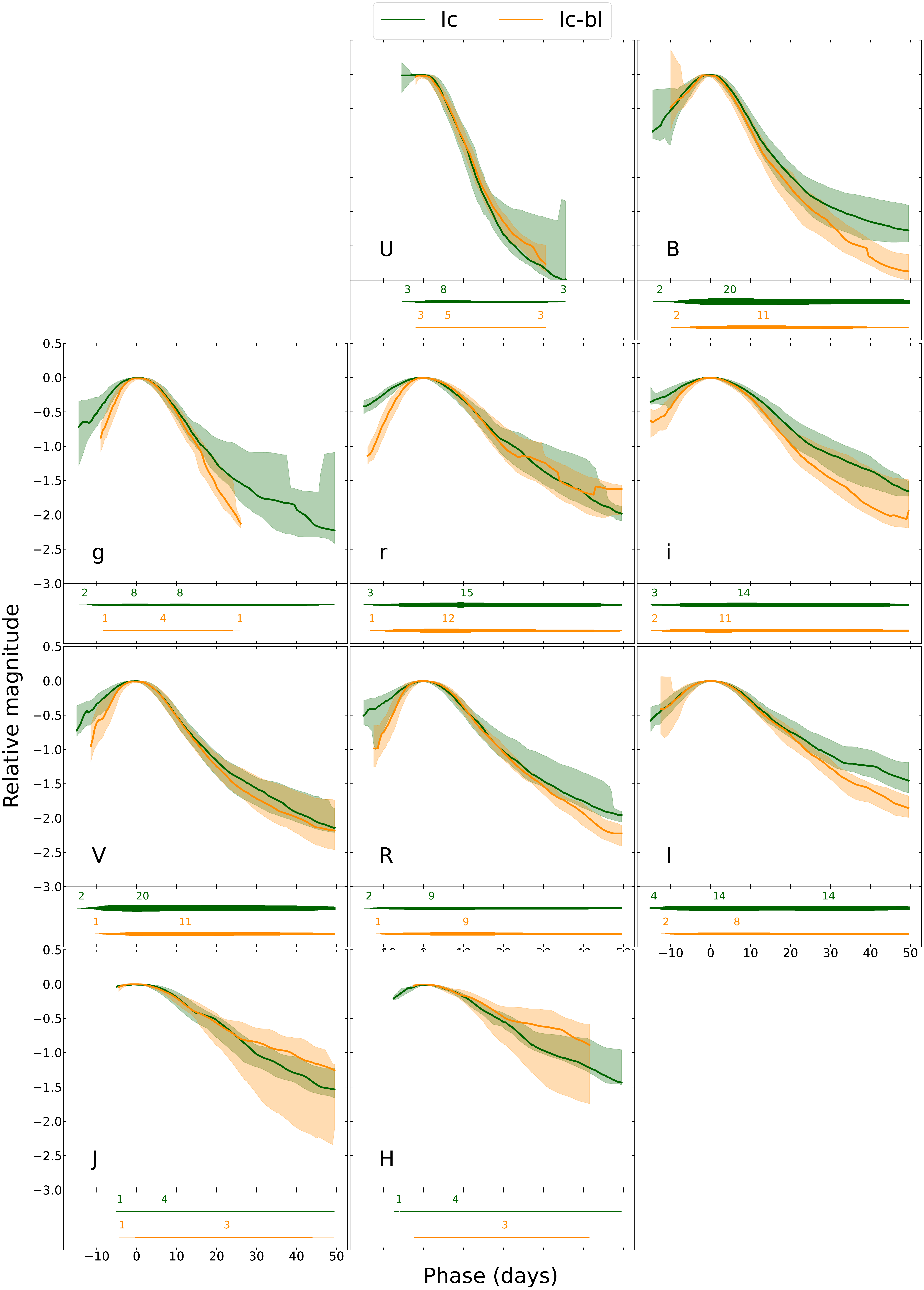}
    \caption{GP templates of subtypes Ic and Ic-bl in all the existing bands. The lower panel in each subplot shows the number of light curves that were used in each time window to create the template. We caution the reader that sharp features in the templates or their uncertainties can be connected to the sudden change in the number of the light curves used to generate the template in that band (\eg\ Ic template near ~40 days after \Vmax).}
    \label{fig:GP_compare_Ic_Ic_bl}
\end{figure*}

\begin{figure*}
    \centering

\includegraphics[width=0.9\textwidth]{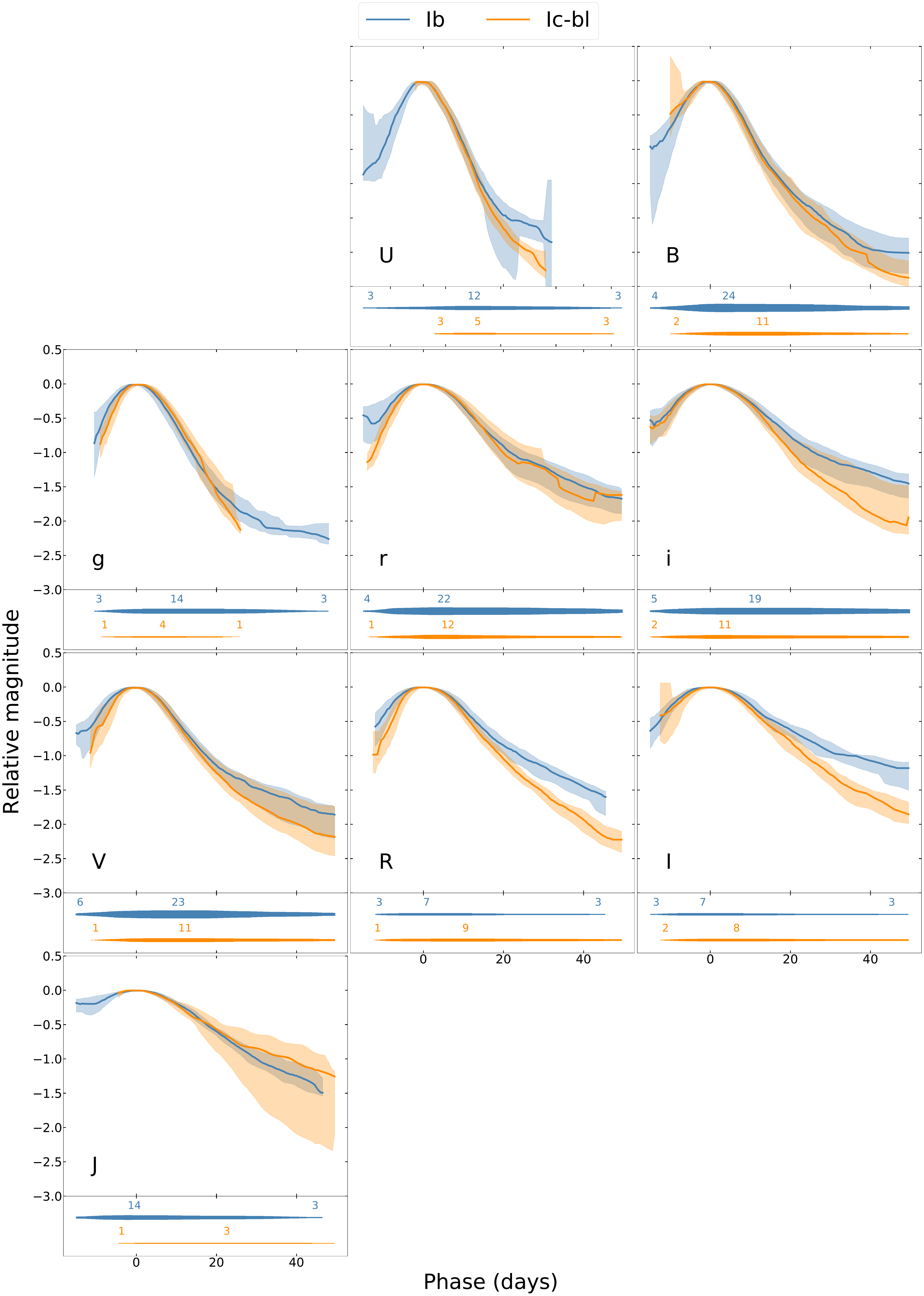}

    \caption{As \autoref{fig:GP_compare_Ic_Ic_bl} for \Ib\ and \blIc\ subtypes.}
    \label{fig:GP_compare_Ib_Ic_bl}
\end{figure*}

\begin{figure*}
    \centering
\includegraphics[width=0.9\textwidth]{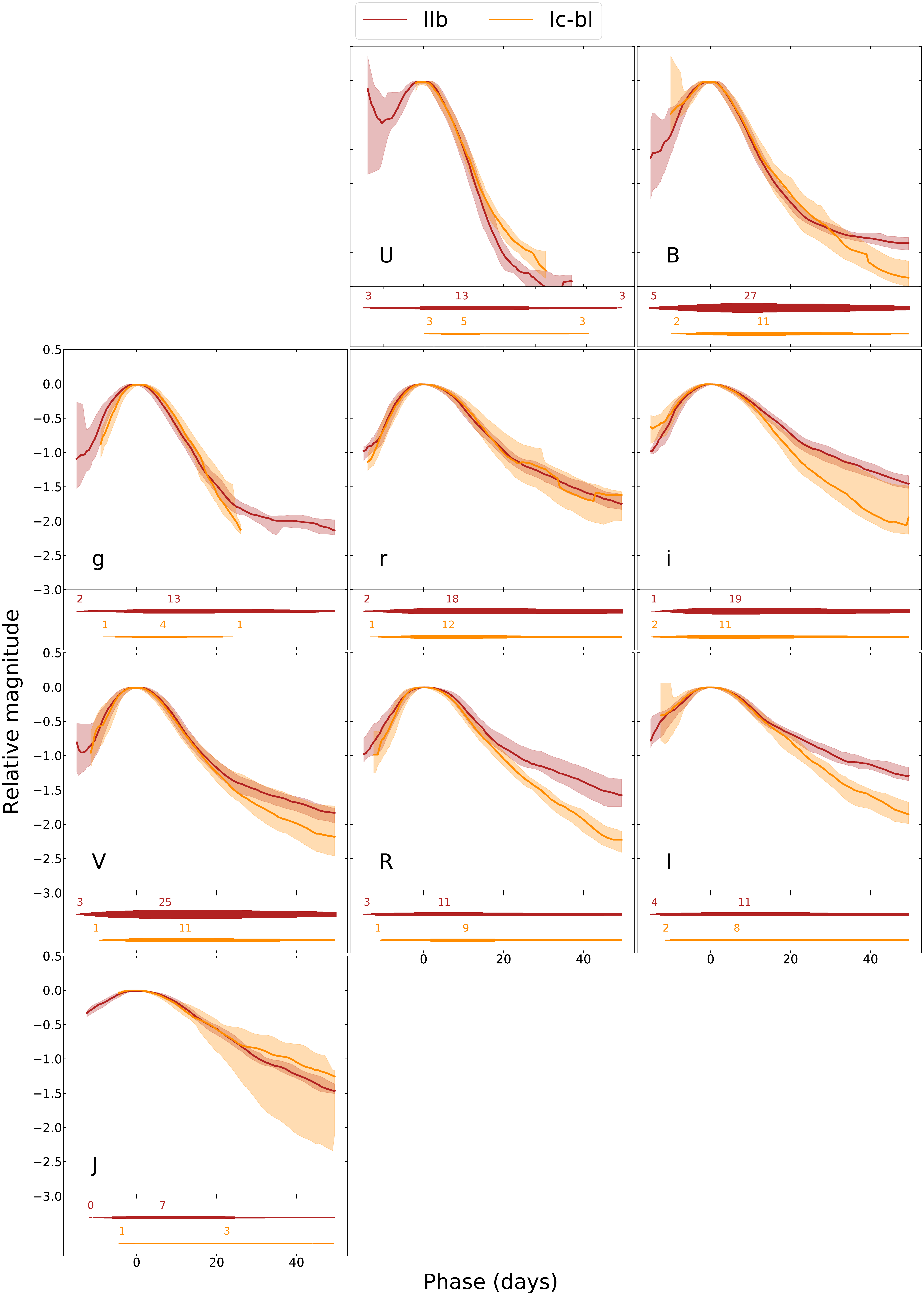}

    \caption{As \autoref{fig:GP_compare_Ic_Ic_bl} for \IIb\ and \blIc\ subtypes.}
    \label{fig:GP_compare_IIb_Ic_bl}
\end{figure*}

\subsection{Comparison with SESN light curves in the \plasticc\ and \elasticc\ dataset}\label{sec:plasticc_compare}
As discussed in \autoref{sec:intro_sec} and \autoref{sec:motiv}, a motivation for our work is the development of templates that can be used to improve photometric classification of \SESN e. To that end, we compare our templates to simulated LSST light curves in order to assess the accuracy of the  synthetic LSST samples that are used to train photometric classifiers. The Photometric LSST Astronomical Time-Series Classification Challenge (\plasticc) \citep{allam2018photometric} was a community-focused data challenge that took place in 2018. The challenge provided simulated light curves of millions of transients and variable stars as will be observed by the Rubin LSST. The dataset has a category of SN Ibc light curves for SESNe. Light curves for the challenge are generated following the process described in \citealt{2019PASP..131i4501K}. Starting with an underlying SED, the SuperNova Analysis (SNANA, \citealt{2009PASP..121.1028K}) is used to forward-model the observing process, simulating
extinction, survey systematics, and observing strategy to arrive at the final synthetic multi-band
photometry.

The \plasticc\ \SESN e light curves are generated using SED time series of 13 well-sampled Ibc (7 Ib and 6 Ic) combined with a parametrization obtained with the MOSFiT package \citep{guillochon2018mosfit}.

We have selected 58 light curves from their Ibc sample that have $z<0.2$ and have at least 2 data points between phases -10 and 10 days in $r$ band. For each band use the light curve only if it has at least four data points in the  $-50 \leq \maxep \leq 100$ days range. We plotted this set of well-sampled simulated light curves in $u, g, r,i$ bands along with our GP templates to investigate their time evolution in \autoref{fig:GPcompare_plasticc}.

To align the \plasticc\ data with the templates in brightness, we interpolate each \plasticc\ light curve with a cubic spline.
In the $u$ band, it is hard to draw a general conclusion due to the large uncertainties. Compared to our templates, the \plasticc\ simulations show significant differences: they have light curves with slower late-time evolution, particularly in the redder bands, as well as a few extremely fast-evolving light curves.


\begin{figure*}
  \begin{center} \includegraphics[width=0.95\textwidth]{ 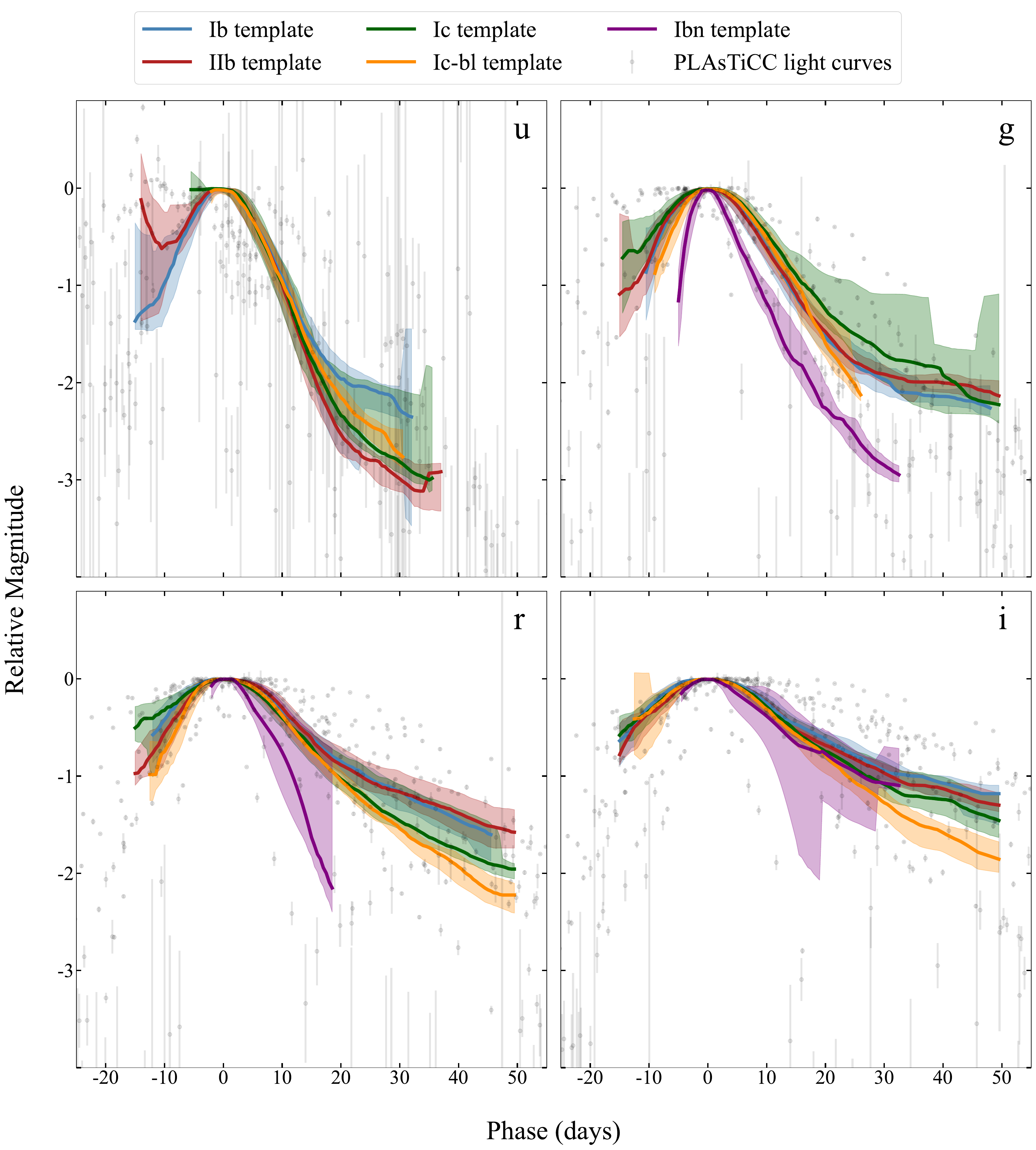} \caption{ Our GP templates (solid lines) are plotted along with the most well-sampled \plasticc\  light curves of SN Ibc in $u, g, r, i$ bands (grey dots). Compared to our templates, the \plasticc\ simulations show significant differences: they have light curves with slower late-time evolution, particularly in the redder bands, as well as a few extremely fast-evolving light curves. } \label{fig:GPcompare_plasticc} \end{center}
\end{figure*}

\begin{figure*}
  \begin{center} \includegraphics[width=0.95\textwidth]{ 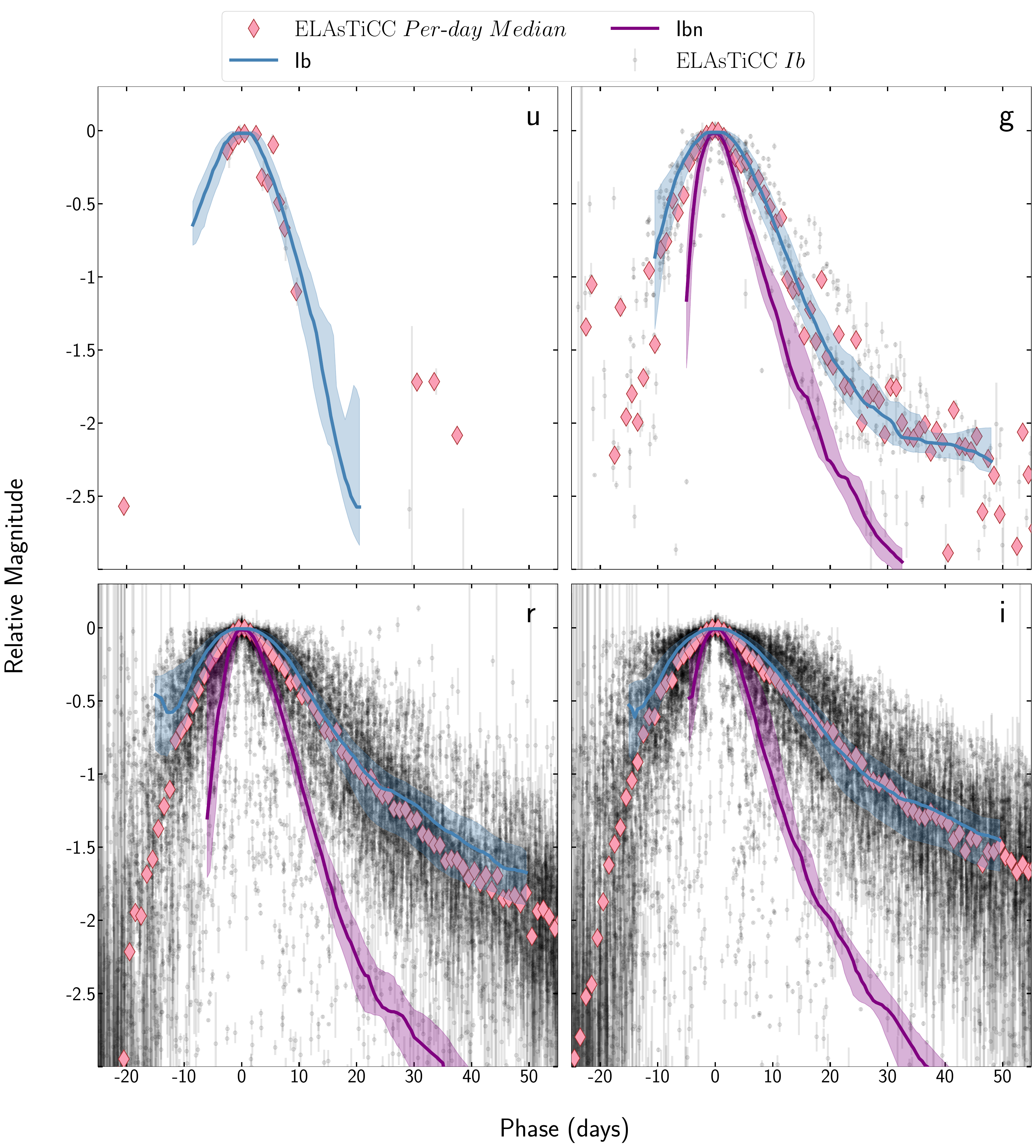} \caption{The GP templates of subtypes Ib (blue curve) and Ibn (purple curve) are plotted along with \elasticc\ light curves of SN Ib in $u, g, r,i$ bands (black dots) and their per-day medians (pink diamonds). The \elasticc\ sample is broadly consistent with the SN Ib behavior observed in our sample, although they display a larger variance. Some of the light curves in the \elasticc\ sample have a rapid evolution broadly consistent with SNe Ibn.} \label{fig:GPcompare_elasticc_Ib} \end{center}
\end{figure*}
\begin{figure*}
  \begin{center} \includegraphics[width=0.95\textwidth]{ 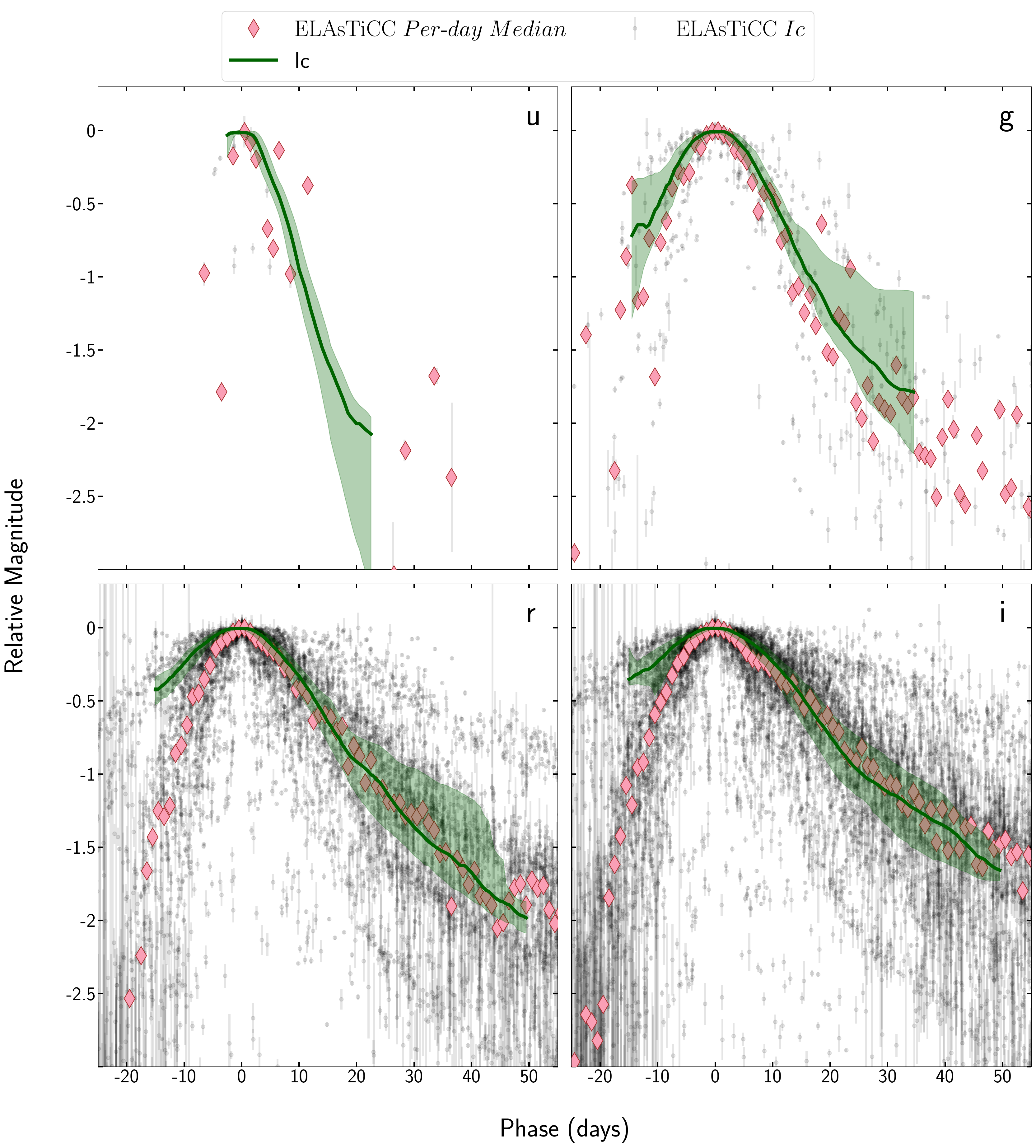} \caption{The GP template of subtype Ic (green curve) is plotted along with \elasticc\ light curves of SN Ic in $u, g, r,i$ bands (black dots) and their per-day medians (pink diamonds). The \elasticc\ sample shows a significantly more rapid rise for SN Ic than observed in our sample in the $r$ and $i$ bands. The fall is broadly consistent with the past-peak behavior of our sample, but like for \autoref{fig:GPcompare_elasticc_Ib}, the \elasticc\ sample shows a larger variance than observed in our data.} \label{fig:GPcompare_elasticc_Ic} \end{center}
\end{figure*}
\begin{figure*}
  \begin{center} \includegraphics[width=0.95\textwidth]{ 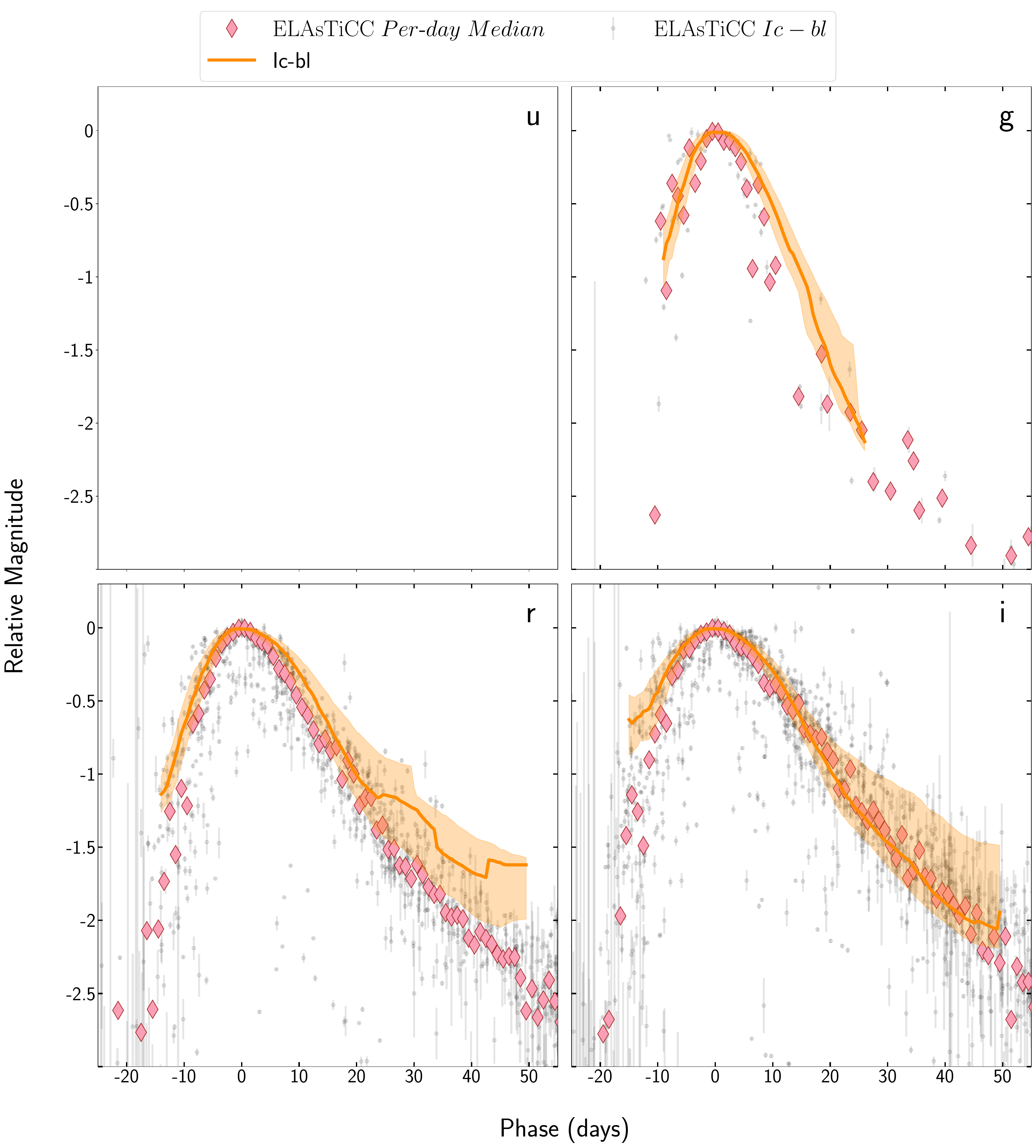} \caption{ The GP template of subtype Ic-bl (orange curve) is plotted along with \elasticc\ light curves of SN Ic-bl in $u, g, r,i$ bands (black dots) and their per-day medians (pink diamonds). The \elasticc\ data shows a behavior broadly consistent with our SNe Ic-bl sample. We note a deviation in late-time behavior in $r$ band (phase $>20$~days), but our templates are rather noisy in this region.} \label{fig:GPcompare_elasticc_Ic-BL} \end{center}
\end{figure*}

In 2022, the next generation of LSST-related light curve simulations was released under the name of The Extended LSST Astronomical Time-Series Classification Challenge (\elasticc, \citealt{2023AAS...24111701N}). The new simulations constitute an upgrade to the \plasticc\ data in many ways \citep{2023MNRAS.520.2887L}, including information about the galaxy host of transients and ``alert''-level information to simulate real-time response to LSST discovery. The SESN sample here is not generated via MOSFiT, and it includes recent CC spectrophotometric templates
from \citealt{vincenzi2019spectrophotometric} (discussed in \autoref{sec:intro_sec}).

In this dataset, the \Ibcs\ are split into their subtypes: \Ibs, \Ics, and \blIcs. We plot these light curves in  \autoref{fig:GPcompare_elasticc_Ib}, \autoref{fig:GPcompare_elasticc_Ic}, and \autoref{fig:GPcompare_elasticc_Ic-BL} respectively. We select objects with $z< 0.2$ with at least one data point between phases $-5 \leq \maxep \leq 5$ days and at least five data points with $\SNR>10$ overall. With more light curves available in this dataset, we also measure the per-day median of all the \elasticc\ light curves in each band for each subtype.
We notice general consistency between our data and the \elasticc's light curve sample, with their daily medians contained within our templates' variance, except for \Ics, where our templates show a slower rise in \rband\ and \iband\ (\elasticc's samples in \uband\ and \gband\ bands are sparse and noisy at early and late times). The difference in the \Ic\ early-time rise between our templates and the \elasticc\ light curves would be problematic in real-time classifications as it raises similarity to other SN types and misleads the classifiers.
We notice however the variance in the evolution of the \elasticc\ light curves for both \Ib\ and \Ic\ types is far larger than the variance in our sample. The \elasticc\ light curves for a number of \Ibs\ and \Ics\ are more rapidly evolving than our observed \Ib\ and \Ic\ templates, and more consistent with \Ibns.
Finally, the \blIc\ \elasticc\ sample is too sparse for a detailed comparison, but it seems to generally reflect the same characteristics as the \Ib\ and \Ic\ samples. 


There is a significant pressure in enabling photometric transient classification to advance time-domain astrophysics in the LSST era, where the spectroscopic samples will be dwarfed by the photometric samples given the lack of large-aperture dedicated spectrographs. If indeed the LSST simulated data on SESNe are inconsistent with the distribution of observed properties of these objects, this inconsistency can bias photometric classifiers trained on these datasets and lead to unrealistic expectations of accuracy. For example: the SCONE classifier \citep{Qu22}, trained on the \plasticc\ data, has a 0.11 (0.07, 0.04) contamination rate of \Ibcs\ in the \Ia\ samples at trigger (5 days after trigger, and 50 days after trigger respectively) and a 0.17 (0.16, 0.08) contamination of \Ibcs\ in the SLSNe samples when spectroscopic information is not available. We note that SLSNe are characteristically slowly evolving and this contamination rate can be explained if the \Ibc\ training sample is biased toward slow evolvers. This might lead to an underestimate of the contamination rate in SN Ia (cosmological) samples. Similarly, \cite{gagliano2023first} has developed a multi-modal SN classifier that uses galaxy information jointly with photometry for early classification of transients, is trained on \elasticc\ and ZTF SN photometry, and we speculate that its poor performance for the \Ibc\ classification may be due to the discrepancy between the \elasticc\ simulations of \Ibc\ and their real behavior.

\begin{figure*}[ht!]
    \centering
    \includegraphics[width=0.9\textwidth]{ 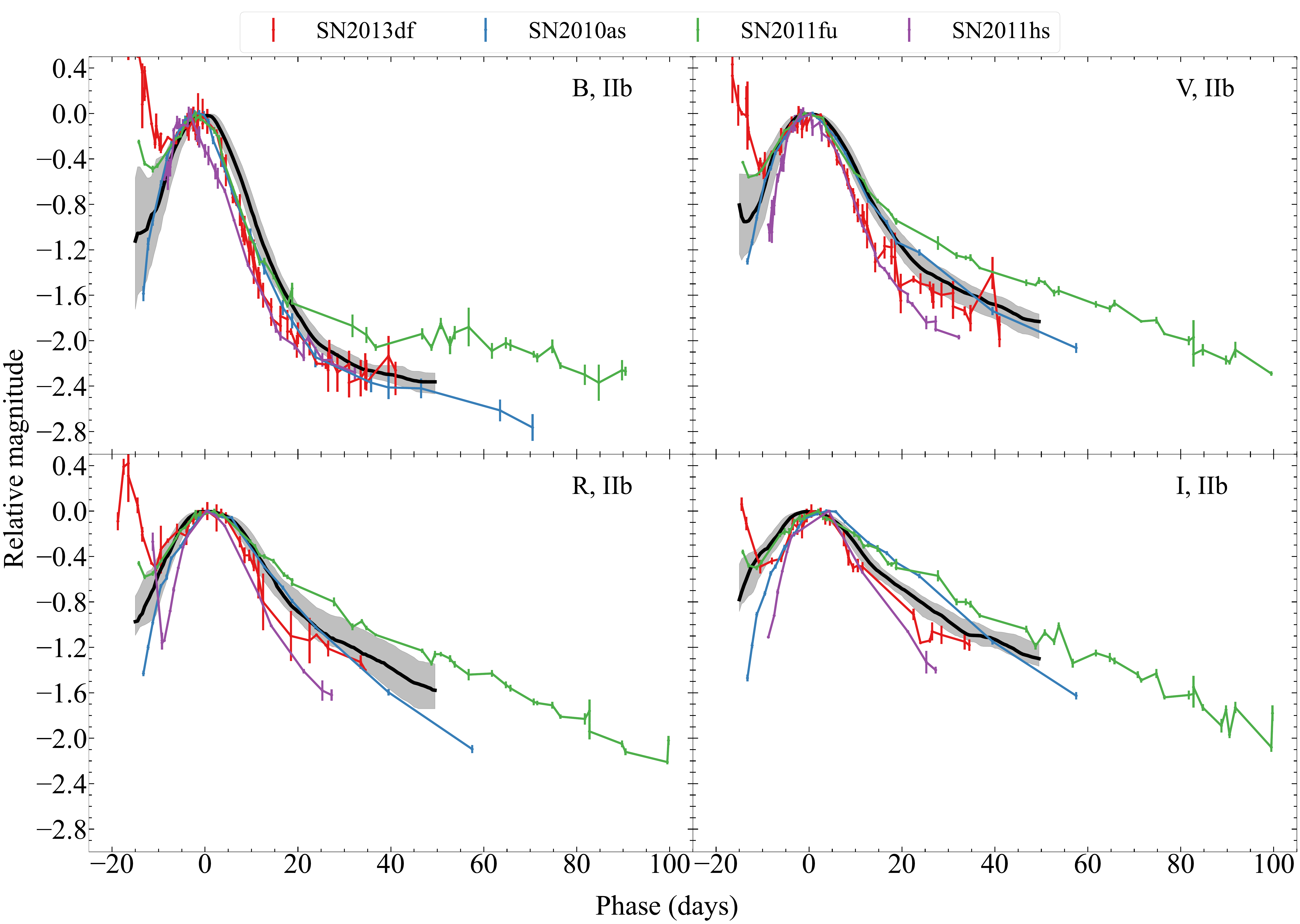}

    \caption{GP templates for the \IIb\ subtype are shown in bands $B$, $V$, $I$ (black curves) along with their IQR (grey shaded area). In each band, light curves of SNe with unusual spectroscopic or photometric features are plotted to compare their photometry with the GP templates. This figure is discussed in \autoref{GP_atypical}}
    \label{fig:GP_atypical_IIb}
\end{figure*}

\begin{figure*}[ht!]
    \centering
 
    \includegraphics[width=1\textwidth]{ 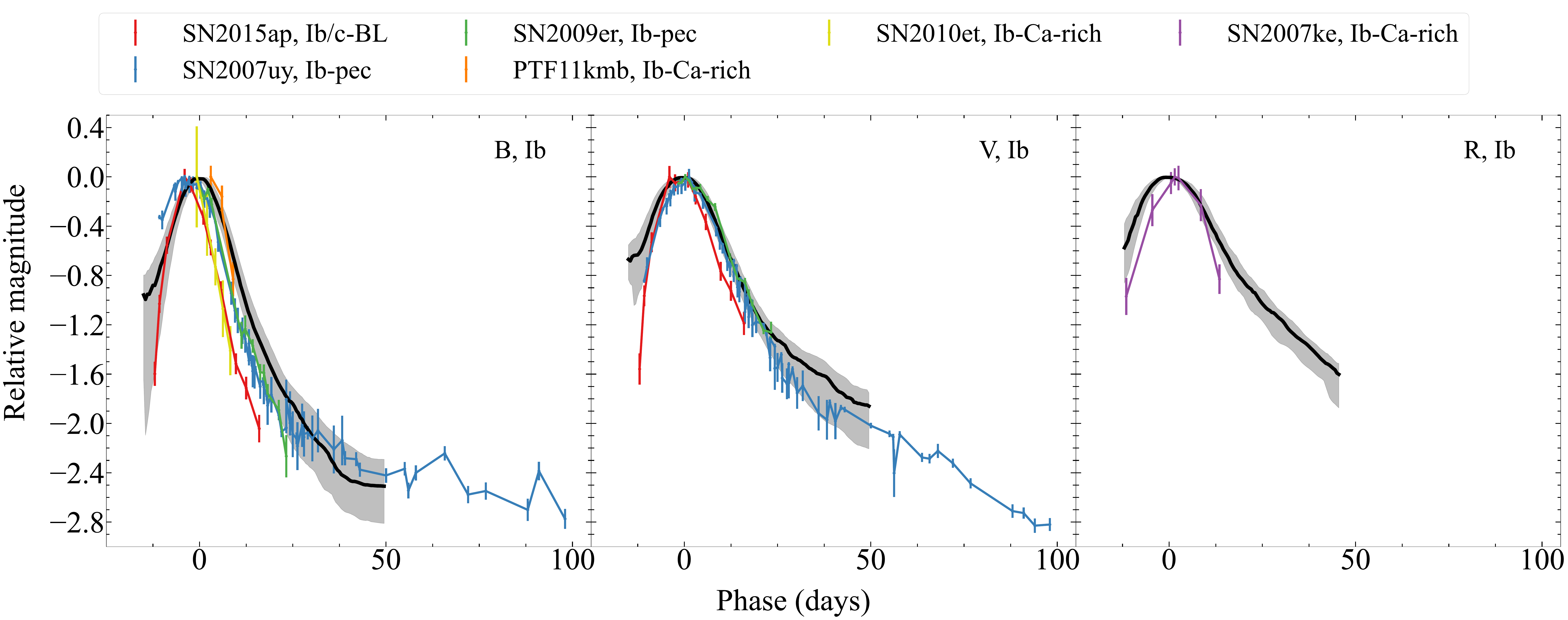}
    \caption{As \autoref{fig:GP_atypical_IIb} for \Ibs}
    \label{fig:GP_atypical_Ib}
\end{figure*}

\begin{figure*}[ht!]
    \centering
    \includegraphics[width=0.9\textwidth]{ 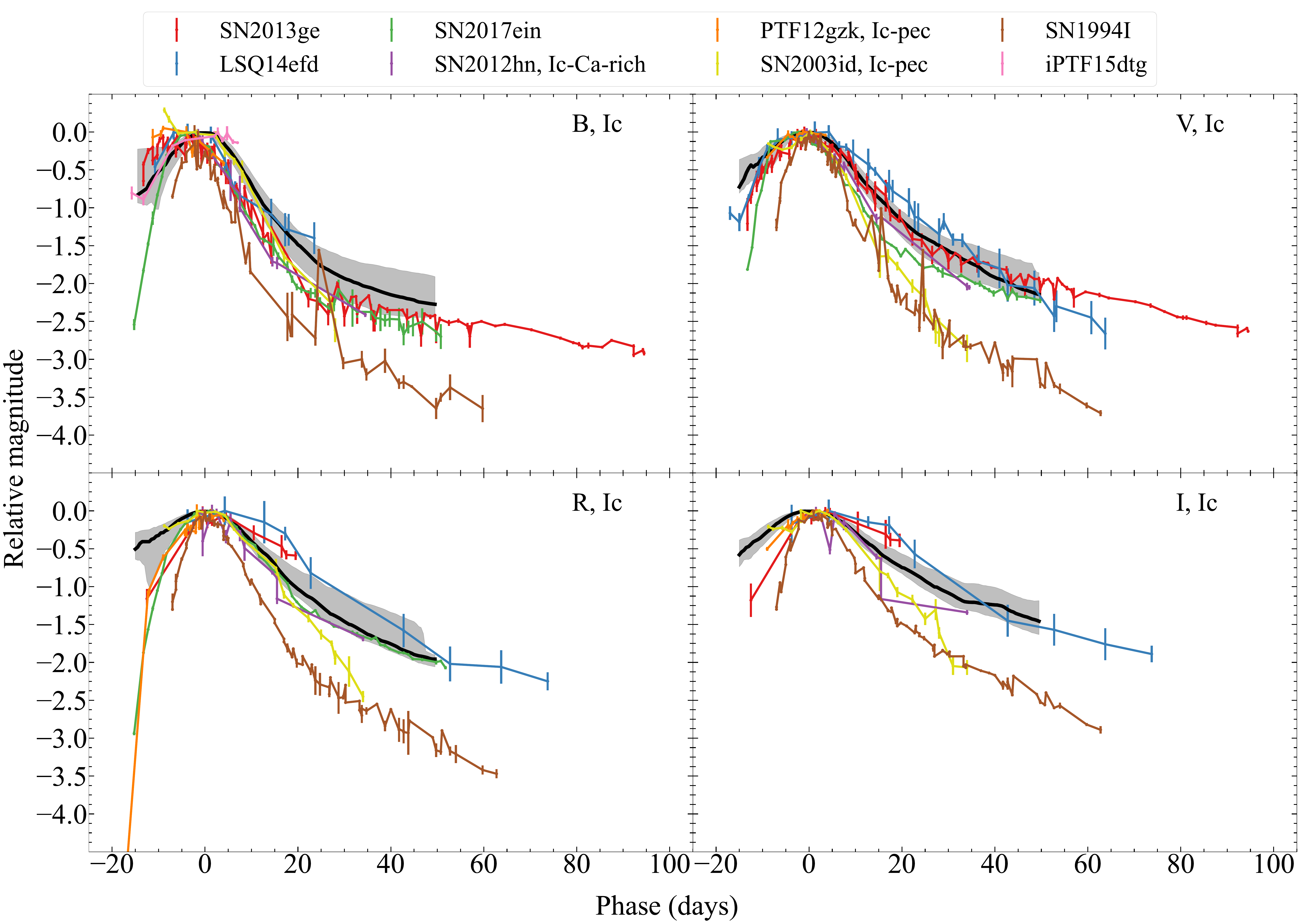}
    \caption{As \autoref{fig:GP_atypical_IIb} for \Ics}
    \label{fig:GP_atypical_Ic}
\end{figure*}

\subsection{Compare GP templates with unusual supernovae}\label{GP_atypical}

In this section, we compare our subtype templates to individual supernovae that have been claimed in the literature to show unusual features, because of their photometric or spectroscopic characteristics. We briefly discuss each \SESN\ and the reason it is considered atypical. For \SESN e identified as spectroscopically peculiar or known as peculiar in any photometric band. In \autoref{fig:GP_atypical_IIb} through \autoref{fig:GP_atypical_Ibn}, we plot their light curves in all bands where photometric measurements are available. Missing light curves in some bands are simply due to a lack of observations in that band.


\autoref{fig:GP_atypical_IIb} shows our GP templates for \IIbs\ in bands \bvri\ (black solid line) along with their 
IQR (grey area). Individual unusual \IIbs\ are plotted in colors and are discussed below.

\begin{itemize}
    \item SN2013df shows flat-bottom \Halpha\ absorption lines that could be an indication of an asymmetrical interaction between the ejecta and the CSM \citep{moralesgaroffolo14}. The photometry of this SN is overall consistent with our templates but their declining curves are close to the lower limit of the templates showing their rather fast decline.

    \item SN2010as is among a group of \IIbs\ that have weak \He\ and \Hy\ lines in their early time spectra and low expansion velocities \citep{folatelli2014supernova}. However, its photometry is consistent with our templates.

    \item SN2011fu is a spectroscopically normal \IIb\ that has been found to have a slow decline in late phases \citep{moralesgaroffolo15}. We confirm these findings in \autoref{fig:GP_atypical_IIb}.

    \item SN2011hs is believed to be a faint \IIb\ with a fast photometric rise and decline seen in \citet{bufano14}. Compared to our templates, it is indeed showing a fast rise in \Bband, \Vband, and \Rband\ and a fast decline in \Vband, \Rband, and \Iband\ bands.


\end{itemize}

\autoref{fig:GP_atypical_Ib} shows GP templates for the \Ib\ subtype in bands \Bband, \Vband, and \Rband\ (black solid line) along with their IQR (grey area). Individual unusual \Ibs\ are plotted in colors and are discussed below.

\begin{itemize}
    \item SN2015ap is classified as \Ib/c-bl and is known to have a fast evolution \citep{gangopadhyay2020optical} which we can confirm when comparing its light curve to our templates in \Bband\ and \Vband\ bands.

    \item SN2007uy and SN2009er are known to be peculiar \Ib\ types since they show broad unusual spectroscopic features compared to the spectra of normal \Ibs\ \citep{modjaz14}. However, the light curves of these SNe in the \Bband\ and \Vband\ filters are consistent with our GP template within the IQR.
    
    \item Calcium-rich (Ca-rich) supernovae are a subclass of SNe with spectra dominated by \Ca\ lines. There is debate over whether these SNe belong to the thermonuclear SN family (\Ias)\ or to the \SESN e family (\Ibcs) \citep{De2021Peculiar, Polin2021Nebular, Das2023Supernovae} as they are often found in the outskirts of the galaxies where \Ias\ are usually found \citep{2023arXiv230812991K}. Additionally, \citet{de2020zwicky} finds Ca-rich SNe with spectroscopic features similar to both \Ibcs\ and \Ias.     Many of these SNe show strong \He\ absorption lines and therefore we compare them with \Ibs\ here. We have plotted photometry for three Ca-rich Ib's in \autoref{fig:GP_atypical_Ib}: SN2007ke, PTF11kmb, and SN2010et. SN2007ke in the \Rband\ band and SN2010et in the \Bband\ band seem to decline faster than our templates. However, PTF11kmb in the \Bband\ band is within the uncertainties of our template (we discussed  SN2012hn, identified as a Ca-rich Ic, later in this section).




\end{itemize}

\begin{figure*}[ht!]
    \centering
    \includegraphics[width=0.9\textwidth]{ 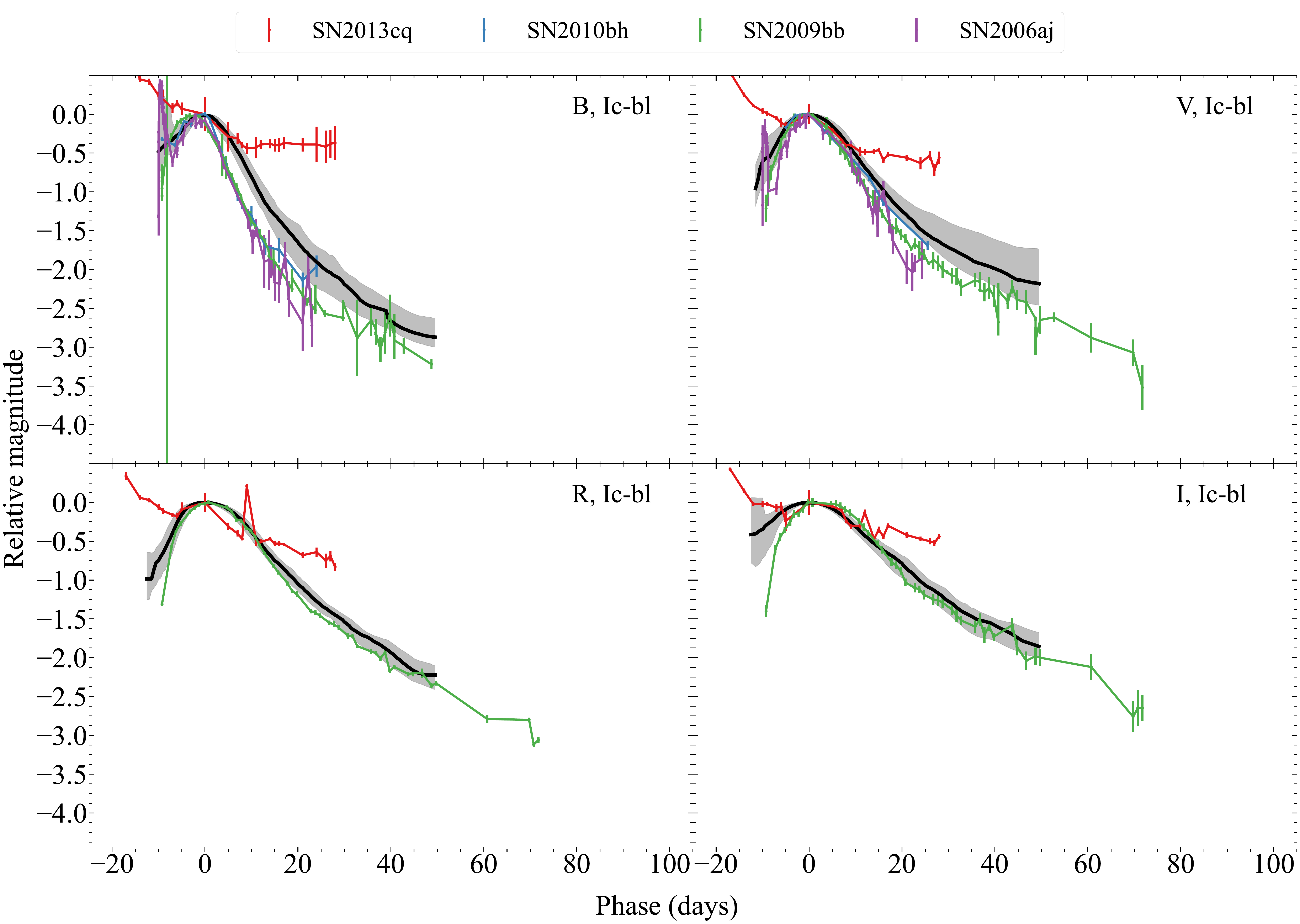}
    \caption{As \autoref{fig:GP_atypical_IIb} for SN Ic-bl.}
    \label{fig:GP_atypical_Ic_bl}
\end{figure*}

\begin{figure*}[t!]
    \centering
    \includegraphics[width=0.9\textwidth]{ 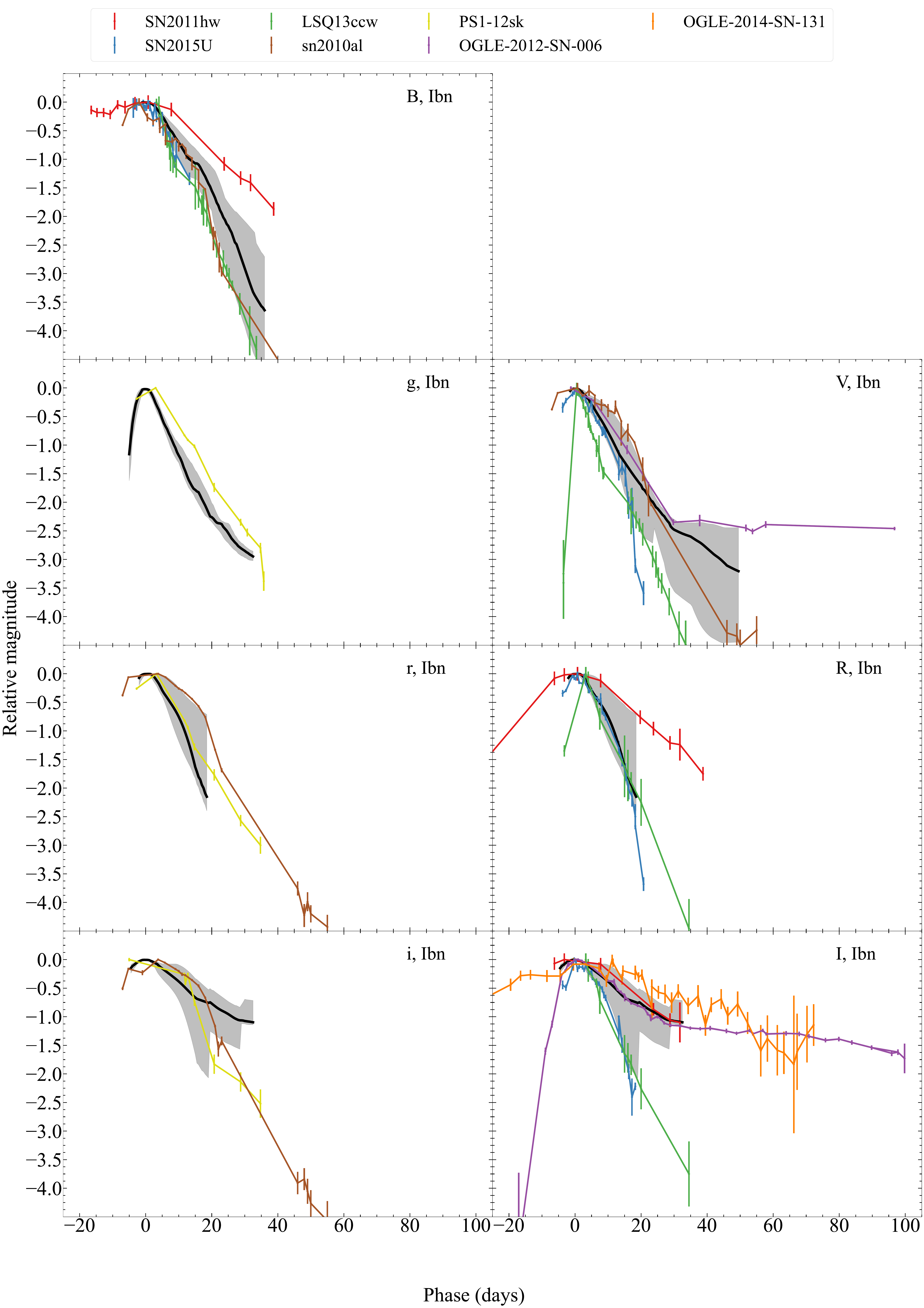}
    \caption{As \autoref{fig:GP_atypical_IIb} for SN Ibn. }
    \label{fig:GP_atypical_Ibn}
\end{figure*}

\autoref{fig:GP_atypical_Ic} shows our GP templates for \Ics\ in bands \ubvri\ (black solid line) along with their IQR (grey area). Individual unusual \Ics\ are plotted in colors and are discussed below.

\begin{itemize}

    \item SN2013ge showed a double-peaked light curve pronounced in the Swift UV bands, but also somewhat in the \uband\ and \Bband\ bands and early spectra with narrow absorption lines with high velocities \citep{drout2016double}. However, the optical photometry of this SN in redder bands (\Vband \Rband \Iband) is generally within the uncertainties of our GP templates except for possible signatures of a slow decline in \Rband.

    \item LSQ14efd shows a variety of unusual spectroscopic features with similarity to \Ics, \blIcs, and late-time \Ias\ spectra \citep{barbarino2017lsq14efd}. Its photometry is however overall consistent with our GP templates.

    \item SN2017ein has narrow spectral lines with high velocities similar to SN2013ge \citep{xiang2019observations}. The photometry of this SN is mostly consistent but at the rapid evolution end of our GP templates with clear signs of a rapid rise in \Bband\ and \Vband, which do not show any signs of the claimed shock breakout emission that the authors claim.
    
     \item SN2012hn is believed to be a faint \Ic\ with absorption features of \Ca\ in its spectra and therefore, it is classified as 
     Ca-rich~Ic \citep{valenti2014pessto}. The photometry of this SN shows a slightly rapid decline compared to our GP templates.
     
     \item PTF12gzk has high expansion velocities but narrow absorption lines that make it a peculiar \Ic\ \citep{ben2012discovery, horesh2013ptf}. Its photometry is consistent with our templates except for the \Bband\ band where, however, the light curve is extremely noisy.

    \item SN2003id exhibits unusual spectroscopic features \citep{morrell2003supernovae}, and its light curves show a fast decline and double-peaked rise compared to our GP templates in all bands.

    \item iPTF15dtg has normal \Ic\ spectra \citep{taddia2016iptf15dtg}. Photometry is only available in the \Bband\ band where we find a wide photometric peak and signs of a double-peaked light curve.

    \item SN1994I is a well-observed  SN Ic that is often claimed as a prototypical \Ic, but compared to our templates, it has very rapid evolution which is on average two to five times the uncertainty region below the GP templates in \bvri\ bands. Both the fast rise and fast decline of SN~1994I have been mentioned before in \citetalias{drout11} and  \citetalias{bianco14}.

\end{itemize}

\autoref{fig:GP_atypical_Ic_bl} shows our GP templates for the \blIc\ subtype in bands \bvri\ (black solid line) along with their IQR (grey area). Individual unusual type Ic-bl \SESN e are plotted in colors and are discussed below.


\begin{itemize}
    \item a GRB afterglow contaminates item SN2013cq in early phases \citep{melandri2014diversity} and it has a flat smooth decline compared to our templates due to the GRB afterglow, which is a power law.

    \item SN2010bh is a spectroscopically normal \blIc\ with GRB but it appears to have higher velocities at late phases compared to other SNe of this type \citep{chornock2010spectroscopic, modjaz2016spectral}. It also appears consistent with, although at the rapid evolution end of, our template in \Bband\ and \Vband\ band where photometry is available.

    \item SN2009bb is known to be a peculiar \blIc\ due to the claimed presence of \He\ in its early spectra (though it is not very convincing) and its associated radio emission \citep{pignata11}. The photometry of this SN shows a slightly faster decline than our template in all bands.

    \item SN2006aj is a \blIc\ associated with GRB 060218 \citep{mirabal2006grb}. Light curves of this SN (available in \Bband\ and \Vband\ only) show the signature of shock breakout and have a faster decline compared to our templates.
\end{itemize}

\begin{figure}
    \centering
 
    \includegraphics[width=0.5\textwidth]{ 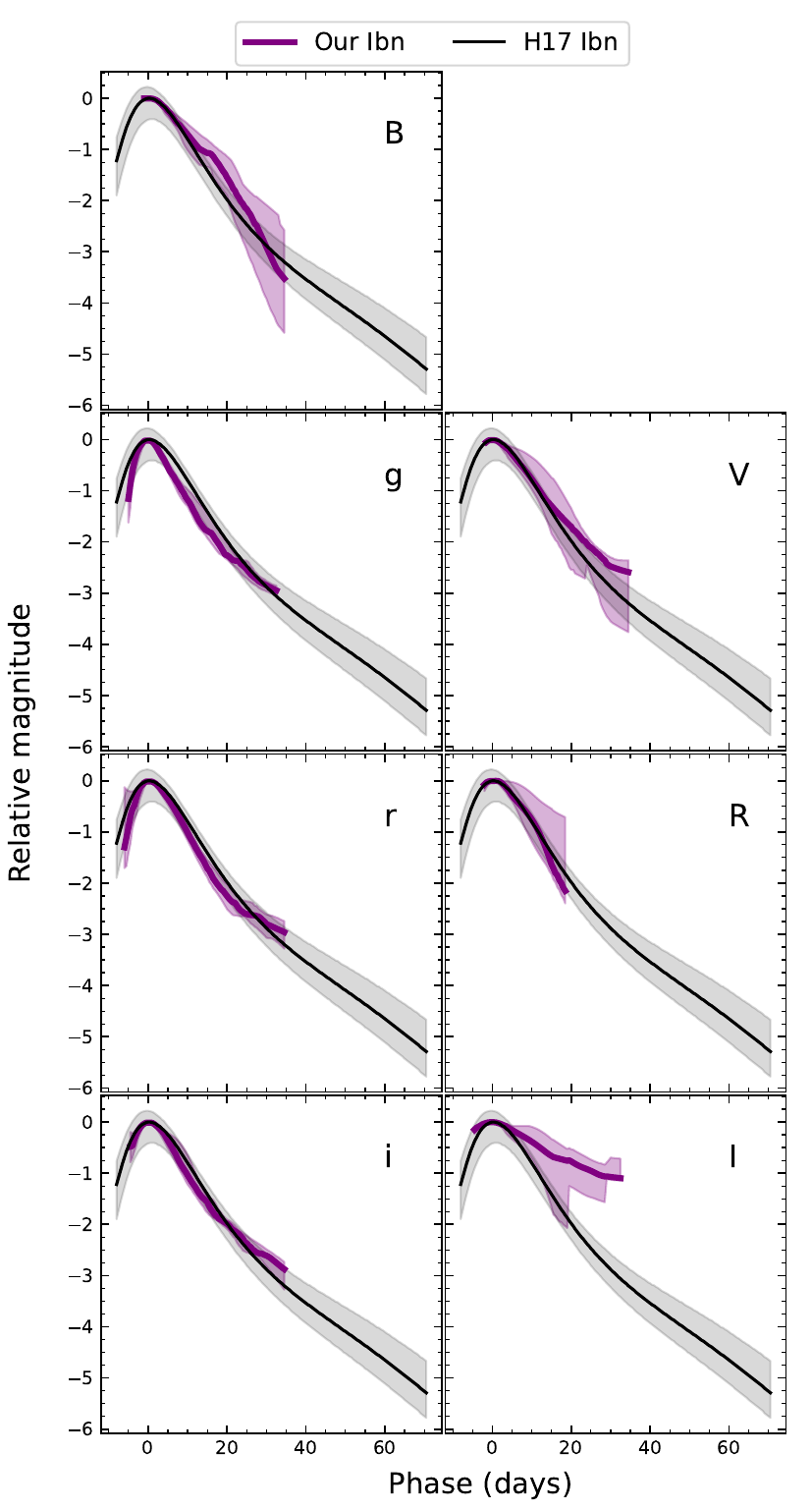}
    \caption{GP templates for \Ibns\ in \bthoughi\ bands  (purple curves) are plotted along with the \citetalias{hosseinzadeh2017type} \Ibn\ template, which is the same in all bands. The behavior of \citetalias{hosseinzadeh2017type}'s \Ibn\ template is mostly consistent with our \Ibn\ template within the uncertainty. A faster rise in \gband\ and \rband, where we have early data, is observed. However, the \gband\ data that generates this template is selected for its rapid evolution (\autoref{sec:compare_ZTF}), introducing a selection bias that may explain the discrepancy.}
    \label{fig:Ibn_H17_compare}
\end{figure}

\subsection{\SESN e Ibn}\label{sec:Ibn_compare}

A small subset of type Ib \SESN e show narrow helium emission lines, thus \Ibn, which are due to interaction between SN ejecta and helium-rich circumstellar material (CSM) (\citealt[][hereafter \citetalias{hosseinzadeh2017type}]{hosseinzadeh2017type}, \citealt{hosseinzadeh2019type}). We created and presented here the first photometric templates for these types of supernovae in 7 bands (a broadband template was presented in \citetalias{hosseinzadeh2017type}). Our sample of \Ibns\ includes 8 \SESN e from the \citetalias{guillochon16} and 6 \SESN e from a sample of ZTF objects that were released in \citetalias{ho2023photometric}. Note that the ZTF sample includes photometry in \gband\ and \rband\ bands, and \iband\ band at a lower cadence, thus our \Ibn\ templates in these bands will be mostly dominated by the ZTF sample.


The ZTF sample includes \SESN e that appear brighter than half of their maximum brightness for less than 12 days (\citetalias{ho2023photometric}). The inclusion of these data in our sample, even those that appeared in the literature after \citetalias{guillochon16} stopped being updated, allows us to increase our \Ibn\ sample size and produce templates for this subtype. However, note that not all objects in this sample are \Ibns. Only spectroscopically confirmed \Ibn\ are used to create our \Ibn\ GP templates (the other light curves in \citetalias{ho2023photometric} are compared with our templates in \autoref{sec:compare_ZTF}).


\autoref{fig:GP_atypical_Ibn} shows GP templates for \Ibns\ in bands \Bband, \gband, \Vband, \Rband, \rband, \Iband, and \iband\ (black solid line) and their IQR (grey area). Along with the templates, we plot all of the \Ibn\ light curves in our sample except for the ZTF objects (\citetalias{ho2023photometric}) since these SNe were explicitly selected for their fast-evolving light curves. We note the relatively slow evolution of \Ibns\ 2011hw, OGLE-2012-SN-006, and OGLE-2014-SN-131, compared to these templates in \gband, \rband, and \iband\ bands. We also see the particularly rapid evolution of LSQ13ccw, faster than our template in all bands available (\bvri), and of SN2015U, which is faster at all times in \Iband\ and at late times in \Vband\ compared to our templates. Note that the \Ibn\ template in the \gband\ band is smoother and with smaller uncertainty regions since the slow-evolving \Ibns\ in our sample did not have \gband\ band data and is dominated by the ZTF sample. This introduces a selection bias, as light curves in this sample are explicitly selected for their rapid evolution.

\citetalias{hosseinzadeh2017type} also presents a \Ibn\ template which is generated by combining all existing photometry, including photometry in different bands, of 18 \Ibns. This average template is created, as ours, fitting the combined light curves with a GP in logarithmic time. The sample includes light curves mostly in \rband/\Rband\ band, and a few objects in \Vband\ (1), \gband\ (1), \Iband\ (1), and \zband\ (1) bands. Their uncertainty regions are created by fitting the positive and negative residuals and selecting the \%95 probability region. For comparison, we plot their template along with our \Ibn\ templates in all of the 7 bands in \autoref{fig:Ibn_H17_compare}. The templates are consistent within the uncertainty, except in the \gband\ band, where we observe a faster rise in our template, and in the \Iband\ band we see a much slower decline in our template compared to the \citetalias{hosseinzadeh2017type} \Ibn\ template, but our uncertainties are very large, and our time baseline very short. 
The selection bias discussed in the previous paragraph explains this discrepancy.

\subsection{Comparison with fast-evolving ZTF objects}
\label{sec:compare_ZTF}

\begin{figure*}
  \begin{center} \includegraphics[width=0.8\textwidth]{ 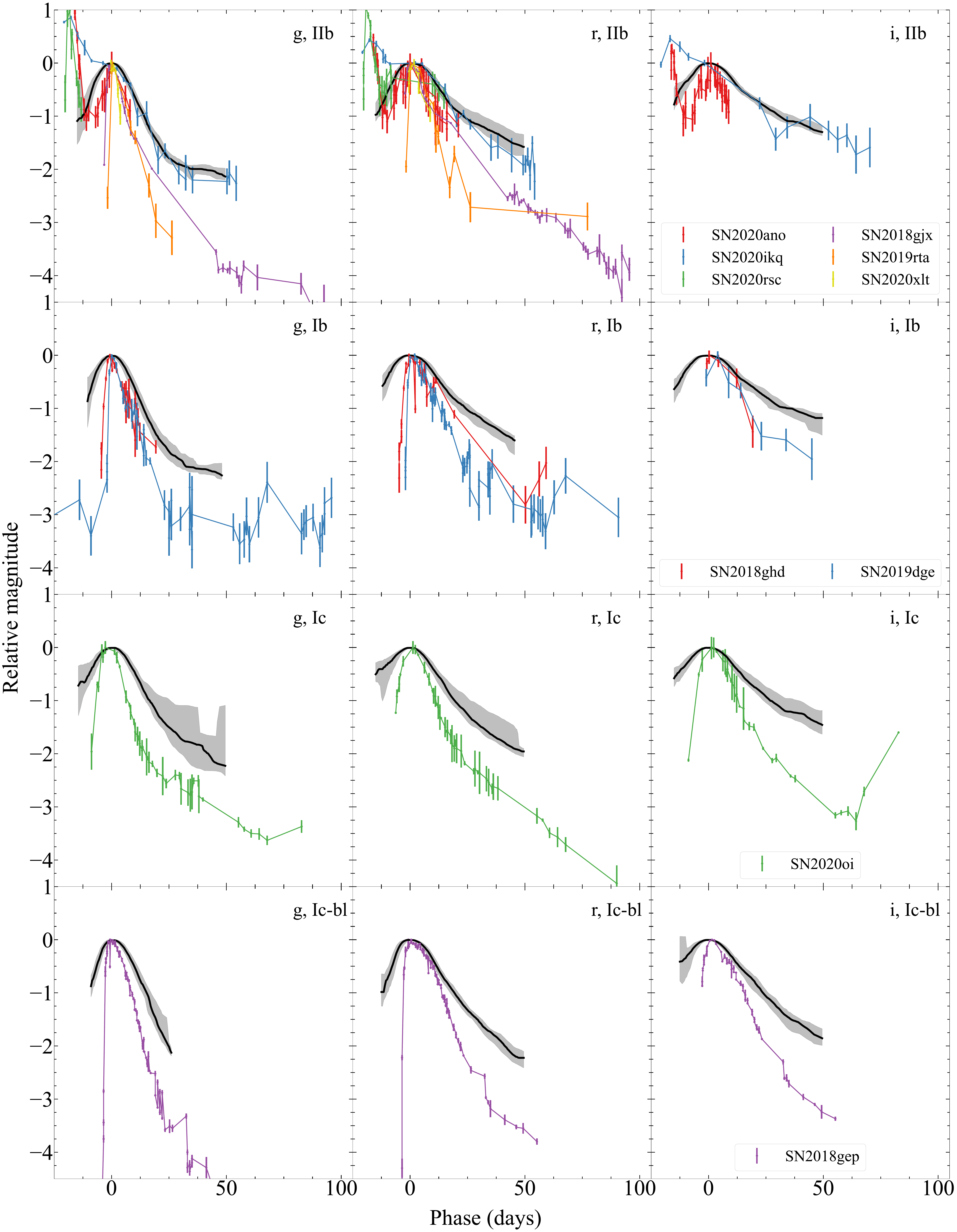} \caption{ Comparison of the fast-evolving \SESN e from \citetalias{ho2023photometric} with our GP templates of the same subtype as the spectroscopic SN classification (from top to bottom, \IIb, \Ib, \Ic, \blIc. This figure is discussed in \autoref{GP_atypical}} \label{fig:ztfcompare_all} \end{center}
\end{figure*}

In \autoref{fig:ztfcompare_all}, we show light curves of the fast-evolving SESN from \citetalias{ho2023photometric} that are not classified as \Ibns, and thus not used to construct our templates and compare them with our templates. In summary, all SNe Ib, Ic, and Ic-bl from \citetalias{ho2023photometric}'s rapidly evolving sample rise and fall more rapidly than any of our GP subtype templates, as expected based on their selection criteria, even though they were spectroscopically classified squarely within the four main subtypes. 
We focus our analysis on the frequently observed presence of two peaks in this sample. We address two questions. We discuss whether the peaks of light curves in  \citetalias{ho2023photometric} are in fact due to shock cooling emission as suggested by \citetalias{ho2023photometric} 
and answer the question of whether the \synNi-powered light curve did in fact have a fast evolution compared to our template. 

The decay of \synNi\ to \synCo\ is commonly believed to be responsible for powering the light curves of most thermonuclear and stripped-envelope SNe, including \SESN e \citep{arnett82}. The peak magnitude of the light curve would then be dependent on the total \synNi\ mass, while the ejecta velocity and mass jointly control the rise and decay rates in this simple model \citep[for applications to \SESN\ see][]{taddia2018carnegie,  Prentice2019Investigating}. However, we occasionally observe two peaks in the \SESN\ light curves, particularly those of \IIbs. Pre-maximum lightcurve peaks are due primarily to two related phenomena: shock breakout, or the associated cooling of an extended envelope heated by the shock \citep[\eg][]{chevalier2008shock, Waxman_2017, sapir2017uv, piro2021shock}. Shock breakout refers to when the shockwave generated by the collapsing star core breaks through the star's surface. This produces an intense, short-lived ($\sim$minutes) flash of radiation at high-frequency wavelengths (X-rays through UV). Shock breakout signatures have been observed for \blIcs\ associated with GRB SN2006aj   \citep{Campana_2006, li2007shock}, \Ib\ SN2008D \citep{soderberg2008extremely, modjaz2009shock}, and \IIb\ SN2016gkg \citep{bersten2018surge}. 

The direct detection of a shock breakout is only possible when the progenitor is compact and observations at short wavelengths are available on time scales of minutes to $\leq 1$ hour after the explosion. The outer layers of the star, heated by the passage of the shock wave, continue to expand and cool down over time emitting radiation in the UV and optical wavelengths that manifests as an early (pre-\maxep ) peak with slower ($\sim $days) evolutionary time scales. This emission is called cooling-envelope or shock cooling and in some cases, it can involve the interaction of the shock with a dense circumstellar wind rather than a stellar envelope \citep[\eg, SN 2020bio][]{pellegrino2023sn}. In what follows, we will refer to it as shock cooling emission to differentiate it from the \synNi-driven peak, without further speculation on the specific mechanisms that generated it. For a detailed discussion of observational shock breakout and cooling-envelope/shock-cooling emission in \SESN e see the review by \citealt{modjaz2019new} and reference therein.
Several of the lightcurves in \citetalias{ho2023photometric} showed two peaks. This prompted us to extend the baseline of the light curves presented in \citetalias{ho2023photometric} to ensure we could probe and investigate the frequency and characteristics of these double-peaked signatures in this sample.

 We note that the \citetalias{ho2023photometric} aligned the light curves on the first peak they observed. 
To ensure that the full light curve evolution is captured (including any potential 2nd peaks), we downloaded the ZTF forced photometry\footnote{\url{https://ztfweb.ipac.caltech.edu/cgi-bin/requestForcedPhotometry.cgi}} of 9 SNe from \citetalias{ho2023photometric} including 6 \IIbs: SN2018gjx, SN2019rta, SN2020ikq, SN2020xlt, SN2020ano, and SN2020rsc, two \Ibs: SN2019dge, SN2018ghd, and one \Ic: SN2020oi. In most cases, the resulting light curves show some differences from \citetalias{ho2023photometric}.\footnote{There is no difference in the  downloaded ZTF forced photometry for \blIc\ SN2018gep from the light curves} published in \citetalias{ho2023photometric}. In the case of SN 2019rta, and SN 2020rsc, for example, the ZTF forced photometry provided us with a longer time baseline and, in several cases, it results in lightcurves with less scatter compared to the \citetalias{ho2023photometric}'s published photometry (\eg, SN 2018ghd, where we also obtained the lightcurve in $i$ band, SN 2018gjx, etc). We plot these light curves along with our GP templates in \autoref{fig:ztfcompare_all}. Each row shows one subtype in bands \gri.  

{\bf\IIbs}: We observed double-peaked light curves for half of the \citetalias{ho2023photometric} sample classified spectroscopically as \IIbs, where two peaks, phenomenologically consistent with a shock cooling followed by \synNi\ evolution are clearly observed for three out of the six \IIbs.
The first row of \autoref{fig:ztfcompare_all} shows \IIbs\ from \citetalias{ho2023photometric} along with our GP \IIb\ templates. 

SN2020ano and SN2020ikq have prominent early peaks that could be associated with a shock cooling emission (as discussed in \citetalias{ho2023photometric}). Our retrieved ZTF forced photometry shows a second peak also for SN2020rsc in \rband, which was not fully visible in \citetalias{ho2023photometric}'s photometry. We have fitted a second-order polynomial to the second peak to estimate its peak time and magnitude. 

We compare the three \IIbs\ with double peaks (SN2020ikq, SN2020ano, SN2020rsc) with our templates aligning them by the \maxep\ derived from the second peak to test if the \synNi-driven part of the light curves shows a rapid evolution. For SN2020ikq the (presumably) \synNi\ evolution is consistent with our \IIb\ template in all bands. This object does not seem to have rapidly-evolving \synNi\ -powered light curve. It does however have an extended shock cooling first peak which in the $i$ band fully hides the second peak. Note that this means that if we only had $i$ band data we would have assumed this is a single peak fast-evolving \IIb. We will return to this point in \autoref{sec:sum_concl}. 
For SN2020rsc, we see both peaks in the \rband\ band where the evolution of the (presumably) \synNi-driven peak is consistent with typical \IIbs\  (we note that in \gband\ we only have data for the shock-cooling portion of the light curve and that the $r$ band photometry we retrieved looks somewhat different from \citetalias{ho2023photometric}'s, in addition to have a longer baseline that straddles the second peak.) SN2020ano is marginally consistent with a typical \IIb\ in the \rband\ band but the photometric uncertainties are large and the evolution is more rapid than our \IIb\ templates in both \gband\ and \iband.

\IIbs\ SN2018gjx, SN2019rta  are confirmed as fast-evolving SNe with no detected second peak out to at least  $\sim80$ days after maximum. 

The light curve of SN2020xlt is too limited to ascertain the presence of a single or double peak. 

In summary, SN2020ano is the only double-peaked SN in this sample where the (presumably) \synNi\ evolution is atypically rapid. Two of the six rapid evolving \IIbs\ in \citetalias{ho2023photometric} appear to be typical \IIbs\ with shock-cooling signatures: SN2020rsc and SN2020ikq are consistent with normal \IIbs\ with cooling envelope emission followed by \synNi\ evolution. The shock-cooling phase is evolving more rapidly than the \synNi\ ones and drives the selection of these objects in the \citetalias{ho2023photometric} sample.



{\bf\Ibs}: For \Ibs\ SN2019dge and SN2018ghd, we find only one peak for each SN. Although the photometry shows a rise at phase $\sim 75$ days for SN2019dge in \gband\ and \rband\ bands, and for SN2018ghd in \rband\ band, we do not think that those data points indicate \synNi\ peak and they are too far from the first peak and the uncertainties are very large. While we cannot determine conclusively whether this single peak is powered by shock cooling emission, by the decay of \synNi\, or by interaction, we nevertheless align the light curves of these \Ibs\ with our \Ib\ template to explore their behavior. We find that the peaks of these two SNe have a faster rise and decay than our \Ib\ templates in all filters.

{\bf \Ics\ and \blIcs}: \Ic\ 2020oi is shown in the third row and \blIc\ 2018gep in the last row of \autoref{fig:ztfcompare_all}. Both show only one peak in their light curves.  We find that these two SNe are more rapidly evolving than the respective templates. Please note that SN2020oi in the original photometry from \citetalias{ho2023photometric} had a couple of data points showing the possible start of a rise in \iband\ band around phase $75$ days. In forced photometry, these points disappeared.

If, with a sufficient baseline and sufficiently dense photometry no second peak is observed, then these findings would be highly interesting: if the single peak is due to shock-cooling emission then the absence of the \synNi-powered peak indicates the production of a small amount of \synNi\ in these rapidly evolving CC SNe (even smaller than seen in SN2020bio, \citealt{Pellegrino2023SN2020bio}). 
For example, \citetalias{ho2023photometric} suggested cooling emission from a shock that broke out of a massive shell of dense CSM as the powering source for \blIc\ 2018gep, as confirmed by \citet{Pritchard2021SN2018gep}. Note the high absolute luminosity of SN 2018gep ($M_V \sim -19.53  \pm 0.23$ mag), which places it between the typical brightness of SLSNe and \blIcs. If the single, rapidly-evolving peak is in fact due to decay of \synNi, then the \citetalias{ho2023photometric} sample may uncover a new population of \SESN e that have very low ejecta masses;  \citet{2021ApJ...908..232R} modeled \Ic\ 2020oi ($M_V\sim-15$, \citealt{2022ApJ...924...55G}) to have very low ejecta and very low \synNi\ masses. Samples of \SESN e selected for their rapid photometric evolution result in collections that, in spite of similar light curve width, include objects with different evolutionary drivers. Note that these objects have very similar photometric evolution to SN 1994I, which, as discussed in \autoref{GP_atypical}, has long been considered a prototypical \Ic\, while in fact its photometry is all but typical. SN 1994I  would be included in fast-evolving SESN samples selected by photometric cuts. 
To better understand the nature of this subclass of \SESN e, a bonafide sample of \Ibcs\ should be a focus of future studies.

\section{Summary and Conclusion}
\label{sec:sum_concl} 

In the era of large surveys, like the ongoing ZTF and upcoming LSST, photometric classification is critical, as spectrographs will only be able to follow up a small fraction of the tens of thousands of transients that LSST will detect every night \citep{ivezic2019lsst, Malz2019Photometric}. With the goal of enhancing our knowledge of photometric behavior of less common SNe to address this need, we have produced two kinds of photometric templates for \SESN e  spanning UV through NIR bands, for subtypes \alltypes\ using a dataset of open-access light curves of \Nlit\ \SESN e, all \SESN e in the \citetalias{guillochon16} that have at least five data points in one band, sampling over the peak, and reliable classification, augmented by a small dataset of \Ibn\ light curves from the ZTF. 

We have generated generic SESN templates, \ie\ \Ibc\ templates, by aggregating all subtypes in each of the \allbands\ bands and computing a weighted rolling median (\autoref{fig:ubtcompare}). We identify a photometric bias towards bright objects that have lower measurement errors in templates generated based on weighted means, as commonly done in the literature (\eg, \citealt{drout11}). This bias produces templates with overestimated magnitudes at late times where the bias is not suppressed by large sample sizes (see the comparison in \autoref{fig:B_compare_temshownplates}). We advocate that templates should be generated from a more reliable statistical aggregate as we have done in this work, namely median and inter-quartile range (IQR,  \autoref{sec:IBCtemplates}).

For the subtype-specific templates generated with GP, we separated the \SESN e family into \alltypes\ and generated templates per band combining Gaussian Processes (GP) fits of the light curves (\autoref{sec:gptempaltes}). GPs provide a flexible statistical framework for modeling light curves, allowing us to place them on a fine grid in time, accounting for uncertainties and gaps in data, and leveraging a physics-informed model for the correlation between the data points. We have developed a robust approach to combine the light curves in our dataset that are observed with different instruments and under different conditions and to fill in gaps in the observations. Our method results in well-fitted GP models for light curves of each \SESN\ in our sample leading to templates that allow us to study time-dependent characteristics of these classes and identify outliers. Our final set of GP templates consists of 54 templates for all subtypes: \IIbs\ (in 12 bands), \Ibs\ (in 13 bands), \Ics\ (in 12 bands), \blIcs\ (in 10 bands), and \Ibns\ (in 7 bands). Using GP fits of individual \SESN\ in our sample allows us to take into account the diversity of all the existing photometric observations. Doing the GP fits in logarithmic time results in better fits to the early time-evolution of the \SESN e. A customized objective function in the GP fit that penalizes rapid variations provides smooth fits that avoid unphysically rapid magnitude changes.
In addition, aggregating the GP fits using a rolling median statistics and defining uncertainty regions using the IQR allow us to prevent biases towards brighter \SESN e with lower uncertainties. 

Comparing subtype templates (\autoref{sec:compare_GP}) reveals that within the currently available data, the photometric evolution of different \SESN e subtypes are mostly consistent with each other within their uncertainties, with the most notable differences being evidence of shock cooling emission in the early-time behavior of \IIbs\  (\eg,      \autoref{fig:GP_compare_Ib_IIb}, \Uband\ band) and the notably fast decline of \blIcs\ compared to SNe of types Ib, IIb, and Ic\ as shown in \autoref{fig:GP_compare_Ic_Ic_bl}, \autoref{fig:GP_compare_Ib_Ic_bl}, and \autoref{fig:GP_compare_IIb_Ic_bl}.

We compare GP templates of each subtype of \SESN\ with the \Ibn\ GP templates that we generated using data from \citetalias{guillochon16} and \citetalias{ho2023photometric} (see \autoref{sec:Ibn_compare}). We find that, formally, in the \rband\ and \iband\ bands, the decline rate as measured by $\Delta m_{15}$ is higher for \Ibns, and inconsistent with the decline rate of the rest of the \SESN\ family at the 2$\sigma$ confidence level: $\Delta m_{15} (r, \mathrm{Ibc}) =  0.60^{+0.19}_
{ -0.17}$ $vs$ $\Delta m_{15} (r, \mathrm{Ibn}) =  1.70^{+0.14}_
{ -0.15}$ and $\Delta m_{15} (i, \mathrm{Ibc}) =  0.36^{+0.11}_
{ -0.10}$ $vs$ $\Delta m_{15} (i, \mathrm{Ibn}) =  1.56^{+0.19}_
{ -0.11}$ (see \autoref{tab:del_m15} and \autoref{tab:del_m10}). 


Our templates are valuable in evaluating the consistency and veracity of the ensemble behavior of simulated light curve samples of \SESN e. This is particularly important when developing classifiers for \SESN\ subtypes in the era of large surveys like LSST when spectroscopic observations for most objects will not be available. Thus, we compared our GP templates with \Ibc\ light curves from the \plasticc\ and \elasticc\ simulated data sets (\autoref{sec:plasticc_compare}), which provide a sandbox dataset to train photometric classification models, and found that in the \plasticc\ sample, a significant number of objects had slower rise and decline than observed in our \SESN e sample. In the more recent \elasticc\ simulations, this issue is largely improved and the per-day median of the simulated light curves is generally consistent with our templates. However, the \elasticc\ light curves of \Ics\ have a slower rise than our GP \Ic\ templates, and those of \blIcs\ have a faster decline at later epochs compared to our GP \blIc\ templates.

Next, we compared both types of our templates (\Ibc\ and GP-based subtype-specific templates) to individual \SESN\ with known unusual photometric and/or spectroscopic characteristics to assess if they appear as outliers compared to our templates (\autoref{sec:compare_Ibc} and \autoref{GP_atypical}). We find that abnormal spectroscopic behavior does not necessarily result in unusual photometric evolution (for \eg, SN2007uy, SN2009er, SN2010as, SN2013ge, LSQ14efd). Our templates also confirm that a well-known and well-observed \SESN\ like SN1994I, which is claimed to be the prototypical \Ic, is in fact photometrically well outside of the uncertainty region of the GP templates for \Ics\ (2-to-8 times the uncertainty region) and \Ibc\ templates (2-to-7 times IQR) altogether in \bvri\ (\autoref{fig:ubtcompareSNe} and \autoref{fig:GP_atypical_Ic}).

When analyzing the sample of rapidly evolving \SESN\ presented in \citetalias{ho2023photometric} and comparing it with our templates we identified at least two normally evolving SNe IIb in the sample, both with shock-cooling emission (SN2020rsc, SN2020ikq) that were not recognized as such in \citetalias{ho2023photometric}. This results from aligning our \IIb\ templates by a second peak in \citetalias{ho2023photometric}'s sample, when present, which in some cases required extending the photometric baseline presented in \citetalias{ho2023photometric} to reveal a second peak consistent with \synNi\ evolutionary time scales for normal \IIbs.
We emphasize the importance of distinguishing the evolution of early peaks, (\eg, due to shock cooling or interaction) from the \synNi\ driven peaks to disentangle the physical processes that may lead to rapid evolution.
When photometry is sparse, limited to some bands, or seasonality only allows for a short light curve to be collected, the presence of two peaks may be difficult to ascertain. Since the duration of the lightcurve can probe physical processes including \synNi\ production and ejecta mass,  misidentifying shock cooling signatures for \synNi\ evolution may skew statistics on SN evolution and progenitor characterization and impair inference on the true nature of the evolutionary drivers. 


While being curated and updated, the \citetalias{guillochon16} was aggregating SN data, including photometry, from all publications into a single publicly accessible repository.  While at the time of writing the \citetalias{guillochon16} still contains most published \SESN e data, many more \SESN e have been observed in the past few years. Some light curves are available only in papers dedicated to one or a few objects, and the very significant effort required to compile a truly complete sample of \SESN e is beyond the scope of this paper. Without claiming completeness, we believe we have leveraged the vast majority of published \SESN\ photometry. Other projects like WISeREP\footnote{\url{https://www.wiserep.org/}} \citep{Yaron_2012} are engaging in similar efforts of collecting and making available published data for SNe. This is a difficult and time-consuming work, which requires infrastructural support that is difficult to obtain and sustain (as demonstrated by the loss of the \citetalias{guillochon16}). We hope our work can highlight the importance of funding and maintaining similar efforts to make open data accessible that are indispensable to enable these kinds of studies.

While at this time, our conclusions remain limited by the availability of data (the templates could not be generated for all \SESN\ subtypes in all bands, and even where they are generated they may be noisy due to the paucity of data), our models and templates are designed to be easily updated to incorporate a growing sample \SESN e. We have made our code and instructions on how to update templates publicly available and accessible.\footnote{\url{https://github.com/fedhere/GPSNtempl}}

Since light curves of different subtypes of \SESN e are used to probe outstanding questions about their progenitors \citep{lyman2016bolometric, taddia2018carnegie, Prentice2019Investigating, Karamehmetoglu2023supernovae}, our work, which includes a larger sample of \SESN e than any of these, sets the stage for probing the diversity of their explosion properties, stripping channels and progenitor rates.

\section{Acknowledgment}\label{sec:thanks}

We used the following {software} packages: \texttt{pandas} \citep{mckinney2011pandas},  \texttt{numpy} \citep{harris2020array}, \texttt{matplotlib} \citep{hunter2007matplotlib}, and \texttt{george} \citep{foreman2015george}.

The authors wish to thank Saurabh Jha for insightful discussions and suggestions and Kevin Krisciunas for providing the unpublished NIR photometry of SN 1999ex. The authors acknowledge the support of the Vera C. Rubin Legacy Survey of Space and Time Science Collaborations\footnote{\url{https://www.lsstcorporation.org/science-collaborations}} and particularly of the Transient and Variable Star Science Collaboration\footnote{\url{https://lsst-tvssc.github.io/}} (TVS SC) that provided opportunities for collaboration and exchange of ideas and knowledge. This publication was made possible through the support of an LSSTC Catalyst Fellowship to S.K., funded through Grant 62192 from the John Templeton Foundation to LSST Corporation. The opinions expressed in this publication are those of the authors and do not necessarily reflect the views of LSSTC or the John Templeton Foundation. M.M. acknowledges support in part from ADAP program grant No. 80NSSC22K0486, from the NSF grant AST-2206657 and from the HST GO program HST-GO-16656. C.L. acknowledges support from the National Science Foundation Graduate Research Fellowship under grant No. DGE-2233066.


This research has made use of the Open SuperNova Catalog (OSNC), NASA's Astrophysics Data
System Bibliographic Services (ADS), the HyperLEDA database and the
NASA/IPAC Extragalactic Database (NED) which is operated by the Jet
Propulsion Laboratory, California Institute of Technology, under
contract with the National Aeronautics and Space Administration.

\clearpage

\appendix
\section{Additional Figures}
\label{app:appA}

\autoref{fig:GP_compare_Ib_IIb}, \autoref{fig:GP_compare_Ib_Ic}, and \autoref{fig:GP_compare_IIb_Ic} are similar to \autoref{fig:GP_compare_Ic_Ic_bl}, \autoref{fig:GP_compare_Ib_Ic_bl} and \autoref{fig:GP_compare_IIb_Ic_bl} and show a comparison of the pairs of (\Ib, \IIb), (\Ib, \Ic), and (\IIb, \Ic) templates in all the bands where both subtypes have a template. These figures indicate that the photometric evolution of these subtype pairs are mostly consistent with each other within the uncertainty regions. 

\begin{figure*}
    \centering
\includegraphics[width=0.9\textwidth]{ 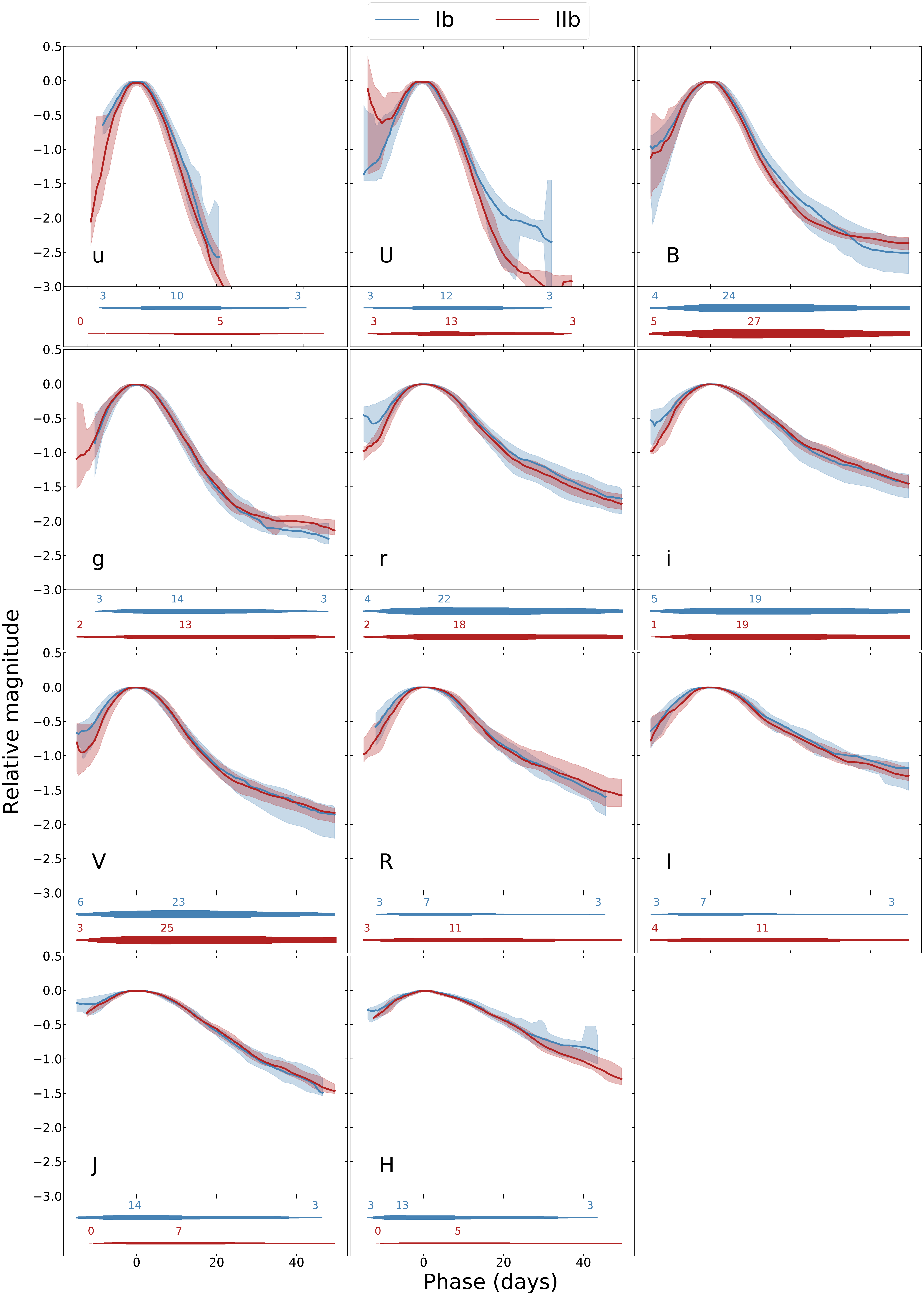}

    \caption{As \autoref{fig:GP_compare_Ic_Ic_bl} for {\it Ib} and {\it IIb} subtypes.}
    \label{fig:GP_compare_Ib_IIb}
\end{figure*}

\begin{figure*}
    \centering

\includegraphics[width=0.9\textwidth]{ 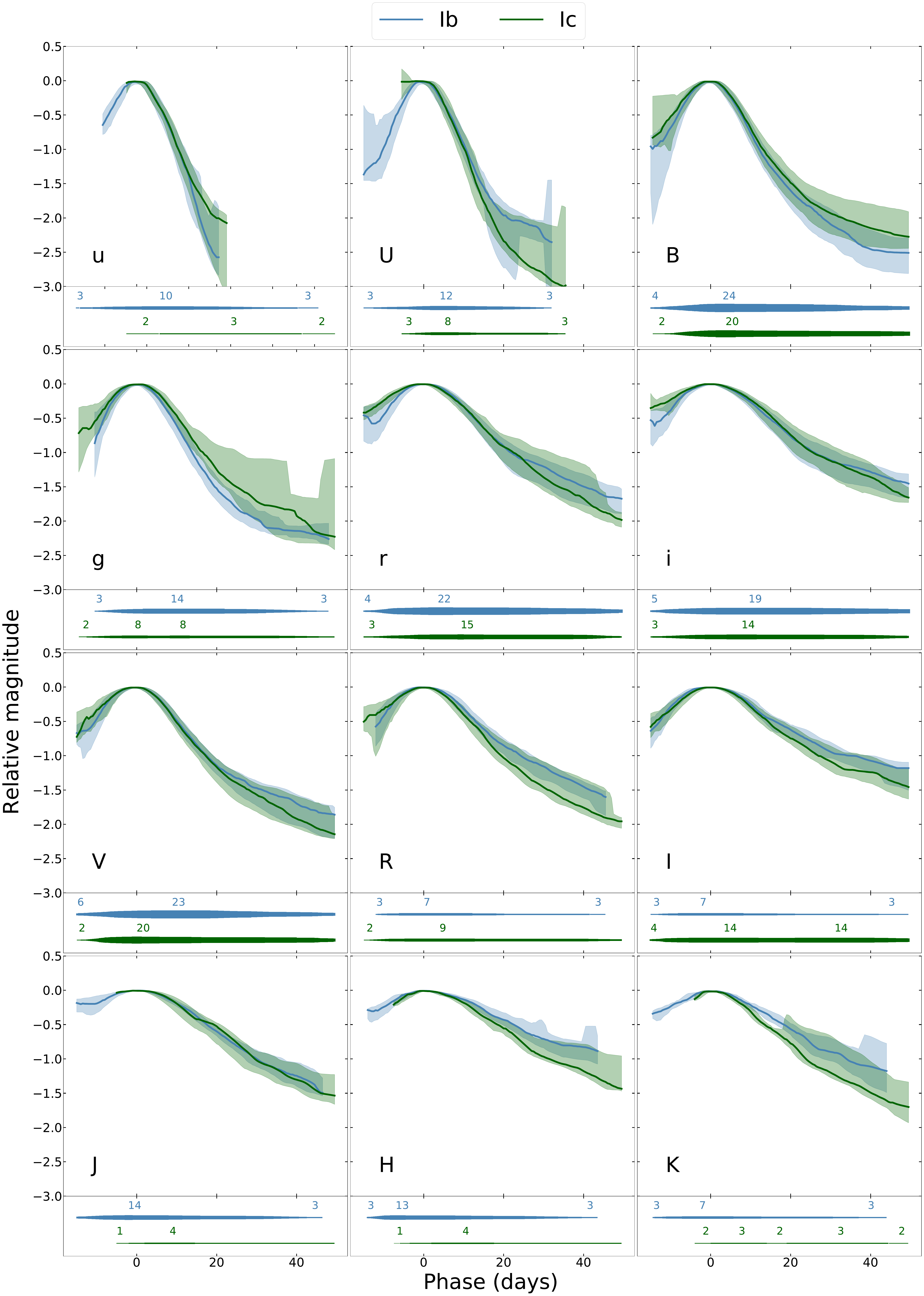}

    \caption{As \autoref{fig:GP_compare_Ic_Ic_bl} for {\it Ib} and {\it Ic} subtypes.}
    \label{fig:GP_compare_Ib_Ic}
\end{figure*}

\begin{figure*}
    \centering
\includegraphics[width=0.9\textwidth]{ 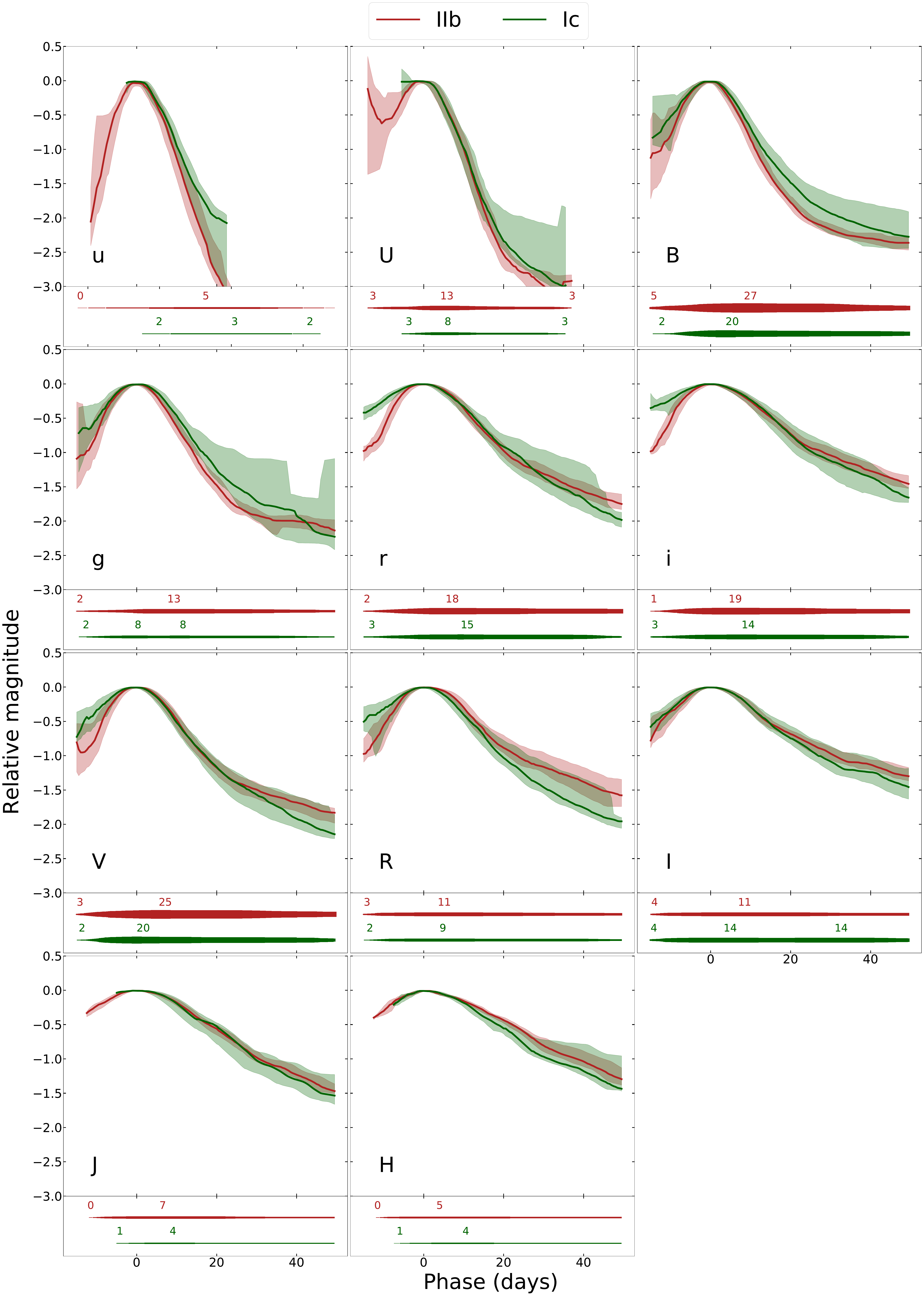}

    \caption{As \autoref{fig:GP_compare_Ic_Ic_bl} for {\it IIb} and {\it Ic} subtypes.}
    \label{fig:GP_compare_IIb_Ic}
\end{figure*}

\autoref{fig:color_evol} shows the evolution of relative colors for each subtype template in phases -10, 0, 10, 20, 30 days since \maxep. The trend in color seems similar for all subtypes and is only different for \Ibn\ subtype in \Vband-\rband\ and \Vband-\iband\ colors, but those have large uncertainties and therefore we cannot draw a definite conclusion from them.

\begin{figure*}
    \centering
\includegraphics[width=0.9\textwidth]{ 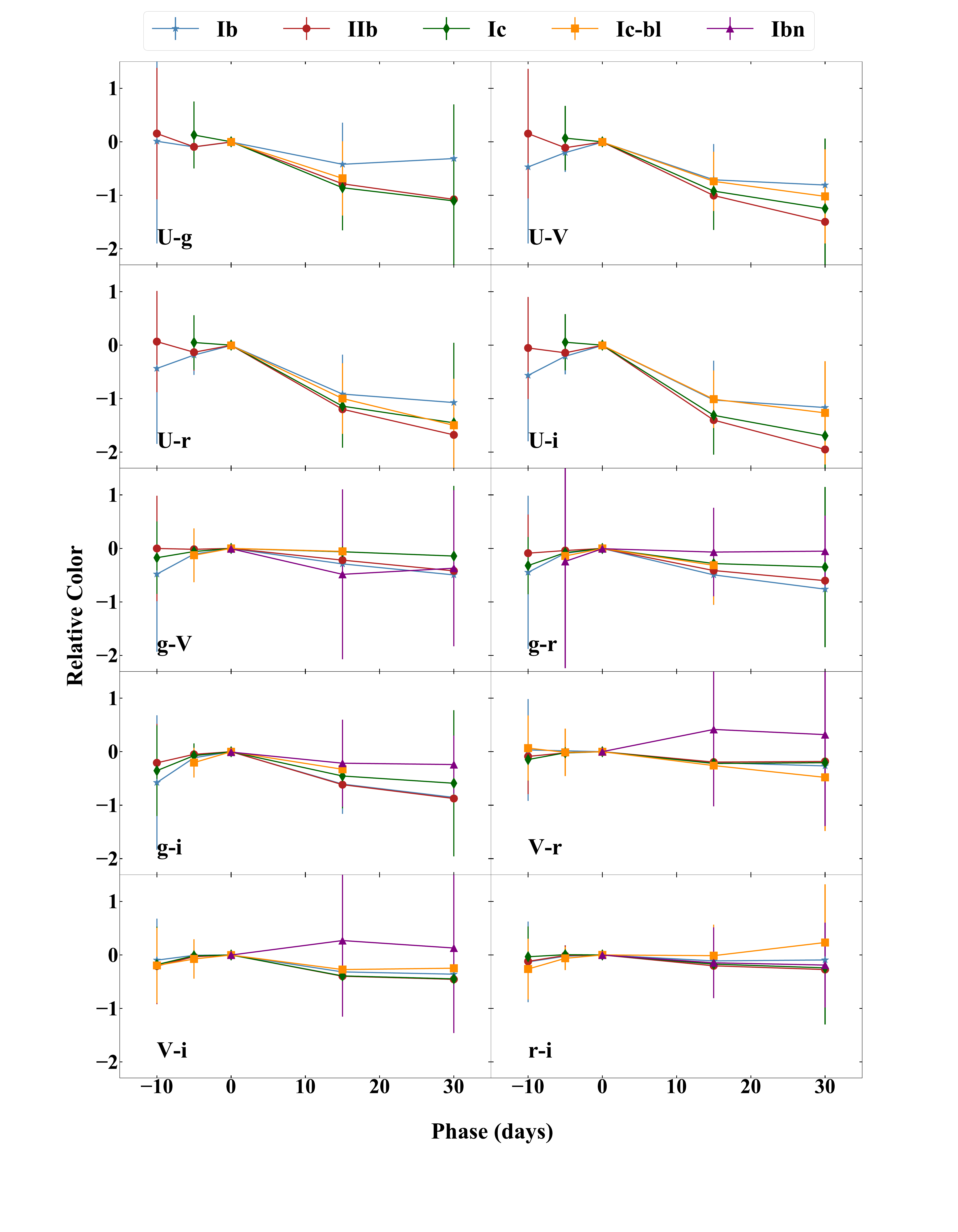}
 
    \caption{}
    \label{fig:color_evol}
\end{figure*}

\clearpage
\section{Additional Tables}\label{app:appB}

\autoref{tab:OPT1}, \autoref{tab:OPT2}, and \autoref{tab:UVNIR} include information about the photometry in our \SESN e sample. For each SN, the type, number of photometric data points per band, minimum and maximum epochs of the photometry per band, and the median \SNR\ per band are shown.

\begin{turnpage}
\begin{center}
\tabletypesize{\tiny}

\input{allphotUBVRI_snr2.tex}
\clearpage

\input{  allphotugri_snr2.tex}
\clearpage


  \input{  AllPhotUVNIRTable.tex}
\end{center}
\end{turnpage}

\bibliography{references}
\end{document}

%% file: photsample.tex
\startlongtable
\begin{deluxetable}{llccc}
\tablecaption{Photometric Templates \SESN e sample}\label{tab:SESNessentials}

\tablehead{\colhead{Name} & \colhead{Type} & \colhead{Vmax} & \colhead{Vmax err} & \colhead{z}}
\label{tab:SESNessentials}
\startdata
\input{photsample_data_sk_add_Ibns}
\enddata
\tablecomments{Classification of  \SESN e denoted with {$^*$} are modified by us after running SNID with the largest \SESN\ template database on their spectra close to maximum light.}
\end{deluxetable}

%% file: photsample_data_sk_add_Ibns.tex
SN1954A          &        Ib &  34850.3 &      3.6 &   0.001 \\
SN1962L          &      Ib/c$^*$ &  38008.3 &      0.1 &   0.004 \\
SN1983N          &        Ib &  45529.4 &      2.6 &   0.003 \\
SN1983V          &        Ic &  45681.4 &      2.0 &   0.006 \\
SN1984I          &        Ib &  45847.8 &      0.6 &   0.011 \\
SN1985F          &      Ib/c$^*$ &  45871.1 &      1.8 &   0.002 \\
SN1991N          &        Ic &  48349.0 &      2.0 &   0.003 \\
SN1993J          &       IIb &  49095.2 &      0.0 &   ----- \\
SN1994I          &        Ic &  49451.9 &      0.0 &   0.002 \\
SN1996cb         &       IIb &  50452.7 &      0.0 &   0.002 \\
SN1997ef         &     Ic-bl &  50793.2 &      0.3 &   0.012 \\
SN1998bw         &     Ic-bl &  50945.5 &      0.0 &   0.009 \\
SN1999dn         &        Ib &  51419.1 &      0.3 &   0.009 \\
SN1999ex         &        Ib &  51501.3 &      0.1 &   0.011 \\
SN2001ig         &       IIb &  52273.4 &      6.4 &   0.003 \\
SN2002ap         &     Ic-bl &  52314.5 &      0.1 &   0.002 \\
SN2002ji         &        Ib &  52620.9 &      1.4 &   0.005 \\
SN2003bg         &       IIb &  52721.0 &      1.3 &   0.005 \\
SN2003dh         &     Ic-bl &  52740.6 &      1.8 &   0.169 \\
SN2003id         &    Ic-pec &  52911.8 &      0.1 &   0.008 \\
SN2003jd         &     Ic-bl &  52943.3 &      0.6 &   0.019 \\
SN2004aw         &        Ic &  53091.5 &      0.7 &   0.016 \\
SN2004dn         &        Ic &  53231.1 &      0.3 &   0.013 \\
SN2004dk         &        Ib &  53243.9 &      1.9 &   0.005 \\
SN2004ex         &       IIb &  53308.4 &      0.3 &   0.018 \\
SN2004ff         &       IIb &  53314.9 &      1.5 &   0.023 \\
SN2004fe         &        Ic &  53319.3 &      0.5 &   0.018 \\
SN2004ge         &        Ic &  53335.6 &      1.6 &   0.016 \\
SN2004gq         &        Ib &  53363.0 &      1.9 &   0.006 \\
SN2004gt         &        Ic &  53363.1 &      0.2 &   0.005 \\
SN2004gv         &        Ib &  53367.4 &      0.2 &   0.020 \\
SN2005az         &        Ic &  53473.9 &      0.8 &   0.009 \\
SN2005bf         &        Ib &  53499.0 &      0.3 &   0.019 \\
SN2005by         &       IIb$^*$ &  53504.0 &      2.4 &   0.027 \\
SN2005fk         &     Ic-bl &  53627.5 &      3.1 &   0.234 \\
SN2005hl         &        Ib &  53632.1 &      2.2 &   0.023 \\
SN2005em         &       IIb$^*$ &  53648.6 &      0.4 &   0.025 \\
SN2005hm         &        Ib &  53652.0 &      2.0 &   0.035 \\
{\myfontsize SDSS-II SN 5339 } &      Ib/c &  53652.4 &      2.0 &   0.138 \\
{\myfontsize SDSS-II SN 4664  }&      Ib/c &  53673.4 &      2.8 &    ----- \\
{\myfontsize SDSS-II SN 6861}  &     Ib/c?$^*$ &  53673.4 &      2.3 &   0.191 \\
{\myfontsize SDSS-II SN 8196 } &      Ib/c &  53680.5 &      3.3 &   0.080 \\
SN2005hg         &        Ib &  53684.4 &      0.3 &   0.021 \\
SN2005kr         &     Ic-bl &  53690.6 &      2.2 &   0.134 \\
SN2005ks         &     Ic-bl$^*$ &  53691.5 &      1.4 &   0.099 \\
SN2005kl         &        Ic &  53703.7 &      0.6 &   0.003 \\
SN2005kz         &        Ic &  53712.9 &      0.4 &   0.027 \\
SN2005mf         &        Ic &  53734.6 &      0.1 &   0.027 \\
SN2006T          &       IIb &  53782.1 &      1.1 &   0.008 \\
SN2006aj         &     Ic-bl &  53794.7 &      0.6 &   0.033 \\
SN2006ba         &       IIb$^*$ &  53825.7 &      1.7 &   0.019 \\
SN2006bf         &       IIb &  53826.5 &      1.3 &   0.024 \\
SN2006cb         &        Ib &  53861.5 &      1.6 &   0.026 \\
SN2006el         &       IIb &  53984.7 &      0.1 &   0.017 \\
SN2006ep         &        Ib &  53988.6 &      0.1 &   0.015 \\
SN2006fo         &        Ib &  54005.6 &      1.3 &   0.021 \\
SN2006ir         &      Ib/c$^*$ &  54015.3 &      2.0 &   0.020 \\
SN2006jo         &        Ib &  54016.6 &      1.3 &   0.077 \\
{\myfontsize SDSS-II SN 14475} &     Ic-bl &  54020.3 &      2.3 &   0.144 \\
SN2006lc         &        Ib &  54041.4 &      2.0 &   0.016 \\
SN2006lv         &        Ib$^*$ &  54045.6 &      1.1 &   0.008 \\
SN2006nx         &     Ic-bl$^*$ &  54056.1 &      2.0 &   0.125 \\
SN2007C          &        Ib &  54117.0 &      0.5 &   0.006 \\
SN2007I          &     Ic-bl &  54118.4 &      2.3 &   0.022 \\
SN2007D          &     Ic-bl &  54123.3 &      0.8 &   0.023 \\
SN2007Y          &        Ib &  54165.1 &      0.2 &   0.006 \\
SN2007ag         &        Ib &  54170.3 &      0.4 &   0.021 \\
SN2007ay         &       IIb &  54200.4 &      0.9 &   0.015 \\
SN2007ce         &     Ic-bl &  54226.7 &      2.0 &   0.046 \\
SN2007cl         &        Ic &  54250.9 &      1.0 &   0.022 \\
SN2007gr         &        Ic &  54339.3 &      0.3 &   0.002 \\
SN2007ke         &     Ib-Ca &  54367.6 &      1.9 &   0.017 \\
SN2007kj         &        Ib &  54382.5 &      0.1 &   0.018 \\
SN2007ms         &        Ic/Ic-bl$^*$ &  54384.8 &      1.3 &   0.039 \\
{\myfontsize SDSS-II SN 19065} &      Ib/c &  54393.7 &      2.7 &   0.154 \\
SN2007nc         &        Ib &  54397.0 &      1.3 &   0.087 \\
SN2007qw         &        Ic$^*$ &  54412.8 &      1.3 &   0.151 \\
{\myfontsize SDSS-II SN 19190} &     Ib/c?$^*$ &  54414.0 &      5.3 &    ----- \\
SN2007qv         &        Ic &  54415.6 &      2.7 &   0.094 \\
SN2007ru         &     Ic-bl &  54440.5 &      0.3 &   0.016 \\
SN2007rz         &        Ic &  54456.5 &      3.1 &   0.013 \\
SN2007uy         &    Ib-pec &  54481.8 &      0.7 &   0.007 \\
SN2008D          &        Ib &  54494.7 &      0.1 &   0.007 \\
SN2008aq         &       IIb &  54532.7 &      0.5 &   0.008 \\
SN2008ax         &       IIb &  54549.9 &      0.1 &   0.002 \\
SN2008bo         &       IIb &  54569.7 &      0.2 &   0.005 \\
SN2008cw         &       IIb &  54620.8 &      1.5 &   0.032 \\
SN2009K          &        IIb$^*$ &  54869.6 &      1.7 &   0.012 \\
SN2009bb         &     Ic-bl &  54923.6 &      0.1 &   0.010 \\
SN2009er         &    Ib-pec &  54982.6 &      0.2 &   0.035 \\
SN2009iz         &        Ib &  55109.4 &      1.4 &   0.014 \\
SN2009jf         &        Ib &  55122.2 &      0.1 &   0.008 \\
SN2009mg         &       IIb &  55189.5 &      0.0 &   0.008 \\
SN2009mk         &       IIb &  55193.2 &      0.3 &   0.005 \\
SN2010X          &       Ic?$^*$ &  55238.8 &      1.3 &   0.015 \\
SN2010ay         &     Ic-bl &  55270.2 &      1.4 &   0.067 \\
SN2010bh         &     Ic-bl &  55282.1 &      0.2 &   0.059 \\
SN2010al         &       Ibn &  55283.9 &      0.7 &   0.017 \\
SN2010as         &       IIb$^*$ &  55289.6 &      2.0 &   0.007 \\
SN2010cn         &        IIb/Ib$^*$ &  55331.4 &      0.4 &   0.026 \\
SN2010et         &        Ib-Ca$^*$ &  55358.7 &      2.2 &   0.023 \\
PTF10qts         &     Ic-bl &  55428.2 &      0.4 &   0.091 \\
SN2010jr         &       IIb$^*$ &  55531.7 &      1.0 &   0.012 \\
SN2011am         &        Ib &  55636.5 &      0.6 &   0.007 \\
SN2011bm         &        Ic &  55679.2 &      1.6 &   0.022 \\
SN2011dh         &       IIb &  55733.1 &      0.2 &   0.002 \\
SN2011ei         &       IIb &  55787.2 &      0.0 &   0.009 \\
PTF11kmb         &        Ib-Ca$^*$ &  55799.7 &      1.3 &   0.017 \\
PTF11qcj         &     Ic-bl$^*$ &  55842.3 &      2.2 &   0.028 \\
SN2011fu         &       IIb &  55847.5 &      0.1 &   0.018 \\
SN2011hg         &       IIb &  55878.9 &      0.6 &   0.024 \\
SN2011hs         &       IIb &  55889.2 &      0.3 &   0.006 \\
SN2011hw         &      Ibn &  55904.7 &      2.0 &   0.023 \\
SN2012P          &       IIb$^*$ &  55951.9 &      2.3 &   0.005 \\
SN2012ap         &     Ic-bl &  55976.6 &      0.5 &   0.012 \\
PS1-12sk         &      Ibn &  56003.7 &      1.7 &   0.054 \\
SN2012au         &        Ib &  56007.3 &      1.0 &   0.005 \\
SN2012hn         &     Ic-pec &  56032.9 &      0.5 &   0.008 \\
SN2012cd         &       IIb &  56034.6 &      1.7 &   0.012 \\
SN2012bz         &       Ic-bl &  56055.8 &      1.5 &   0.283 \\
PTF12gzk         &    Ic-pec &  56149.9 &      0.8 &   0.014 \\
PTF12hni         &        Ic &  56170.7 &      2.6 &   0.106 \\
{\myfontsize OGLE-2012-sn-006} &      Ibn &  56215.9 &      1.5 &   0.060 \\
SN2013ak         &       IIb &  56369.1 &      0.2 &   0.004 \\
SN2013cu         &        II &  56424.4 &      1.2 &   0.025 \\
SN2013cq         &     Ic-bl &  56433.5 &      1.5 &   0.339 \\
SN2013df         &       IIb &  56470.0 &      0.2 &   0.002 \\
iPTF13bvn        &        Ib &  56476.3 &      0.4 &   0.004 \\
SN2013dk         &        Ic &  56476.5 &      0.2 &   0.005 \\
LSQ13ccw         &      Ibn &  56539.0 &      1.2 &   0.060 \\
{\myfontsize OGLE-2013-SN-091} &        Ic &  56578.0 &      2.5 &   0.080 \\
{\myfontsize OGLE-2013-SN-134} &  Ic &  56578.0 &      2.5 &   0.039 \\
SN2013ge         &        Ic &  56621.4 &      0.2 &   0.004 \\
SN2014C          &        Ib &  56671.2 &      0.1 &   0.003 \\
SN2014L          &        Ic &  56695.7 &      0.1 &   0.008 \\
{\myfontsize OGLE-2014-SN-014} &        Ib &  56697.4 &      2.4 &   0.043 \\
SN2014ad         &     Ic-bl &  56741.3 &      0.2 &   0.006 \\
LSQ14efd         &        Ib$^*$ &  56900.0 &      0.6 &   0.080 \\
{\myfontsize OGLE-2014-SN-131} &       Ibn &  56969.1 &      2.6 &   0.085 \\
ASASSN-14ms      &        Ib &  57024.0 &      0.8 &   0.065 \\
SN2015U          &      Ibn &  57071.4 &      0.2 &   0.014 \\
OGLE15eo         &        Ic &  57263.7 &      2.0 &   0.064 \\
SN2015ap         &   Ib/c-bl &  57287.4 &      0.5 &   0.011 \\
OGLE15jy         &     Ib/c? &  57298.1 &      2.0 &   0.060 \\
iPTF15dld        &      Ibn$^*$ &  57308.0 &      1.9 &   0.047 \\
OGLE15rb         &        Ic &  57341.0 &      1.6 &   0.028 \\
iPTF15dtg        &        Ic &  57353.9 &      2.0 &   0.052 \\
OGLE15vk         &        Ic &  57357.4 &      1.7 &   0.050 \\
SN2016bau        &        Ib &  57478.8 &      1.3 &   0.004 \\
DES16S1kt        &       IIb &  57641.7 &      2.2 &   0.069 \\
SN2016gkg        &       IIb &  57678.7 &      5.5 &   0.005 \\
OGLE16ekf        &       IIb &  57680.0 &      1.0 &   0.068 \\
SN2016hgs        &        Ib &  57690.9 &      2.0 &   0.017 \\
SN2017ein        &        Ic &  57914.6 &      1.9 &   0.003 \\
SN2018bcc        &       Ibn &  58232.4 &      2.0 &   0.064 \\
SN2019bjv        &        Ic &  58530.6 &      1.5 &   0.027 \\
SN2019all        &      Ib/c &  58538.1 &      1.7 &   0.037 \\
SN2019gqd        &        Ic &  58646.5 &      1.5 &   0.036 \\
SN2019ilo        &        Ic &  58679.2 &      2.3 &   0.034 \\
SN2019pik        &        Ic &  58734.9 &      1.6 &   0.118 \\
SN2019myn        &       Ibn &  58706.0 &      2.3 &   0.100 \\
SN2019php        &       Ibn &  58730.3 &      1.3 &   0.087 \\
SN2019rii        &       Ibn &  58755.9 &      1.6 &   0.123 \\
SN2019aajs       &       Ibn &  58540.9 &      2.2 &   0.036 \\
SN2019deh        &       Ibn &  58587.1 &      2.4 &   0.055 \\

%% file: del_m_15_10_data2.tex
\renewcommand*{\arraystretch}{1.2}
\begin{longtable*}{lll|lllll}
  \caption{${\Delta m}_{15}$ values for measured from photometry generated in this work. ${\Delta m}_{15}$ is the defined as the difference between magnitude at \Vmax\ and 15 days after to \Vmax, as measured in the specific photometric band. Positive values indicate dimming with time. {Values marked with * are negative, contrary to the expectation that brightness will decrease past peak, but we note that the values are consistent with 0 and that the location of the peak is delayed with respect to \Vmax.}
  }\label{tab:del_m15}\\

\toprule
{} &            &          Ibc &                      Ib &                     IIb &                      Ic &                   Ic-bl &                     Ibn \\
\midrule
\endhead
\midrule
\multicolumn{8}{r}{{Continued on next page}} \\
\midrule
\endfoot

\bottomrule
\endlastfoot
		& w2  &    $0.88_{-0.09}^{+0.61}$ &                     --- &                     --- &                     --- &                     --- &                     --- \\
& m2  &     $1.19_{-1.16}^{+1.00}$ &                     --- &                     --- &                     --- &                     --- &                     --- \\
		& w1  &    $1.13_{-0.22}^{+0.87}$ &                     --- &  $1.59_{-0.63}^{+1.18}$ &                     --- &                     --- &                     --- \\
\hline

		& u  &    $2.13_{-0.49}^{+0.14}$ &  $1.86_{-0.58}^{+0.18}$ &   $2.07_{-0.41}^{+0.20}$ &   $1.57_{-0.16}^{+0.5}$ &                     --- &                     --- \\
		& U   &     $2.01_{-0.52}^{+0.10}$ &   $1.54_{-0.16}^{+0.30}$ &   $1.88_{-0.20}^{+0.22}$ &  $1.78_{-0.26}^{+0.19}$ &  $1.65_{-0.12}^{+0.17}$ &                     --- \\
  & B   &    $1.43_{-0.18}^{+0.08}$ &  $1.24_{-0.14}^{+0.19}$ &    $1.40_{-0.10}^{+0.08}$ &  $1.14_{-0.16}^{+0.24}$ &  $1.28_{-0.11}^{+0.25}$ &  $1.07_{-0.26}^{+0.27}$ \\

		& g   &      $1.20_{-0.26}^{+0.10}$ &  $1.13_{-0.17}^{+0.12}$ &   $1.10_{-0.14}^{+0.11}$ &  $0.92_{-0.24}^{+0.08}$ &  $0.97_{-0.15}^{+0.23}$ &  $1.76_{-0.33}^{+0.12}$ \\
 ${\Delta m}_{15}$ 		& V   &    $0.87_{-0.17}^{+0.14}$ &   $0.84_{-0.08}^{+0.10}$ &  $0.88_{-0.09}^{+0.08}$ &   $0.86_{-0.1}^{+0.15}$ &  $0.92_{-0.12}^{+0.11}$ &  $1.28_{-0.54}^{+0.52}$ \\
		& r  &     $0.60_{-0.17}^{+0.19}$ &  $0.63_{-0.09}^{+0.16}$ &  $0.69_{-0.07}^{+0.06}$ &  $0.64_{-0.17}^{+0.14}$ &  $0.66_{-0.18}^{+0.16}$ &   $1.70_{-0.15}^{+0.14}$ \\
			& R   &    $0.64_{-0.19}^{+0.11}$ &  $0.61_{-0.08}^{+0.13}$ &  $0.64_{-0.14}^{+0.08}$ &   $0.73_{-0.16}^{+0.1}$ &   $0.73_{-0.06}^{+0.1}$ &  $1.63_{-1.08}^{+0.18}$ \\
& i  &     $0.36_{-0.10}^{+0.11}$ &  $0.52_{-0.11}^{+0.14}$ &  $0.49_{-0.06}^{+0.06}$ &  $0.47_{-0.11}^{+0.16}$ &   $0.65_{-0.13}^{+0.10}$ &  $1.56_{-0.11}^{+0.19}$ \\
& I   &    $0.43_{-0.11}^{+0.04}$ &  $0.46_{-0.05}^{+0.15}$ &  $0.53_{-0.04}^{+0.04}$ &  $0.56_{-0.14}^{+0.06}$ &  $0.57_{-0.08}^{+0.11}$ &  $0.66_{-0.33}^{+0.99}$ \\
\hline
& J   &      $0.10_{-0.02}^{+0.00}$ &  $0.38_{-0.06}^{+0.06}$ &  $0.38_{-0.04}^{+0.05}$ &  $0.43_{-0.12}^{+0.14}$ &  $0.41_{-0.16}^{+0.15}$ &                     --- \\
		& H   &  $-0.03_{-0.02}^{+0.05^*}$ &   $0.30_{-0.09}^{+0.05}$ &  $0.29_{-0.02}^{+0.02}$ &  $0.38_{-0.09}^{+0.07}$ &   $0.29_{-0.10}^{+0.29}$ &                     --- \\
		& K   &   $0.02_{-0.02}^{+0.12}$ &  $0.39_{-0.09}^{+0.05}$ &                     --- &  $0.55_{-0.02}^{+0.03}$ &                     --- &                     --- \\

\bottomrule
\end{longtable*}

\begin{longtable*}{lll|lllll}
  \caption{${\Delta m}_{-10}$ values measured from photometric templates generated in this work. ${\Delta m}_{-10}$ is defined as the difference between magnitude at \Vmax\ and 10 days prior to \Vmax, as measured in the specific photometric band. Positive values indicate brightening with time. Values marked with * are negative, contrary to the expectation that brightness will increase toward peak, but we note that the location of the peak is earlier in bluer wavelenghts compared to \Vmax.}\label{tab:del_m10}
\\
\toprule
{} &            &          Ibc &                      Ib &                     IIb &                      Ic &                   Ic-bl &                     Ibn \\
\midrule
\endhead
\midrule
\multicolumn{8}{r}{{Continued on next page}} \\
\midrule
\endfoot

\bottomrule
\endlastfoot
\\
		& w2 &  $-0.19_{-0.04}^{+0.08^*}$ &                     --- &                     --- &                     --- &                     --- &                     --- \\
		& m2 &    $0.45_{-0.45}^{+0.15}$ &                     --- &                     --- &                     --- &                     --- &                     --- \\
		& w1 &   $-0.18_{-0.37}^{+0.10^*}$ &                     --- &                     --- &                     --- &                     --- &                     --- \\

\hline
	& u &    $0.08_{-0.06}^{+0.05}$ &   $0.74_{-0.10}^{+0.26}$ &  $1.89_{-0.77}^{+0.34}$ &                     --- &                     --- &                     --- \\
        & U  &    $0.02_{-0.05}^{+0.05}$ &  $0.97_{-0.39}^{+0.54}$ &  $0.59_{-0.26}^{+0.45}$ &                     --- &                     --- &                     --- \\
		& B  &    $0.27_{-0.11}^{+0.18}$ &  $0.71_{-0.18}^{+0.52}$ &  $0.84_{-0.31}^{+0.26}$ &   $0.54_{-0.35}^{+0.5}$ &                     --- &                     --- \\
		& g  &     $0.29_{-0.10}^{+0.33}$ &  $0.98_{-0.64}^{+0.32}$ &    $0.75_{-0.20}^{+0.30}$ &  $0.57_{-0.22}^{+0.18}$ &                     --- &                     --- \\
${\Delta m}_{-10}	$	& V  &     $0.40_{-0.15}^{+0.26}$ &   $0.50_{-0.19}^{+0.29}$ &  $0.75_{-0.27}^{+0.21}$ &   $0.4_{-0.15}^{+0.11}$ &  $0.64_{-0.11}^{+0.25}$ &                     --- \\
		& r &    $0.46_{-0.15}^{+0.27}$ &   $0.54_{-0.30}^{+0.16}$ &   $0.66_{-0.10}^{+0.11}$ &  $0.25_{-0.05}^{+0.07}$ &   $0.71_{-0.10}^{+0.13}$ &                     --- \\
		& R  &    $0.51_{-0.22}^{+0.35}$ &  $0.45_{-0.17}^{+0.22}$ &  $0.55_{-0.14}^{+0.24}$ &  $0.33_{-0.13}^{+0.55}$ &   $0.89_{-0.30}^{+0.22}$ &                     --- \\
		& i &    $0.51_{-0.13}^{+0.29}$ &  $0.41_{-0.14}^{+0.14}$ &   $0.54_{-0.10}^{+0.12}$ &  $0.22_{-0.06}^{+0.37}$ &   $0.45_{-0.13}^{+0.2}$ &                     --- \\
  		& I  &      $0.50_{-0.10}^{+0.25}$ &  $0.28_{-0.07}^{+0.13}$ &  $0.36_{-0.05}^{+0.13}$ &  $0.33_{-0.08}^{+0.11}$ &  $0.38_{-0.32}^{+0.51}$ &                     --- \\
\hline
& J  &    $0.44_{-0.04}^{+0.04}$ &   $0.19_{-0.10}^{+0.17}$ &  $0.27_{-0.05}^{+0.06}$ &                     --- &                     --- &                     --- \\
		& H  &     $0.44_{-0.01}^{+0.12}$ &  $0.27_{-0.09}^{+0.11}$ &  $0.34_{-0.03}^{+0.03}$ &                     --- &                     --- &                     --- \\
		& K  &     $0.49_{-0.02}^{+0.01}$ &  $0.25_{-0.06}^{+0.05}$ &                     --- &                     --- &                     --- &                     --- \\
\bottomrule

\end{longtable*}

%% file: allphotUBVRI_snr2.tex
\begin{longrotatetable}
\begin{longtable}{llccccc}
 \caption{ Templates \SESN e sample \emph{UBVRI} photometric epochs details}
 \label{tab:OPT1}\\
\toprule
{} &     Type &       U &       B &       V &       R &        I \\
Name             &          &    N, [min,max], SNR                        &     N, [min,max], SNR                       &    N, [min,max], SNR                        &    N, [min,max], SNR                        &    N, [min,max], SNR                         \\
\midrule
\endhead
\midrule
\multicolumn{7}{r}{{Continued on next page}} \\
\midrule
\endfoot

\bottomrule
\endlastfoot
SN1954A          &       Ib &     5, [52.5,109.8], 575.7 &    58, [-7.8,343.5], 575.7 &    44, [44.5,109.8], 575.7 &                          - &                           - \\
SN1962L          &     Ib-c &     11, [-9.8,18.0], 575.7 &    93, [-15.8,78.0], 575.7 &     11, [-9.8,18.0], 575.7 &                          - &                           - \\
SN1983N          &       Ib &      6, [11.2,18.2], 575.7 &    27, [-10.2,18.2], 575.7 &   56, [-10.8,170.8], 575.7 &                          - &                           - \\
SN1983V          &       Ic &                          - &    8, [-15.8,237.5], 575.7 &     5, [-15.8,29.2], 575.7 &                          - &                           - \\
SN1984I          &       Ib &     11, [-4.0,31.8], 575.7 &    18, [-17.0,20.8], 575.7 &     14, [-4.0,20.8], 575.7 &                          - &                           - \\
SN1985F          &     Ib-c &                          - &   18, [-16.8,321.2], 575.7 &   20, [253.8,444.5], 575.7 &                          - &                           - \\
SN1991N          &       Ic &                          - &       3, [2.5,15.5], 575.7 &      9, [-4.0,18.5], 575.7 &                          - &                           - \\
SN1993J          &      IIb &  191, [-17.8,737.2], 575.7 &  401, [-18.8,737.2], 575.7 &  459, [-18.8,737.2], 575.7 &  398, [-18.5,737.2], 575.7 &   365, [-18.5,642.5], 575.7 \\
SN1994I          &       Ic &      11, [-5.0,9.8], 575.7 &     52, [-7.0,59.8], 287.8 &    111, [-7.0,62.8], 575.7 &     98, [-7.0,62.8], 575.7 &      90, [-7.0,62.8], 575.7 \\
SN1996cb         &      IIb &                          - &    34, [-14.2,70.5], 575.7 &   45, [-14.2,139.2], 575.7 &   46, [-14.2,139.2], 575.7 &                           - \\
SN1997ef         &    Ic-bl &                          - &     6, [-11.8,23.0], 575.7 &   43, [-12.2,143.0], 191.9 &    8, [-15.2,110.2], 575.7 &                           - \\
SN1998bw         &    Ic-bl &    22, [-7.5,401.2], 575.7 &  131, [-14.2,486.5], 575.7 &  167, [-15.5,521.5], 575.7 &   65, [-15.5,521.5], 575.7 &    140, [-7.8,521.5], 575.7 \\
SN1999dn         &       Ib &     11, [-2.8,104.8], 95.9 &    14, [-2.8,121.8], 115.1 &    16, [-2.8,370.0], 143.9 &    16, [-2.8,370.0], 115.1 &      13, [-2.8,121.8], 95.9 \\
SN1999ex         &       Ib &     86, [-19.8,8.2], 457.1 &   134, [-19.8,20.2], 509.2 &   138, [-19.8,20.2], 479.7 &   136, [-19.8,20.2], 549.5 &    134, [-19.8,20.2], 479.7 \\
SN2001ig         &      IIb &                          - &    1, [-18.0,-18.0], 575.7 &    4, [-18.0,124.8], 575.7 &     7, [-7.2,243.0], 575.7 &                           - \\
SN2002ap         &    Ic-bl &    105, [-9.8,29.0], 575.7 &   227, [-9.8,306.2], 575.7 &   739, [-9.8,325.2], 575.7 &   331, [-9.8,306.2], 575.7 &    242, [-9.8,306.2], 575.7 \\
SN2002ji         &       Ib &                          - &                          - &                          - &     11, [-8.0,71.8], 575.7 &                           - \\
SN2003bg         &      IIb &                          - &   19, [-20.5,298.8], 239.9 &   23, [-18.5,298.8], 250.3 &   11, [-19.8,298.8], 287.8 &    17, [-10.5,298.8], 250.3 \\
SN2003dh         &    Ic-bl &    55, [-12.0,-6.0], 287.8 &   333, [-12.5,26.8], 119.9 &   384, [-12.5,26.8], 191.9 &  1644, [-12.8,67.0], 122.5 &    350, [-12.5,19.0], 112.9 \\
SN2003id         &   Ic-pec &                          - &      12, [-8.8,28.0], 91.7 &     20, [-8.8,34.0], 153.9 &     21, [-8.8,34.0], 255.8 &      20, [-8.8,34.0], 164.7 \\
SN2003jd         &    Ic-bl &                          - &     12, [-0.5,71.2], 120.0 &     16, [-1.5,74.2], 185.9 &      15, [1.5,74.2], 147.6 &      18, [-1.5,74.2], 165.0 \\
SN2003lw         &    Ic-bl &                          - &                          - &      4, [-4.8,68.0], 335.8 &     10, [-9.8,68.0], 385.5 &      11, [-9.8,66.2], 383.8 \\
SN2004aw         &       Ic &      16, [-5.8,85.2], 51.4 &     56, [-5.8,255.2], 78.0 &    64, [-5.8,255.2], 115.1 &    66, [-5.8,345.2], 143.9 &     56, [-5.8,255.2], 129.5 \\
SN2004dn         &       Ic &                          - &     15, [-10.2,52.8], 35.8 &     15, [-10.0,82.0], 48.9 &    15, [-10.0,82.0], 143.5 &     15, [-10.0,82.0], 112.4 \\
SN2004dk         &       Ib &                          - &   16, [-24.2,293.8], 265.0 &    16, [-24.2,293.8], 22.3 &   16, [-24.2,293.8], 246.4 &    15, [-24.2,293.8], 415.3 \\
SN2004ex         &      IIb &                          - &    22, [-15.0,48.0], 338.6 &    24, [-15.0,55.0], 397.5 &                          - &                           - \\
SN2004ff         &      IIb &                          - &     17, [-4.2,68.8], 190.0 &     15, [-4.2,57.5], 338.6 &                          - &                           - \\
SN2004fe         &       Ic &                          - &     25, [-9.5,43.2], 205.6 &     26, [-9.5,43.2], 287.8 &                          - &                           - \\
SN2004gq         &       Ib &     10, [-10.2,83.5], 38.7 &    40, [-10.2,75.8], 274.1 &    44, [-10.2,83.5], 479.7 &                          - &                           - \\
SN2004gt         &       Ic &                          - &    24, [-6.2,134.8], 169.3 &    28, [-6.2,134.8], 287.8 &                          - &                           - \\
SN2004gv         &       Ib &                          - &    18, [-10.5,32.2], 413.3 &    21, [-10.5,32.2], 411.2 &                          - &                           - \\
SN2005az         &       Ic &                          - &       5, [-8.0,76.0], 72.9 &     11, [-8.0,88.0], 143.9 &                          - &                           - \\
SN2005bf         &       Ib &   100, [-27.2,27.8], 116.3 &   116, [-31.2,70.5], 274.1 &   129, [-31.2,70.5], 303.0 &                          - &                           - \\
SN2005by         &      IIb &                          - &      10, [-2.2,28.8], 15.2 &      19, [-2.2,69.8], 17.1 &      18, [-2.2,66.8], 28.0 &       18, [-2.2,77.8], 30.3 \\
SN2005em         &      IIb &                          - &      9, [-3.8,36.2], 159.9 &     11, [-3.8,59.2], 191.9 &                          - &                           - \\
SN2005hg         &       Ib &     30, [-12.8,25.2], 60.0 &     31, [-12.8,82.2], 99.2 &   44, [-12.8,127.0], 151.6 &                          - &                           - \\
SN2005kl         &       Ic &                          - &     14, [-3.8,154.2], 17.7 &     33, [-3.8,138.0], 43.3 &                          - &                           - \\
SN2005kz         &       Ic &                          - &       6, [-2.5,16.5], 20.0 &       8, [-2.5,17.5], 95.2 &                          - &                           - \\
SN2005mf         &       Ic &         4, [0.5,5.5], 26.9 &      19, [-0.5,65.2], 25.4 &      23, [-0.5,65.2], 63.3 &                          - &                           - \\
SN2006T          &      IIb &       4, [-4.2,21.8], 21.9 &   58, [-14.2,108.5], 274.1 &   62, [-14.2,108.5], 411.2 &                          - &                           - \\
SN2006aj         &    Ic-bl &     62, [-10.0,22.2], 79.9 &     77, [-10.0,23.0], 76.8 &     71, [-10.0,24.2], 54.3 &                          - &                           - \\
SN2006ba         &      IIb &                          - &      12, [-2.2,36.8], 65.0 &      22, [-2.2,65.8], 90.0 &                          - &                           - \\
SN2006bf         &      IIb &                          - &       6, [-2.8,22.2], 67.6 &     17, [-2.8,48.2], 106.6 &                          - &                           - \\
SN2006cb         &       Ib &                          - &         1, [3.2,3.2], 60.6 &        5, [2.2,19.2], 52.8 &                          - &                           - \\
SN2006el         &      IIb &                          - &       15, [0.2,46.0], 47.6 &      32, [0.2,110.8], 82.9 &                          - &                           - \\
SN2006ep         &       Ib &                          - &     24, [-8.8,68.0], 120.4 &    39, [-8.8,106.2], 115.6 &                          - &                           - \\
SN2006fo         &       Ib &                          - &     20, [-1.8,87.2], 130.9 &     34, [-1.8,133.0], 99.3 &                          - &                           - \\
SN2006ir         &     Ib-c &                          - &       6, [2.5,70.2], 212.8 &      22, [1.5,85.2], 177.2 &                          - &                           - \\
SN2006lc         &       Ib &       1, [-7.0,-7.0], 18.6 &      13, [-7.8,38.2], 79.9 &     22, [-7.8,45.2], 138.7 &                          - &                           - \\
SN2007C          &       Ib &       2, [-3.0,-1.2], 56.8 &    37, [-3.0,126.5], 125.1 &    51, [-3.0,136.8], 179.9 &                          - &                           - \\
SN2007I          &    Ic-bl &                          - &       8, [-0.5,40.2], 41.8 &      21, [-0.5,70.2], 81.1 &                          - &                           - \\
SN2007d          &    Ic-bl &                          - &       6, [-5.5,15.2], 29.0 &       9, [-5.5,21.2], 29.8 &                          - &                           - \\
SN2007Y          &       Ib &     18, [-12.5,20.8], 70.7 &    37, [-12.5,32.5], 111.8 &     37, [-12.5,39.5], 87.2 &                          - &                           - \\
SN2007ag         &       Ib &                          - &       9, [-0.5,29.2], 67.7 &     14, [-0.5,40.5], 163.0 &                          - &                           - \\
SN2007ay         &      IIb &                          - &      3, [-5.8,-1.8], 230.3 &      5, [-6.8,15.2], 274.1 &                          - &                           - \\
SN2007ce         &    Ic-bl &                          - &       9, [0.0,18.0], 137.1 &      17, [0.0,54.0], 191.9 &                          - &                           - \\
SN2007cl         &       Ic &                          - &       6, [-6.0,27.2], 60.3 &     23, [-6.0,63.0], 179.9 &                          - &                           - \\
SN2007gr         &       Ic &    54, [-8.8,153.0], 205.6 &   128, [-8.8,182.2], 239.9 &   129, [-8.8,401.2], 261.7 &    80, [-8.8,401.2], 230.3 &     76, [-8.8,401.2], 230.3 \\
SN2007ke         &    Ib-Ca &                          - &        4, [2.2,35.2], 21.6 &        4, [2.2,37.5], 35.3 &      7, [-11.5,13.5], 48.0 &                           - \\
SN2007kj         &       Ib &                          - &      28, [-6.0,51.2], 66.1 &     32, [-6.0,51.2], 110.9 &                          - &                           - \\
SN2007ru         &    Ic-bl &      9, [-3.5,29.5], 143.9 &     44, [-3.5,59.5], 143.9 &    36, [-3.5,210.8], 221.4 &    22, [-3.5,210.8], 256.0 &     19, [-3.5,150.0], 274.7 \\
SN2007rz         &       Ic &                          - &     15, [-10.8,40.2], 40.0 &    21, [-10.8,64.2], 112.9 &                          - &                           - \\
SN2007uy         &   Ib-pec &     17, [-10.0,13.2], 61.2 &    79, [-10.8,134.0], 81.1 &   83, [-10.8,134.0], 101.0 &                          - &                           - \\
SN2008d          &       Ib &     39, [-18.0,12.5], 30.8 &   110, [-19.8,121.0], 92.9 &  115, [-19.8,100.0], 122.5 &    18, [-19.8,12.2], 575.7 &     18, [-19.8,12.2], 575.7 \\
SN2008aq         &      IIb &        9, [-7.2,7.8], 74.8 &    36, [-7.2,120.8], 209.8 &    38, [-7.2,120.8], 281.0 &                          - &                           - \\
SN2008ax         &      IIb &     28, [-19.5,33.0], 68.7 &   76, [-19.5,305.0], 143.9 &   98, [-18.5,305.0], 191.9 &   41, [-17.5,305.0], 191.9 &    33, [-17.5,305.0], 191.9 \\
SN2008bo         &      IIb &       9, [-8.5,11.0], 38.1 &     61, [-9.8,107.0], 57.0 &     68, [-9.8,107.0], 82.8 &                          - &                           - \\
SN2008cw         &      IIb &                          - &        7, [2.2,28.0], 37.1 &        7, [2.2,28.0], 60.0 &                          - &                           - \\
SN2009K          &       II &                          - &   13, [-20.8,258.5], 442.8 &   13, [-20.8,258.5], 523.3 &                          - &                           - \\
SN2009bb         &    Ic-bl &        2, [-9.2,4.5], 16.8 &     64, [-9.2,48.8], 115.1 &     87, [-9.2,113.8], 86.1 &     47, [-9.2,71.8], 191.9 &       47, [-9.2,71.8], 95.9 \\
SN2009er         &   Ib-pec &                          - &      13, [-0.8,23.2], 88.6 &     16, [-1.5,23.2], 185.9 &                          - &                           - \\
SN2009iz         &       Ib &      4, [-12.0,-7.8], 52.8 &     35, [-13.2,72.5], 92.8 &   35, [-13.2,135.5], 221.4 &                          - &                           - \\
SN2009jf         &       Ib &    13, [-18.0,27.8], 143.9 &   63, [-18.0,247.0], 213.2 &   62, [-18.0,247.0], 319.8 &   33, [-18.0,247.0], 261.7 &    33, [-18.0,247.0], 205.6 \\
SN2009mg         &      IIb &     18, [-13.0,52.2], 42.4 &     22, [-13.0,52.2], 63.3 &     22, [-13.0,52.2], 68.5 &                          - &                           - \\
SN2009mk         &      IIb &       9, [-10.5,5.5], 20.2 &     13, [-10.5,13.8], 38.9 &     18, [-10.5,34.0], 41.9 &                          - &                           - \\
SN2010ay         &    Ic-bl &                          - &                          - &                          - &      6, [-18.0,26.8], 86.3 &                           - \\
SN2010bh         &    Ic-bl &                          - &     14, [-9.2,24.0], 105.5 &     11, [-5.0,25.5], 191.9 &                          - &                           - \\
SN2010al         &      Ibn &       10, [0.5,20.5], 72.0 &      26, [-7.0,46.0], 83.8 &      28, [-7.0,55.0], 63.0 &                          - &                           - \\
SN2010as         &      IIb &                          - &    31, [-13.2,70.5], 133.9 &   25, [-13.2,100.5], 274.1 &   25, [-13.2,100.5], 338.6 &    24, [-13.2,100.5], 319.8 \\
SN2010cn         &       Ic &        6, [-7.8,3.2], 31.8 &        7, [-7.8,4.8], 37.1 &        7, [-7.8,4.8], 24.7 &                          - &                           - \\
SN2010et         &    Ib-Ca &                          - &        5, [-0.8,8.2], 25.0 &                          - &                          - &                           - \\
PTF10qts         &    Ic-bl &                          - &       4, [-4.5,22.5], 57.1 &                          - &    24, [-14.5,55.5], 167.9 &                           - \\
SN2010jr         &      IIb &     22, [-15.5,53.0], 45.5 &     31, [-15.5,51.5], 48.0 &     30, [-15.5,51.5], 40.1 &                          - &                           - \\
SN2011am         &       Ib &      10, [-6.5,11.0], 37.6 &      15, [-6.5,31.2], 55.3 &      15, [-6.5,31.2], 58.2 &                          - &                           - \\
SN2011bm         &       Ic &     27, [-15.8,81.2], 48.0 &    43, [-16.8,269.5], 64.0 &    45, [-16.8,269.5], 64.0 &    45, [-16.8,269.5], 72.0 &     45, [-16.8,269.5], 82.2 \\
SN2011dh         &      IIb &   90, [-17.2,152.2], 116.3 &  158, [-17.2,222.5], 143.9 &  821, [-19.0,289.2], 221.4 &   78, [-17.0,289.2], 287.8 &    81, [-17.0,220.2], 191.9 \\
PTF11iqb         &      IIn &                          - &     10, [1.5,117.2], 130.9 &     10, [1.5,117.2], 123.9 &  255, [-12.8,552.0], 575.7 &      51, [-9.8,138.2], 91.4 \\
SN2011ei         &      IIb &      14, [-10.5,9.0], 51.7 &     21, [-10.5,28.0], 36.9 &     21, [-10.5,28.0], 32.9 &                          - &                           - \\
PTF11kmb         &    Ib-Ca &                          - &         4, [3.0,9.0], 60.8 &                          - &                          - &                           - \\
PTF11qcj         &    Ic-bl &                          - &                          - &                          - &    54, [24.8,110.5], 575.7 &                           - \\
SN2011fu         &      IIb &     21, [-14.2,65.8], 95.9 &   45, [-14.2,105.5], 191.9 &   54, [-14.2,141.5], 287.8 &   58, [-14.2,150.5], 287.8 &    60, [-14.2,150.5], 191.9 \\
SN2011hg         &      IIb &        5, [-9.0,3.2], 33.7 &        8, [-9.0,8.8], 44.0 &        8, [-9.2,8.8], 32.9 &                          - &                           - \\
SN2011hs         &      IIb &      17, [-8.5,27.2], 66.2 &    41, [-8.8,210.8], 115.1 &    44, [-8.8,210.8], 287.8 &   21, [-11.2,210.8], 575.7 &     14, [-8.8,210.8], 575.7 \\
SN2011hw         &      Ibn &      5, [-16.5,-8.5], 81.1 &     13, [-16.5,38.8], 52.3 &      5, [-16.5,-8.5], 49.2 &  80, [-154.0,350.0], 575.7 &        6, [-6.2,31.8], 52.3 \\
SN2012P          &      IIb &       6, [-11.2,4.0], 42.8 &     32, [-7.0,152.8], 63.6 &        4, [-0.8,4.0], 69.1 &                          - &                           - \\
SN2012ap         &    Ic-bl &        7, [-6.8,7.5], 28.4 &      26, [-8.8,22.2], 51.7 &      26, [-8.8,22.2], 77.1 &     17, [-8.8,20.2], 143.9 &      15, [-2.8,22.2], 143.9 \\
PS1-12sk         &      Ibn &         1, [9.8,9.8], 57.6 &         1, [9.8,9.8], 57.6 &         1, [9.8,9.8], 44.3 &                          - &                           - \\
SN2012au         &       Ib &     12, [-5.0,31.8], 105.5 &     12, [-5.0,31.8], 115.1 &     12, [-5.0,31.8], 143.9 &                          - &                           - \\
SN2012hn         &    Ic-Ca &        4, [-1.5,7.5], 28.8 &      18, [-1.5,34.5], 92.9 &    30, [-1.5,184.8], 110.9 &     40, [-1.5,234.8], 82.2 &     34, [-1.5,234.8], 133.2 \\
SN2012cd         &      IIb &                          - &                          - &                          - &    26, [-12.8,13.2], 575.7 &                           - \\
SN2012bz         &      IIb &                          - &                          - &                          - &       2, [-6.2,-1.2], 76.8 &      1, [-15.2,-15.2], 72.0 \\
PTF12gzk         &   Ic-pec &       1, [-2.8,-2.8], 61.9 &      13, [-11.2,4.0], 99.2 &      13, [-9.0,3.5], 159.9 &     35, [-17.2,4.0], 383.8 &       16, [-9.0,4.0], 191.9 \\
PTF12hni         &       Ic &                          - &        7, [6.0,75.0], 24.0 &                          - &                          - &                           - \\
OGLE-2012-sn-006 &      Ibn &      9, [96.8,185.5], 64.0 &     11, [15.8,180.5], 64.0 &    46, [-1.2,186.5], 105.5 &     34, [15.8,180.5], 95.9 &  192, [-104.0,370.8], 143.9 \\
SN2013ak         &      IIb &      15, [-5.8,30.0], 68.5 &      15, [-5.8,30.0], 92.8 &      15, [-5.8,30.0], 97.6 &                          - &                           - \\
SN2013cu         &       II &       9, [-7.0,12.8], 76.8 &       8, [-2.0,21.5], 70.7 &       8, [-2.0,21.5], 43.0 &      3, [-8.5,-8.2], 575.7 &                           - \\
SN2013cq         &    Ic-bl &                          - &     24, [-17.0,28.0], 72.0 &    24, [-17.0,28.0], 143.9 &    24, [-17.0,28.0], 143.9 &     24, [-17.0,28.0], 143.9 \\
SN2013df         &      IIb &     66, [-13.5,41.0], 70.7 &    84, [-16.5,230.2], 82.2 &    86, [-16.5,606.0], 95.9 &   46, [-18.8,606.0], 115.1 &     37, [-14.5,606.0], 95.9 \\
iPTF13bvn        &       Ib &     27, [-14.8,42.8], 57.6 &    96, [-15.0,199.8], 92.8 &   94, [-15.0,199.8], 143.9 &   70, [-14.8,262.8], 191.9 &    71, [-14.8,262.8], 143.9 \\
SN2013dk         &       Ic &        6, [-8.5,4.8], 27.3 &        7, [-8.5,8.8], 51.9 &        7, [-8.5,8.8], 61.2 &                          - &                           - \\
LSQ13ccw         &      Ibn &                          - &       22, [3.2,33.5], 31.1 &      30, [-3.5,33.5], 35.2 &      10, [-3.2,34.5], 30.3 &         9, [3.2,34.5], 24.0 \\
OGLE-2013-sn-091 &       Ic &                          - &                          - &                          - &                          - &    26, [-222.5,137.8], 21.4 \\
SN2013ge         &       Ic &     45, [-14.0,36.2], 91.4 &  104, [-13.2,159.2], 143.9 &  113, [-13.2,159.2], 115.1 &     12, [-12.5,19.5], 95.9 &      12, [-12.5,19.5], 83.9 \\
OGLE-2013-sn-134 &       Ic &                          - &                          - &                          - &                          - &    31, [-283.2,112.5], 46.0 \\
SN2014C          &       Ib &       12, [-7.5,5.8], 37.4 &       12, [-4.2,5.8], 76.8 &       11, [-4.2,5.8], 87.2 &                          - &                           - \\
SN2014L          &       Ic &    26, [-10.2,56.5], 115.1 &     53, [-10.5,80.5], 82.2 &   67, [-10.5,141.2], 191.9 &   74, [-10.5,141.2], 143.9 &    74, [-10.5,141.2], 191.9 \\
OGLE-2014-sn-014 &       Ib &                          - &                          - &                          - &                          - &    34, [-111.5,251.5], 44.8 \\
SN2014ad         &    Ic-bl &    30, [-10.8,79.0], 224.4 &    29, [-10.8,79.0], 319.8 &    29, [-10.8,79.0], 442.8 &    28, [-10.8,79.0], 303.0 &     28, [-10.8,79.0], 303.0 \\
LSQ14efd         &       Ic &                          - &     10, [-10.8,23.5], 38.4 &     34, [-17.0,63.8], 38.4 &       9, [-3.8,73.8], 27.4 &        9, [-3.8,73.8], 30.3 \\
OGLE-2014-sn-131 &      Ibn &                          - &                          - &                          - &                          - &     46, [-61.5,249.8], 38.4 \\
ASASSN-14ms      &       Ib &       3, [-1.0,1.0], 137.1 &       3, [-1.0,1.0], 122.5 &      26, [-6.0,25.0], 81.1 &                          - &                           - \\
SN2015U          &      Ibn &                          - &      22, [-3.8,13.2], 58.5 &     29, [-3.8,20.8], 111.7 &     31, [-3.8,20.8], 140.6 &      27, [-3.8,18.2], 124.2 \\
OGLE15eo         &       Ic &                          - &                          - &                          - &                          - &    56, [-376.0,173.8], 53.3 \\
SN2015ap         &  Ib/c-bl &     11, [-12.0,16.0], 72.9 &     11, [-12.0,16.0], 82.2 &     11, [-12.0,16.0], 71.1 &                          - &                           - \\
OGLE15jy         &     Ib-c &                          - &                          - &                          - &                          - &    22, [-218.2,133.8], 25.1 \\
OGLE15rb         &       Ic &                          - &                          - &                          - &                          - &     26, [-361.2,97.8], 30.2 \\
iPTF15dtg        &       Ic &                          - &      10, [-15.8,7.0], 94.4 &                          - &                          - &                           - \\
OGLE15vk         &       Ic &                          - &                          - &                          - &                          - &     23, [-471.5,54.2], 60.6 \\
SN2016gkg        &      IIb &     28, [-19.0,39.8], 78.9 &    89, [-19.8,81.5], 111.8 &    169, [-20.5,81.5], 72.0 &    27, [-18.5,81.5], 143.9 &     27, [-19.5,81.5], 143.9 \\
OGLE16ekf        &      IIb &                          - &                          - &                          - &                          - &    31, [-269.2,187.5], 24.8 \\
SN2017ein        &       Ic &                          - &    51, [-15.2,50.8], 117.5 &    49, [-13.2,49.8], 262.2 &    52, [-15.2,51.8], 383.8 &      20, [22.8,50.8], 442.8 \\
SN2019all        &     Ib-c &                          - &                          - &                          - &                          - &    33, [-323.8,373.2], 18.6 \\
SN2019aajs       &      Ibn &                          - &        4, [7.0,28.2], 56.0 &        4, [7.0,28.2], 28.8 &                          - &                           - \\
SN2019pik        &       Ic &                          - &                          - &                          - &                          - &     21, [-57.8,149.8], 47.2 \\
\end{longtable}
\end{longrotatetable}

%% file: allphotugri_snr2.tex
\begin{longrotatetable}
\begin{longtable}{llcccc}
 \caption{ Templates \SESN e sample \emph{u'g'r'i'} photometric epochs details}
 \label{tab:OPT2}\\
\toprule
{} &    Type &    u' &     g' &     r' &     i' \\
Name             &         &            N, [min,max], SNR              &            N, [min,max], SNR               &   N, [min,max], SNR                        &      N, [min,max], SNR                     \\
\midrule
\endhead
\midrule
\multicolumn{6}{r}{{Continued on next page}} \\
\midrule
\endfoot

\bottomrule
\endlastfoot
SN2004ex         &     IIb &  18, [-15.0,12.2], 151.5 &   26, [-15.0,55.0], 371.8 &   26, [-15.0,55.0], 411.2 &   26, [-15.0,55.0], 359.8 \\
SN2004ff         &     IIb &   10, [-4.2,12.8], 128.5 &    19, [-4.2,68.8], 303.0 &    19, [-4.2,57.5], 411.2 &    20, [-4.2,68.8], 411.2 \\
SN2004fe         &      Ic &   15, [-9.5,21.2], 205.6 &    20, [-9.5,43.2], 359.8 &    28, [-9.5,43.2], 359.8 &    28, [-9.5,43.2], 383.8 \\
SN2004gq         &      Ib &  25, [-10.2,47.8], 133.9 &   31, [-10.2,53.8], 479.7 &   43, [-10.2,83.5], 274.1 &   44, [-10.2,83.5], 239.9 \\
SN2004gt         &      Ic &     9, [-6.2,14.8], 83.4 &    18, [-6.2,44.8], 221.4 &   31, [-6.2,169.8], 319.8 &   25, [-6.2,150.8], 221.4 \\
SN2004gv         &      Ib &  13, [-10.5,17.2], 185.7 &   18, [-10.5,32.2], 390.7 &   20, [-10.5,32.2], 479.7 &   19, [-10.5,32.2], 427.0 \\
SN2005az         &      Ic &                        - &                         - &    16, [-0.8,76.0], 213.2 &    20, [-8.0,88.0], 191.9 \\
SN2005bf         &      Ib &  35, [-31.2,31.5], 359.8 &   36, [-31.2,31.5], 383.8 &  112, [-31.2,28.5], 274.1 &  105, [-31.2,66.5], 230.3 \\
SN2005fk         &   Ic-bl &                        - &      3, [-1.8,42.0], 18.2 &       5, [-4.8,9.2], 42.0 &       5, [-4.8,9.2], 41.1 \\
SN2005hl         &      Ib &     3, [-9.5,-4.5], 65.4 &     6, [-9.5,34.2], 320.8 &    11, [-9.5,42.2], 383.8 &    13, [-9.5,55.5], 319.8 \\
SN2005em         &     IIb &   12, [-7.5,17.5], 102.6 &    28, [-7.5,59.2], 179.9 &    29, [-7.5,59.2], 261.7 &    32, [-7.5,88.2], 230.3 \\
SN2005hm         &      Ib &     5, [-23.2,2.8], 72.0 &   13, [-23.2,41.8], 104.7 &   17, [-23.2,52.5], 155.6 &   17, [-23.2,52.5], 143.9 \\
SDSS-II SN 5339  &    Ib-c &                        - &      4, [-6.8,17.0], 36.0 &     10, [-6.8,48.0], 61.9 &      6, [-6.8,35.0], 46.0 \\
SDSS-II SN 4664  &    Ib-c &                        - &     10, [-32.5,8.5], 23.1 &    15, [-32.5,32.2], 30.6 &    16, [-15.8,32.2], 36.5 \\
SDSS-II SN 6861  &    Ib-c &                        - &       7, [-6.8,8.2], 42.6 &      8, [-8.8,14.2], 40.4 &       7, [-8.8,8.2], 36.4 \\
SDSS-II SN 8196  &    Ib-c &     2, [-6.8,-4.8], 28.4 &      3, [-6.8,-2.8], 72.0 &      5, [-6.8,13.2], 51.4 &      6, [-6.8,24.2], 47.4 \\
SN2005hg         &      Ib &                        - &                         - &  51, [-12.8,127.0], 221.4 &  49, [-12.8,114.2], 191.9 \\
SN2005kr         &   Ic-bl &    2, [-10.5,-8.5], 33.7 &     8, [-13.5,15.2], 73.4 &     8, [-13.5,15.2], 97.7 &     8, [-13.5,15.2], 77.6 \\
SN2005ks         &   Ic-bl &    18, [-67.8,13.0], 4.9 &                         - &     19, [-67.8,13.0], 6.9 &     19, [-67.8,13.0], 6.8 \\
SN2005kl         &      Ic &                        - &                         - &   37, [-3.8,154.2], 159.9 &   45, [-3.8,154.2], 174.4 \\
SN2005kz         &      Ic &                        - &                         - &     8, [-2.5,17.5], 155.6 &     8, [-2.5,17.5], 103.7 \\
SN2005mf         &      Ic &                        - &                         - &     32, [-0.5,77.2], 46.6 &   37, [-0.5,4855.2], 71.1 \\
SN2006T          &     IIb &  19, [-13.2,20.8], 198.5 &  36, [-14.2,108.5], 442.8 &  62, [-14.2,108.5], 359.8 &  61, [-14.2,108.5], 383.8 \\
SN2006aj         &   Ic-bl &                        - &                         - &    18, [-6.0,25.8], 177.5 &    20, [-4.2,25.8], 149.5 \\
SN2006ba         &     IIb &     2, [-2.2,-1.2], 45.0 &     11, [-2.2,45.8], 67.7 &    22, [-2.2,65.8], 157.7 &    22, [-2.2,65.8], 171.9 \\
SN2006bf         &     IIb &                        - &    10, [-2.8,36.0], 115.7 &    22, [-2.8,65.2], 162.4 &    24, [-2.8,65.2], 125.2 \\
SN2006cb         &      Ib &                        - &                         - &      5, [2.2,30.2], 143.9 &       6, [2.2,30.2], 84.2 \\
SN2006el         &     IIb &                        - &                         - &    31, [0.2,110.8], 130.8 &    29, [0.2,110.8], 130.8 \\
SN2006ep         &      Ib &     7, [-8.8,15.2], 53.8 &    14, [-8.8,80.0], 329.2 &   41, [-8.8,107.2], 169.3 &   44, [-8.8,107.2], 185.9 \\
SN2006fo         &      Ib &    9, [-10.5,7.2], 143.9 &   19, [-10.5,87.2], 281.0 &  58, [-10.5,133.0], 250.3 &  59, [-10.5,133.0], 239.9 \\
SN2006ir         &    Ib-c &       2, [2.5,3.2], 58.1 &      8, [2.5,70.2], 329.2 &     21, [1.5,85.2], 165.4 &     23, [1.5,85.2], 211.1 \\
SN2006jo         &      Ib &     2, [-7.8,-5.8], 32.1 &      7, [-7.8,21.2], 85.9 &    11, [-7.8,46.2], 108.6 &     10, [-7.8,37.2], 95.9 \\
SDSS-II SN 14475 &   Ic-bl &                        - &    13, [-11.5,16.5], 36.0 &    17, [-11.5,28.5], 53.8 &    18, [-10.5,40.5], 45.7 \\
SN2006lc         &      Ib &      8, [-9.5,7.5], 62.6 &   18, [-11.5,38.2], 311.4 &   32, [-11.5,45.2], 281.0 &   31, [-11.5,45.2], 250.3 \\
SN2006nx         &   Ic-bl &                        - &    12, [-15.0,13.0], 85.3 &   12, [-15.0,12.8], 115.6 &    11, [-14.0,12.8], 89.9 \\
SN2007C          &      Ib &      5, [1.8,10.8], 95.9 &    18, [1.8,126.5], 295.4 &   47, [-2.0,136.8], 287.8 &   49, [-2.0,136.8], 213.2 \\
SN2007I          &   Ic-bl &                        - &                         - &    21, [-0.5,70.2], 115.1 &    20, [-0.5,70.2], 129.4 \\
SN2007d          &   Ic-bl &                        - &                         - &      9, [-5.5,21.2], 60.0 &      7, [-5.5,21.2], 53.3 \\
SN2007Y          &      Ib &   11, [-10.5,9.5], 205.6 &   16, [-10.5,32.5], 523.3 &   17, [-10.5,29.5], 575.7 &   17, [-10.5,29.5], 523.3 \\
SN2007ag         &      Ib &                        - &     5, [-0.5,29.2], 250.3 &    16, [-0.5,76.2], 217.9 &    16, [-0.5,76.2], 195.2 \\
SN2007ay         &     IIb &                        - &                         - &     5, [-6.8,15.2], 274.1 &     5, [-6.8,15.0], 191.9 \\
SN2007ce         &   Ic-bl &                        - &                         - &     17, [0.0,54.0], 213.2 &     18, [0.0,54.0], 178.2 \\
SN2007cl         &      Ic &                        - &                         - &    26, [-6.0,63.0], 197.9 &    25, [-6.0,63.0], 198.5 \\
SN2007gr         &      Ic &                        - &                         - &    51, [8.8,182.2], 359.8 &    52, [9.8,182.2], 359.8 \\
SN2007ke         &   Ib-Ca &                        - &                         - &       5, [2.2,37.5], 43.6 &       5, [2.2,37.5], 40.5 \\
SN2007kj         &      Ib &     6, [-6.0,4.2], 160.4 &    14, [-6.0,41.0], 250.3 &    31, [-6.0,75.2], 164.5 &    31, [-6.0,75.2], 185.7 \\
SN2007ms         &      Ic &     7, [-28.0,9.0], 36.0 &     7, [-28.0,9.0], 230.3 &   15, [-18.0,40.8], 213.2 &                         - \\
SDSS-II SN 19065 &    Ib-c &                        - &      4, [-7.8,15.0], 50.0 &     12, [-7.8,31.0], 45.0 &     15, [-7.8,31.0], 37.4 \\
SN2007nc         &      Ib &   1, [-10.2,-10.2], 19.7 &       4, [-8.2,9.8], 68.5 &       6, [-8.2,9.8], 78.9 &       4, [-3.2,9.8], 60.3 \\
SN2007qw         &      Ic &                        - &      2, [-4.0,-1.0], 99.4 &      4, [-4.0,6.0], 106.6 &      6, [-4.0,12.0], 73.8 \\
SDSS-II SN 19190 &    Ib-c &                        - &     11, [-67.2,2.8], 23.8 &    15, [-40.2,19.8], 25.8 &    13, [-47.2,19.8], 25.1 \\
SN2007qv         &      Ic &      6, [-7.0,9.0], 43.5 &      6, [-7.0,9.0], 155.1 &      6, [-7.0,9.0], 126.3 &       6, [-7.0,9.0], 98.5 \\
SN2007ru         &   Ic-bl &                        - &                         - &    14, [-1.0,46.0], 287.8 &    14, [-1.0,51.0], 287.8 \\
SN2007rz         &      Ic &                        - &    8, [-10.8,13.0], 158.5 &   22, [-10.8,64.2], 102.5 &   24, [-10.8,64.2], 151.2 \\
SN2007uy         &  Ib-pec &                        - &                         - &  45, [-10.8,134.0], 338.6 &  45, [-10.8,134.0], 287.8 \\
SN2008d          &      Ib &                        - &                         - &  44, [-18.8,121.0], 179.9 &  40, [-18.8,121.0], 191.9 \\
SN2008aq         &     IIb &    4, [-5.0,11.0], 139.2 &   18, [-5.0,120.8], 442.8 &   33, [-5.8,120.8], 274.1 &   33, [-5.8,120.8], 303.0 \\
SN2008ax         &     IIb &  29, [-16.5,49.5], 151.5 &   32, [-19.2,46.8], 103.1 &   104, [-20.2,49.5], 76.2 &   60, [-19.2,49.5], 147.6 \\
SN2008bo         &     IIb &                        - &                         - &   40, [-9.8,107.0], 225.8 &   41, [-9.8,107.0], 198.5 \\
SN2008cw         &     IIb &                        - &                         - &       9, [2.2,28.0], 85.9 &       9, [2.2,29.0], 88.6 \\
SN2009K          &      II &    9, [-20.8,5.0], 221.4 &     9, [-20.8,5.0], 523.3 &    12, [-20.8,5.0], 441.5 &  13, [-20.8,258.5], 361.2 \\
SN2009bb         &   Ic-bl &    17, [-9.2,28.8], 61.9 &    21, [-9.2,48.8], 205.6 &    18, [-9.2,48.8], 261.7 &    21, [-9.2,52.8], 274.1 \\
SN2009er         &  Ib-pec &                        - &                         - &    16, [-1.5,23.2], 245.1 &    17, [-1.5,23.2], 230.3 \\
SN2009iz         &      Ib &   13, [-13.2,10.8], 43.6 &                         - &  31, [-13.2,135.5], 261.7 &  32, [-13.2,135.5], 245.1 \\
SN2009jf         &      Ib &    7, [-2.5,31.5], 140.4 &                         - &    30, [-2.5,86.2], 359.8 &    30, [-2.5,86.2], 303.0 \\
SN2010X          &      Ic &                        - &      2, [13.0,16.0], 24.5 &     11, [-4.2,17.0], 52.3 &      3, [13.0,18.0], 41.1 \\
SN2010bh         &   Ic-bl &                        - &                         - &                         - &    16, [-9.5,47.0], 191.9 \\
SN2010al         &     Ibn &   11, [-7.0,21.0], 147.6 &                         - &    18, [-7.0,55.0], 284.3 &    20, [-7.0,55.0], 215.2 \\
SN2010as         &     IIb &                        - &   29, [-13.8,63.0], 261.7 &   26, [-12.5,81.0], 397.5 &    26, [-9.8,81.0], 359.8 \\
SN2010et         &   Ib-Ca &                        - &     14, [-5.0,26.0], 34.0 &   39, [-11.0,274.5], 52.3 &    21, [-5.0,128.0], 57.6 \\
PTF10qts         &   Ic-bl &                        - &                         - &      7, [-4.5,87.2], 72.0 &     9, [-10.5,87.2], 24.0 \\
SN2011bm         &      Ic &   17, [-15.8,81.2], 33.9 &   22, [-29.8,248.5], 72.0 &   22, [-32.8,248.5], 77.1 &  22, [-32.8,248.5], 105.5 \\
PTF11iqb         &     IIn &                        - &                         - &   52, [30.2,552.0], 575.7 &                         - \\
PTF11kmb         &   Ib-Ca &                        - &     31, [-4.0,30.8], 64.0 &    44, [-10.0,34.8], 52.3 &      11, [3.0,33.8], 95.9 \\
SN2011hs         &     IIb &     6, [-5.8,16.2], 34.3 &    23, [-8.8,51.2], 287.8 &    20, [-8.8,53.2], 575.7 &    19, [-8.8,29.2], 287.8 \\
SN2012P          &     IIb &                        - &   54, [-7.0,195.5], 133.9 &  85, [-15.0,195.5], 287.8 &   61, [-7.0,195.5], 164.5 \\
PS1-12sk         &     Ibn &    3, [11.8,14.8], 115.1 &    10, [-2.8,35.8], 143.9 &     8, [-2.8,34.8], 143.9 &     8, [-5.0,34.8], 107.9 \\
SN2012au         &      Ib &                        - &                         - &    2, [308.8,334.8], 50.9 &                         - \\
SN2012hn         &   Ic-Ca &                        - &      7, [-1.5,15.5], 67.7 &      4, [-0.5,15.5], 48.2 &      2, [4.5,15.5], 162.3 \\
SN2012cd         &     IIb &                        - &                         - &    14, [19.2,204.5], 44.3 &                         - \\
SN2012bz         &     IIb &    5, [-16.0,-3.2], 14.8 &    18, [-15.2,23.0], 77.1 &    37, [-15.8,28.8], 82.2 &    46, [-16.0,28.8], 68.0 \\
PTF12gzk         &  Ic-pec &                        - &    18, [-11.0,3.0], 179.9 &    39, [-11.2,3.5], 191.9 &    17, [-11.2,3.0], 191.9 \\
PTF12hni         &      Ic &                        - &     17, [6.0,111.8], 27.4 &   35, [-24.0,111.8], 57.6 &     21, [6.0,111.8], 48.0 \\
SN2013cq         &   Ic-bl &                        - &                         - &    16, [-17.5,4.0], 575.7 &                         - \\
SN2013df         &     IIb &    4, [-8.0,26.0], 185.0 &                         - &   13, [-8.0,490.0], 143.9 &   10, [-8.0,230.2], 287.8 \\
iPTF13bvn        &      Ib &                        - &  86, [-15.2,303.2], 442.8 &  95, [-16.5,337.2], 411.2 &  86, [-15.2,245.5], 479.7 \\
SN2013ge         &      Ic &                        - &                         - &   10, [-5.5,442.5], 191.9 &    8, [-5.5,442.5], 239.9 \\
SN2014L          &      Ic &                        - &                         - &    6, [-11.8,15.5], 425.0 &                         - \\
LSQ14efd         &      Ic &                        - &    12, [-10.8,49.0], 29.7 &    16, [-10.8,79.8], 32.1 &    17, [-10.8,79.8], 27.4 \\
OGLE15eo         &      Ic &                        - &        2, [1.0,1.0], 79.1 &       1, [1.0,1.0], 119.9 &        1, [1.0,1.0], 89.9 \\
iPTF15dld        &   Ic-bl &                        - &                         - &     50, [-9.2,92.2], 28.8 &     26, [11.5,92.2], 28.8 \\
iPTF15dtg        &      Ic &                        - &   82, [-19.8,817.5], 82.2 &   73, [-18.0,761.8], 44.6 &  44, [-18.0,817.5], 148.0 \\
SN2016bau        &      Ib &                        - &                         - &   15, [-14.8,45.0], 403.1 &                         - \\
DES16s1kt        &     IIb &                        - &   17, [-15.0,140.8], 32.1 &    24, [-16.0,97.8], 50.2 &    22, [-15.0,97.8], 64.6 \\
SN2016gkg        &     IIb &                        - &  40, [-18.8,-14.8], 230.3 &  37, [-18.8,-14.8], 338.6 &  46, [-18.8,336.8], 281.0 \\
SN2016hgs        &      Ib &                        - &     16, [-9.2,10.8], 43.2 &     23, [-9.2,37.8], 64.0 &     14, [-3.5,37.8], 98.7 \\
SN2018bcc        &     Ibn &                        - &   158, [-3.2,23.8], 575.7 &     32, [-7.2,57.5], 24.6 &      5, [-4.2,22.8], 32.0 \\
SN2019bjv        &      Ic &                        - &     6, [-18.5,42.5], 79.9 &    6, [-18.5,45.5], 105.5 &                         - \\
SN2019aajs       &     Ibn &      4, [7.0,28.2], 57.6 &     49, [-1.0,61.5], 64.0 &     39, [-1.2,46.8], 38.4 &      22, [0.5,57.8], 55.0 \\
SN2019deh        &     Ibn &                        - &    38, [-6.0,50.8], 115.1 &    32, [-6.2,47.5], 129.5 &    14, [-5.0,47.5], 129.5 \\
SN2019gqd        &      Ic &                        - &      5, [-13.8,4.2], 36.0 &       3, [-7.8,4.2], 64.0 &   30, [-14.5,31.5], 101.1 \\
SN2019ilo        &      Ic &                        - &    10, [-15.5,11.5], 48.0 &     5, [-11.5,13.5], 64.0 &     1, [-3.5,-3.5], 191.9 \\
SN2019myn        &     Ibn &       1, [4.5,4.5], 28.8 &     69, [-5.0,32.0], 48.0 &     40, [-4.0,27.0], 37.5 &      8, [-4.0,17.0], 31.7 \\
SN2019php        &     Ibn &                        - &      17, [-3.5,9.5], 82.2 &     25, [-3.5,38.5], 28.8 &                         - \\
SN2019rii        &     Ibn &                        - &     36, [-6.2,15.8], 57.6 &     52, [-5.2,19.8], 32.0 &        4, [1.8,9.8], 52.9 \\
\end{longtable}
\end{longrotatetable}

%% file: AllPhotUVNIRTable.tex
\begin{turnpage}
\tabletypesize{\tiny}
\input{allphotJHKUV_snr2.tex}

\end{turnpage}


%% file: allphotJHKUV_snr2.tex
\begin{longrotatetable}
\begin{longtable}{llcccccc}
 \caption{Templates \SESN e sample \emph{JH$K_s$w1w2m2} photometric epochs details}\label{tab:UVNIR}\\
\toprule
{} &     Type &     J &     H &     K &   w1 &   w2 &   m2 \\
Name             &          &   N, [min,max], SNR                       &       N, [min,max], SNR                   &   N, [min,max], SNR                       &   N, [min,max], SNR                      &   N, [min,max], SNR                      &    N, [min,max], SNR                     \\
\midrule
\endhead
\midrule
\multicolumn{8}{r}{{Continued on next page}} \\
\midrule
\endfoot

\bottomrule
\endlastfoot
SN1999dn         &       Ib &    3, [23.0,121.8], 38.4 &    3, [23.0,121.8], 57.6 &    3, [23.0,121.8], 19.2 &                       - &                       - &                       - \\
SN2003dh         &    Ic-bl &    8, [-9.2,-1.2], 167.9 &    7, [-9.2,-3.0], 143.9 &                        - &                       - &                       - &                       - \\
SN2003lw         &    Ic-bl &   5, [-20.8,65.0], 169.3 &   4, [-20.8,65.0], 169.4 &   8, [-20.8,65.0], 143.9 &                       - &                       - &                       - \\
SN2004ex         &      IIb &   16, [-7.8,67.0], 179.9 &   16, [-7.8,67.0], 135.5 &                        - &                       - &                       - &                       - \\
SN2004ff         &      IIb &    8, [-3.2,27.8], 306.4 &    4, [-3.2,27.8], 198.5 &                        - &                       - &                       - &                       - \\
SN2004gq         &       Ib &   9, [-0.2,340.8], 274.1 &  13, [-4.2,340.8], 383.8 &   1, [340.8,340.8], 18.4 &                       - &                       - &                       - \\
SN2004gt         &       Ic &    25, [0.8,161.8], 63.3 &    25, [0.8,161.8], 52.8 &  11, [102.8,161.8], 26.2 &                       - &                       - &                       - \\
SN2004gv         &       Ib &   12, [-8.8,19.2], 198.5 &   11, [-8.8,19.2], 221.4 &                        - &                       - &                       - &                       - \\
SN2005az         &       Ic &    23, [28.0,83.0], 99.2 &    25, [28.0,83.0], 78.9 &    20, [28.0,77.0], 32.9 &                       - &                       - &                       - \\
SN2005bf         &       Ib &  54, [-17.2,27.2], 130.8 &   51, [-17.2,26.8], 84.7 &   50, [-17.2,27.2], 72.9 &                       - &                       - &                       - \\
SN2005em         &      IIb &    12, [-5.8,29.2], 87.5 &    10, [-4.8,29.2], 56.7 &                        - &                       - &                       - &                       - \\
SN2005hg         &       Ib &   33, [-8.0,49.2], 133.9 &    32, [-8.0,49.2], 80.0 &    28, [-8.0,49.2], 52.4 &                       - &                       - &                       - \\
SN2005kl         &       Ic &   27, [-2.8,71.2], 110.7 &    22, [-2.8,69.2], 36.0 &    23, [-2.8,69.2], 42.6 &                       - &                       - &                       - \\
SN2005mf         &       Ic &    16, [0.2,25.2], 119.0 &     17, [0.2,25.2], 71.1 &     16, [0.2,25.2], 35.9 &                       - &                       - &                       - \\
SN2006T          &      IIb &   49, [-8.2,86.5], 230.3 &   47, [-8.2,86.5], 205.6 &                        - &                       - &                       - &                       - \\
SN2006aj         &    Ic-bl &     6, [-6.0,4.0], 285.0 &      6, [-6.0,4.0], 88.9 &      5, [-6.0,4.0], 61.2 &  32, [-10.0,12.2], 74.8 &   55, [-10.0,3.0], 65.4 &                       - \\
SN2006ba         &      IIb &   16, [-5.0,38.8], 133.9 &     7, [-5.0,18.8], 68.5 &                        - &                       - &                       - &                       - \\
SN2006bf         &      IIb &     11, [3.2,20.2], 83.4 &      8, [0.2,20.2], 44.6 &                        - &                       - &                       - &                       - \\
SN2006ep         &       Ib &   26, [-7.8,61.0], 178.5 &   10, [-7.8,35.2], 118.3 &                        - &                       - &                       - &                       - \\
SN2006fo         &       Ib &    42, [0.2,112.2], 32.9 &    29, [0.2,105.2], 22.7 &     24, [0.2,64.2], 26.3 &                       - &                       - &                       - \\
SN2006ir         &     Ib-c &    20, [1.2,59.2], 140.4 &     20, [1.5,59.2], 90.8 &                        - &                       - &                       - &                       - \\
SN2006lc         &       Ib &     8, [-3.8,7.2], 225.8 &     8, [-3.8,7.2], 329.2 &                        - &                       - &                       - &                       - \\
SN2007C          &       Ib &  61, [-7.2,107.5], 108.6 &   58, [-7.2,117.8], 51.2 &   33, [-1.0,106.8], 32.9 &                       - &                       - &                       - \\
SN2007I          &    Ic-bl &  42, [10.5,106.2], 119.9 &   39, [10.5,106.2], 55.9 &   31, [10.5,105.2], 48.8 &                       - &                       - &                       - \\
SN2007d          &    Ic-bl &     8, [-6.8,57.2], 23.0 &     2, [-6.8,56.2], 19.2 &     4, [-6.8,51.2], 19.4 &    2, [-5.0,-4.2], 23.1 &                       - &                       - \\
SN2007Y          &       Ib &  17, [-14.5,33.5], 303.0 &  15, [-14.5,33.5], 143.9 &                        - &  16, [-12.5,19.0], 36.5 &    9, [-12.5,4.5], 34.7 &                       - \\
SN2007ag         &       Ib &     9, [-1.5,38.2], 53.3 &      4, [0.5,23.5], 43.5 &                        - &                       - &                       - &                       - \\
SN2007ce         &    Ic-bl &    10, [2.0,199.2], 65.8 &      7, [2.0,60.0], 45.3 &      2, [2.0,17.0], 40.4 &                       - &                       - &                       - \\
SN2007gr         &       Ic &   68, [8.8,167.2], 198.7 &   66, [8.8,167.2], 195.2 &   62, [9.8,167.2], 117.5 &                       - &                       - &                       - \\
SN2007kj         &       Ib &    11, [-4.8,45.0], 85.9 &    13, [-4.8,45.0], 52.3 &                        - &                       - &                       - &                       - \\
SN2007rz         &       Ic &   8, [-11.8,10.2], 217.3 &   9, [-11.8,36.0], 130.8 &                        - &                       - &                       - &                       - \\
SN2007uy         &   Ib-pec &   42, [-7.0,98.0], 126.5 &   44, [-7.0,98.0], 104.7 &    41, [-7.0,98.0], 84.7 &   12, [-10.0,8.5], 34.7 &   10, [-10.0,4.2], 14.5 &                       - \\
SN2008d          &       Ib &  18, [-19.0,16.0], 115.1 &   17, [-19.0,16.0], 82.2 &   18, [-19.0,16.0], 82.2 &  5, [-18.0,-17.2], 18.0 &                       - &                       - \\
SN2008aq         &      IIb &   16, [-2.0,85.0], 274.1 &   14, [-2.0,67.0], 274.7 &                        - &     6, [-7.2,2.0], 31.6 &    2, [-4.0,-3.0], 11.0 &                       - \\
SN2008ax         &      IIb &  33, [-10.0,251.0], 32.9 &  32, [-12.0,242.0], 32.9 &  36, [-12.0,251.0], 32.9 &  11, [-19.5,25.5], 43.0 &    7, [-9.8,28.8], 22.9 &                       - \\
SN2008bo         &      IIb &                        - &                        - &                        - &     5, [-6.8,1.5], 16.6 &     2, [-6.8,1.5], 10.8 &                       - \\
SN2009K          &       II &   39, [-22.8,29.2], 32.9 &   38, [-22.8,29.2], 32.9 &   14, [-19.8,29.2], 32.9 &                       - &                       - &                       - \\
SN2009bb         &    Ic-bl &   31, [-4.0,52.0], 397.5 &   22, [-4.0,52.0], 178.6 &                        - &                       - &                       - &                       - \\
SN2009er         &   Ib-pec &    12, [3.2,20.2], 174.6 &     11, [3.2,20.2], 45.3 &      4, [4.2,15.2], 48.8 &                       - &                       - &                       - \\
SN2009iz         &       Ib &   31, [-9.5,46.5], 202.5 &   30, [-9.5,46.5], 154.2 &    18, [-8.5,46.5], 32.9 &   4, [-12.0,-7.8], 22.6 &                       - &                       - \\
SN2009jf         &       Ib &   20, [-8.5,64.2], 261.7 &   20, [-8.5,64.2], 217.9 &   17, [-8.5,64.2], 147.6 &                       - &                       - &                       - \\
SN2009mg         &      IIb &                        - &                        - &                        - &  10, [-11.0,15.5], 28.5 &     4, [-8.5,0.0], 17.4 &                       - \\
SN2010al         &      Ibn &     15, [0.0,31.0], 32.9 &     15, [0.0,32.0], 32.9 &      8, [1.0,28.0], 32.9 &   10, [-4.5,16.0], 65.4 &    9, [-4.5,14.0], 50.5 &                       - \\
SN2010cn         &       Ic &                        - &                        - &                        - &    2, [-7.8,-5.0], 21.4 &    5, [-7.8,28.2], 18.7 &                       - \\
SN2010jr         &      IIb &                        - &                        - &                        - &  32, [-15.5,51.5], 29.0 &  27, [-15.5,48.2], 19.6 &                       - \\
SN2011am         &       Ib &                        - &                        - &                        - &     5, [-6.5,4.0], 17.3 &     3, [-6.5,1.8], 15.9 &                       - \\
SN2011dh         &      IIb &                        - &                        - &                        - &  38, [-17.2,55.8], 71.1 &  39, [-17.2,55.8], 42.0 &                       - \\
PTF11iqb         &      IIn &                        - &                        - &                        - &  10, [-12.0,29.8], 99.0 &  11, [-12.0,29.8], 94.4 &                       - \\
SN2011ei         &      IIb &                        - &                        - &                        - &   10, [-10.5,4.8], 24.7 &   5, [-10.5,-4.8], 18.0 &                       - \\
SN2011hs         &      IIb &                        - &                        - &                        - &     7, [-8.0,2.8], 35.8 &     6, [-8.0,2.8], 22.2 &                       - \\
SN2011hw         &      Ibn &                        - &                        - &                        - &   5, [-16.5,-8.5], 75.7 &   5, [-16.5,-8.5], 59.3 &                       - \\
SN2012P          &      IIb &                        - &                        - &                        - &    6, [-11.2,4.0], 17.8 &                       - &                       - \\
SN2012ap         &    Ic-bl &                        - &                        - &                        - &    3, [-4.8,-0.5], 19.4 &                       - &                       - \\
PS1-12sk         &      Ibn &      5, [9.8,25.0], 57.6 &     2, [18.0,25.0], 83.9 &     2, [18.0,25.0], 95.9 &     6, [9.8,29.2], 48.3 &     6, [9.8,29.2], 25.8 &     6, [9.8,29.2], 31.3 \\
SN2012au         &       Ib &                        - &                        - &                        - &   11, [-5.0,31.8], 95.9 &   12, [-5.0,31.8], 72.0 &   12, [-5.0,31.8], 82.2 \\
SN2012bz         &      IIb &   10, [-16.0,10.0], 45.3 &   6, [-16.0,-13.0], 55.0 &     1, [-1.0,-1.0], 41.1 &  3, [-16.0,-15.8], 15.6 &  2, [-16.0,-16.0], 20.4 &  3, [-16.0,-15.8], 12.2 \\
PTF12gzk         &   Ic-pec &     7, [-4.5,1.5], 174.4 &     7, [-4.5,1.5], 155.6 &     7, [-4.5,1.5], 143.9 &    6, [-10.5,1.0], 59.2 &     4, [-3.2,1.0], 39.9 &                       - \\
OGLE-2012-sn-006 &      Ibn &    7, [15.8,184.5], 82.2 &    6, [15.8,184.5], 69.9 &                        - &                       - &                       - &                       - \\
SN2013ak         &      IIb &                        - &                        - &                        - &   14, [-5.8,30.0], 44.9 &   10, [-5.8,12.0], 33.1 &                       - \\
SN2013cu         &       II &                        - &                        - &                        - &    9, [-7.2,12.8], 82.2 &    7, [-2.0,12.8], 66.2 &                       - \\
SN2013df         &      IIb &                        - &                        - &                        - &  43, [-13.0,41.0], 72.0 &  47, [-13.0,41.0], 58.7 &  21, [-13.0,41.0], 64.0 \\
iPTF13bvn        &       Ib &                        - &                        - &                        - &   12, [-15.0,6.0], 31.1 &    7, [-11.0,2.0], 19.9 &                       - \\
SN2013dk         &       Ic &                        - &                        - &                        - &    1, [-8.5,-8.5], 17.0 &    1, [-8.5,-8.5], 12.3 &                       - \\
SN2013ge         &       Ic &                        - &                        - &                        - &  40, [-14.0,14.2], 72.0 &  20, [-13.2,-4.0], 68.0 &                       - \\
SN2014C          &       Ib &                        - &                        - &                        - &     7, [-6.5,5.5], 22.1 &                       - &                       - \\
SN2014ad         &    Ic-bl &                        - &                        - &                        - &   10, [-5.5,12.8], 44.5 &    9, [-5.5,12.8], 36.0 &   10, [-5.5,12.8], 48.3 \\
ASASSN-14ms      &       Ib &                        - &                        - &                        - &    3, [-1.0,1.0], 127.9 &    3, [-1.0,1.0], 137.1 &    3, [-1.0,1.0], 127.9 \\
SN2015ap         &  Ib/c-bl &                        - &                        - &                        - &    8, [-12.0,5.8], 42.2 &   4, [-10.8,-2.5], 24.5 &                       - \\
SN2016gkg        &      IIb &                        - &                        - &                        - &  35, [-19.0,39.8], 61.9 &   24, [-19.0,9.8], 44.8 &                       - \\
\end{longtable}
\end{longrotatetable}

%% file: main.bbl
\begin{thebibliography}{}
\expandafter\ifx\csname natexlab\endcsname\relax\def\natexlab#1{#1}\fi
\providecommand{\url}[1]{\href{#1}{#1}}
\providecommand{\dodoi}[1]{doi:~\href{http://doi.org/#1}{\nolinkurl{#1}}}
\providecommand{\doeprint}[1]{\href{http://ascl.net/#1}{\nolinkurl{http://ascl.net/#1}}}
\providecommand{\doarXiv}[1]{\href{https://arxiv.org/abs/#1}{\nolinkurl{https://arxiv.org/abs/#1}}}

\bibitem[{Aigrain \& Foreman-Mackey(2023)}]{aigrain2023gaussian}
Aigrain, S., \& Foreman-Mackey, D. 2023, Annual Review of Astronomy and Astrophysics, 61, 329

\bibitem[{Allam~Jr {et~al.}(2018)Allam~Jr, Bahmanyar, Biswas, Dai, Galbany, Hlo{\v{z}}ek, Ishida, Jha, Jones, Kessler, {et~al.}}]{allam2018photometric}
Allam~Jr, T., Bahmanyar, A., Biswas, R., {et~al.} 2018, arXiv preprint arXiv:1810.00001

\bibitem[{{Ambikasaran} {et~al.}(2014){Ambikasaran}, {Foreman-Mackey}, {Greengard}, {Hogg}, \& {O'Neil}}]{Ambikasaran14}
{Ambikasaran}, S., {Foreman-Mackey}, D., {Greengard}, L., {Hogg}, D.~W., \& {O'Neil}, M. 2014, ArXiv e-prints.
\newblock \doarXiv{1403.6015}

\bibitem[{{Arcavi}(2017)}]{2017hsn..book..239A}
{Arcavi}, I. 2017, in Handbook of Supernovae, ed. A.~W. {Alsabti} \& P.~{Murdin}, 239, \dodoi{10.1007/978-3-319-21846-5_39}

\bibitem[{Arcavi {et~al.}(2011)Arcavi, Gal-Yam, Yaron, Sternberg, Rabinak, Waxman, Kasliwal, Quimby, Ofek, Horesh, Kulkarni, Filippenko, Silverman, Cenko, Li, Bloom, Sullivan, Nugent, Poznanski, Gorbikov, Fulton, Howell, Bersier, Riou, Lamotte-Bailey, Griga, Cohen, Hachinger, Polishook, Xu, Ben-Ami, Manulis, Walker, Maguire, Pan, Matheson, Mazzali, Pian, Fox, Gehrels, Law, James, Marchant, Smith, Mottram, Barnsley, Kandrashoff, \& Clubb}]{Arcavi_2011}
Arcavi, I., Gal-Yam, A., Yaron, O., {et~al.} 2011, The Astrophysical Journal Letters, 742, L18, \dodoi{10.1088/2041-8205/742/2/L18}

\bibitem[{{Armstrong} {et~al.}(2021){Armstrong}, {Tucker}, {Rest}, {Ridden-Harper}, {Zenati}, {Piro}, {Hinton}, {Lidman}, {Margheim}, {Narayan}, {Shaya}, {Garnavich}, {Kasen}, {Villar}, {Zenteno}, {Arcavi}, {Drout}, {Foley}, {Wheeler}, {Anais}, {Campillay}, {Coulter}, {Dimitriadis}, {Jones}, {Kilpatrick}, {Mu{\~n}oz-Elgueta}, {Rojas-Bravo}, {Vargas-Gonz{\'a}lez}, {Bulger}, {Chambers}, {Huber}, {Lowe}, {Magnier}, {Shappee}, {Smartt}, {Smith}, {Barclay}, {Barentsen}, {Dotson}, {Gully-Santiago}, {Hedges}, {Howell}, {Cody}, {Auchettl}, {B{\'o}di}, {Bogn{\'a}r}, {Brimacombe}, {Brown}, {Cseh}, {Galbany}, {Hiramatsu}, {Holoien}, {Howell}, {Jha}, {K{\"o}nyves-T{\'o}th}, {Kriskovics}, {McCully}, {Milne}, {Mu{\~n}oz}, {Pan}, {P{\'a}l}, {Sai}, {S{\'a}rneczky}, {Smith}, {S{\'o}dor}, {Szab{\'o}}, {Szak{\'a}ts}, {Valenti}, {Vink{\'o}}, {Wang}, {Zhang}, \& {Zsidi}}]{2021MNRAS.507.3125A}
{Armstrong}, P., {Tucker}, B.~E., {Rest}, A., {et~al.} 2021, \mnras, 507, 3125, \dodoi{10.1093/mnras/stab2138}

\bibitem[{{Arnett}(1982)}]{arnett82}
{Arnett}, W.~D. 1982, \apj, 253, 785, \dodoi{10.1086/159681}

\bibitem[{Barbarino {et~al.}(2017)Barbarino, Botticella, Dall'Ora, Della~Valle, Benetti, Lyman, Smartt, Arcavi, Baltay, Bersier, {et~al.}}]{barbarino2017lsq14efd}
Barbarino, C., Botticella, M.~T., Dall'Ora, M., {et~al.} 2017, Monthly Notices of the Royal Astronomical Society, 471, 2463

\bibitem[{Barbarino {et~al.}(2021)Barbarino, Sollerman, Taddia, Fremling, Karamehmetoglu, Arcavi, Gal-Yam, Laher, Schulze, Wozniak, {et~al.}}]{barbarino2021type}
Barbarino, C., Sollerman, J., Taddia, F., {et~al.} 2021, Astronomy \& Astrophysics, 651, A81

\bibitem[{Bellm {et~al.}(2018)Bellm, Kulkarni, Graham, Dekany, Smith, Riddle, Masci, Helou, Prince, Adams, {et~al.}}]{bellm2018zwicky}
Bellm, E.~C., Kulkarni, S.~R., Graham, M.~J., {et~al.} 2018, Publications of the Astronomical Society of the Pacific, 131, 018002

\bibitem[{Ben-Ami {et~al.}(2012)Ben-Ami, Gal-Yam, Filippenko, Mazzali, Modjaz, Yaron, Arcavi, Cenko, Horesh, Howell, {et~al.}}]{ben2012discovery}
Ben-Ami, S., Gal-Yam, A., Filippenko, A.~V., {et~al.} 2012, The Astrophysical Journal Letters, 760, L33

\bibitem[{Benetti {et~al.}(2011)Benetti, Turatto, Valenti, Pastorello, Cappellaro, Botticella, Bufano, Ghinassi, Harutyunyan, Inserra, {et~al.}}]{benetti2011type}
Benetti, S., Turatto, M., Valenti, S., {et~al.} 2011, Monthly Notices of the Royal Astronomical Society, 411, 2726

\bibitem[{{Bernstein} {et~al.}(2012){Bernstein}, {Kessler}, {Kuhlmann}, {Biswas}, {Kovacs}, {Aldering}, {Crane}, {D'Andrea}, {Finley}, {Frieman}, {Hufford}, {Jarvis}, {Kim}, {Marriner}, {Mukherjee}, {Nichol}, {Nugent}, {Parkinson}, {Reis}, {Sako}, {Spinka}, \& {Sullivan}}]{bernstein12}
{Bernstein}, J.~P., {Kessler}, R., {Kuhlmann}, S., {et~al.} 2012, \apj, 753, 152, \dodoi{10.1088/0004-637X/753/2/152}

\bibitem[{{Bersten} {et~al.}(2018){Bersten}, {Folatelli}, {Garc{\'\i}a}, {van Dyk}, {Benvenuto}, {Orellana}, {Buso}, {S{\'a}nchez}, {Tanaka}, {Maeda}, {Filippenko}, {Zheng}, {Brink}, {Cenko}, {de Jaeger}, {Kumar}, {Moriya}, {Nomoto}, {Perley}, {Shivvers}, \& {Smith}}]{2018Natur.554..497B}
{Bersten}, M.~C., {Folatelli}, G., {Garc{\'\i}a}, F., {et~al.} 2018, \nat, 554, 497, \dodoi{10.1038/nature25151}

\bibitem[{Bersten {et~al.}(2018)Bersten, Folatelli, Garc{\'\i}a, Van~Dyk, Benvenuto, Orellana, Buso, S{\'a}nchez, Tanaka, Maeda, {et~al.}}]{bersten2018surge}
Bersten, M.~C., Folatelli, G., Garc{\'\i}a, F., {et~al.} 2018, Nature, 554, 497

\bibitem[{{Bianco} {et~al.}(2014){Bianco}, {Modjaz}, {Hicken}, {Friedman}, {Kirshner}, {Bloom}, {Challis}, {Marion}, {Wood-Vasey}, \& {Rest}}]{bianco14}
{Bianco}, F.~B., {Modjaz}, M., {Hicken}, M., {et~al.} 2014, \apjs, 213, 19, \dodoi{10.1088/0067-0049/213/2/19}

\bibitem[{{Blondin} {et~al.}(2007){Blondin}, {Modjaz}, {Kirshner}, {Challis}, \& {Calkins}}]{blondin07}
{Blondin}, S., {Modjaz}, M., {Kirshner}, R., {Challis}, P., \& {Calkins}, M. 2007, Central Bureau Electronic Telegrams, 800

\bibitem[{Boone(2019)}]{boone2019avocado}
Boone, K. 2019, The Astronomical Journal, 158, 257

\bibitem[{{Bufano} {et~al.}(2012){Bufano}, {Pian}, {Sollerman}, {Benetti}, {Pignata}, {Valenti}, {Covino}, {D'Avanzo}, {Malesani}, {Cappellaro}, {Della Valle}, {Fynbo}, {Hjorth}, {Mazzali}, {Reichart}, {Starling}, {Turatto}, {Vergani}, {Wiersema}, {Amati}, {Bersier}, {Campana}, {Cano}, {Castro-Tirado}, {Chincarini}, {D'Elia}, {de Ugarte Postigo}, {Deng}, {Ferrero}, {Filippenko}, {Goldoni}, {Gorosabel}, {Greiner}, {Hammer}, {Jakobsson}, {Kaper}, {Kawabata}, {Klose}, {Levan}, {Maeda}, {Masetti}, {Milvang-Jensen}, {Mirabel}, {M{\o}ller}, {Nomoto}, {Palazzi}, {Piranomonte}, {Salvaterra}, {Stratta}, {Tagliaferri}, {Tanaka}, {Tanvir}, \& {Wijers}}]{2012ApJ...753...67B}
{Bufano}, F., {Pian}, E., {Sollerman}, J., {et~al.} 2012, \apj, 753, 67, \dodoi{10.1088/0004-637X/753/1/67}

\bibitem[{{Bufano} {et~al.}(2014){Bufano}, {Pignata}, {Bersten}, {Mazzali}, {Ryder}, {Margutti}, {Milisavljevic}, {Morelli}, {Benetti}, {Cappellaro}, {Gonzalez-Gaitan}, {Romero-Ca{\~n}izales}, {Stritzinger}, {Walker}, {Anderson}, {Contreras}, {de Jaeger}, {F{\"o}rster}, {Gutierrez}, {Hamuy}, {Hsiao}, {Morrell}, {Olivares E.}, {Paillas}, {Parker}, {Pian}, {Pickering}, {Sanders}, {Stockdale}, {Turatto}, {Valenti}, {Fesen}, {Maza}, {Nomoto}, {Phillips}, \& {Soderberg}}]{bufano14}
{Bufano}, F., {Pignata}, G., {Bersten}, M., {et~al.} 2014, \mnras, 439, 1807, \dodoi{10.1093/mnras/stu065}

\bibitem[{Campana {et~al.}(2006)Campana, Mangano, Blustin, Brown, Burrows, Chincarini, Cummings, Cusumano, Valle, Malesani, Mészáros, Nousek, Page, Sakamoto, Waxman, Zhang, Dai, Gehrels, Immler, Marshall, Mason, Moretti, O’Brien, Osborne, Page, Romano, Roming, Tagliaferri, Cominsky, Giommi, Godet, Kennea, Krimm, Angelini, Barthelmy, Boyd, Palmer, Wells, \& White}]{Campana_2006}
Campana, S., Mangano, V., Blustin, A.~J., {et~al.} 2006, Nature, 442, 1008–1010, \dodoi{10.1038/nature04892}

\bibitem[{{Cano}(2013{\natexlab{a}})}]{cano13}
{Cano}, Z. 2013{\natexlab{a}}, \mnras, 434, 1098, \dodoi{10.1093/mnras/stt1048}

\bibitem[{{Cano}(2013{\natexlab{b}})}]{2013MNRAS.434.1098C}
---. 2013{\natexlab{b}}, \mnras, 434, 1098, \dodoi{10.1093/mnras/stt1048}

\bibitem[{{Cano}(2014)}]{cano14}
---. 2014, \apj, 794, 121, \dodoi{10.1088/0004-637X/794/2/121}

\bibitem[{Chevalier \& Fransson(2008)}]{chevalier2008shock}
Chevalier, R.~A., \& Fransson, C. 2008, The Astrophysical Journal, 683, L135

\bibitem[{Chornock {et~al.}(2010)Chornock, Berger, Levesque, Soderberg, Foley, Fox, Frebel, Simon, Bochanski, Challis, {et~al.}}]{chornock2010spectroscopic}
Chornock, R., Berger, E., Levesque, E.~M., {et~al.} 2010, arXiv preprint arXiv:1004.2262

\bibitem[{Ciabattari {et~al.}(2011)Ciabattari, Mazzoni, Jin, Gao, Tomasella, Valenti, Ochner, Benetti, Cappellaro, \& Pastorello}]{ciabattari2011supernova}
Ciabattari, F., Mazzoni, E., Jin, Z., {et~al.} 2011, Central Bureau Electronic Telegrams, 2827, 1

\bibitem[{Ciabattari {et~al.}(2013)Ciabattari, Mazzoni, Donati, Petroni, Foglia, Galli, Cenko, Clubb, Zheng, Kelly, {et~al.}}]{ciabattari2013supernova}
Ciabattari, F., Mazzoni, E., Donati, S., {et~al.} 2013, Central Bureau Electronic Telegrams, 3557, 1

\bibitem[{Clocchiatti \& Wheeler(1997)}]{clocchiatti1997light}
Clocchiatti, A., \& Wheeler, J. 1997, The Astrophysical Journal, 491, 375

\bibitem[{{Das} {et~al.}(2023){Das}, {Kasliwal}, {Fremling}, {Yang}, {Schulze}, {Sollerman}, {Sit}, {De}, {Tzanidakis}, {Perley}, {Anand}, {Andreoni}, {Barbarino}, {Brudge}, {Drake}, {Gal-Yam}, {Laher}, {Karambelkar}, {Kulkarni}, {Masci}, {Medford}, {Polin}, {Reedy}, {Riddle}, {Sharma}, {Smith}, {Yan}, {Yang}, \& {Yao}}]{Das2023Supernovae}
{Das}, K.~K., {Kasliwal}, M.~M., {Fremling}, C., {et~al.} 2023, \apj, 959, 12, \dodoi{10.3847/1538-4357/acfeeb}

\bibitem[{{De} {et~al.}(2021){De}, {Fremling}, {Gal-Yam}, {Yaron}, {Kasliwal}, \& {Kulkarni}}]{De2021Peculiar}
{De}, K., {Fremling}, U.~C., {Gal-Yam}, A., {et~al.} 2021, \apjl, 907, L18, \dodoi{10.3847/2041-8213/abd627}

\bibitem[{De {et~al.}(2020)De, Kasliwal, Tzanidakis, Fremling, Adams, Aloisi, Andreoni, Bagdasaryan, Bellm, Bildsten, {et~al.}}]{de2020zwicky}
De, K., Kasliwal, M.~M., Tzanidakis, A., {et~al.} 2020, The Astrophysical Journal, 905, 58

\bibitem[{Drout {et~al.}(2016)Drout, Milisavljevic, Parrent, Margutti, Kamble, Soderberg, Challis, Chornock, Fong, Frank, {et~al.}}]{drout2016double}
Drout, M., Milisavljevic, D., Parrent, J., {et~al.} 2016, The Astrophysical Journal, 821, 57

\bibitem[{Drout {et~al.}(2011)Drout, Soderberg, Gal-Yam, Cenko, Fox, Leonard, Sand, Moon, Arcavi, \& Green}]{drout11}
Drout, M.~R., Soderberg, A.~M., Gal-Yam, A., {et~al.} 2011, The Astrophysical Journal, 741, 97, \dodoi{10.1088/0004-637X/741/2/97}

\bibitem[{{Fan} {et~al.}(2011){Fan}, {Zhang}, {Xu}, {Liang}, \& {Zhang}}]{2011ApJ...726...32F}
{Fan}, Y.-Z., {Zhang}, B.-B., {Xu}, D., {Liang}, E.-W., \& {Zhang}, B. 2011, \apj, 726, 32, \dodoi{10.1088/0004-637X/726/1/32}

\bibitem[{Filippenko(1997)}]{filippenko97}
Filippenko, A.~V. 1997, Annual Review of Astronomy and Astrophysics, 35, 309, \dodoi{10.1146/annurev.astro.35.1.309}

\bibitem[{Filippenko {et~al.}(1993)Filippenko, Matheson, \& Ho}]{filippenko93}
Filippenko, A.~V., Matheson, T., \& Ho, L.~C. 1993, The Astrophysical Journal, 415, L103, \dodoi{10.1086/187043}

\bibitem[{Folatelli {et~al.}(2014)Folatelli, Bersten, Kuncarayakti, Estay, Anderson, Holmbo, Maeda, Morrell, Nomoto, Pignata, {et~al.}}]{folatelli2014supernova}
Folatelli, G., Bersten, M.~C., Kuncarayakti, H., {et~al.} 2014, The Astrophysical Journal, 792, 7

\bibitem[{Foreman-Mackey(2015)}]{foreman2015george}
Foreman-Mackey, D. 2015, Astrophysics Source Code Library, ascl

\bibitem[{Gagliano {et~al.}(2023)Gagliano, Contardo, Mackey, Malz, \& Aleo}]{gagliano2023first}
Gagliano, A., Contardo, G., Mackey, D.~F., Malz, A.~I., \& Aleo, P.~D. 2023, arXiv preprint arXiv:2305.08894

\bibitem[{{Gagliano} {et~al.}(2022){Gagliano}, {Izzo}, {Kilpatrick}, {Mockler}, {Jacobson-Gal{\'a}n}, {Terreran}, {Dimitriadis}, {Zenati}, {Auchettl}, {Drout}, {Narayan}, {Foley}, {Margutti}, {Rest}, {Jones}, {Aganze}, {Aleo}, {Burgasser}, {Coulter}, {Gerasimov}, {Gall}, {Hjorth}, {Hsu}, {Magnier}, {Mandel}, {Piro}, {Rojas-Bravo}, {Siebert}, {Stacey}, {Stroh}, {Swift}, {Taggart}, {Tinyanont}, \& {Tinyanont}}]{2022ApJ...924...55G}
{Gagliano}, A., {Izzo}, L., {Kilpatrick}, C.~D., {et~al.} 2022, \apj, 924, 55, \dodoi{10.3847/1538-4357/ac35ec}

\bibitem[{Gal-Yam(2016)}]{gal2016observational}
Gal-Yam, A. 2016, arXiv preprint arXiv:1611.09353

\bibitem[{{Gal-Yam} {et~al.}(2002){Gal-Yam}, {Ofek}, \& {Shemmer}}]{2002MNRAS.332L..73G}
{Gal-Yam}, A., {Ofek}, E.~O., \& {Shemmer}, O. 2002, \mnras, 332, L73, \dodoi{10.1046/j.1365-8711.2002.05535.x}

\bibitem[{Gangopadhyay {et~al.}(2020)Gangopadhyay, Misra, Sahu, Wang, Kumar, Li, Anupama, Dastidar, Elias-Rosa, Kumar, {et~al.}}]{gangopadhyay2020optical}
Gangopadhyay, A., Misra, K., Sahu, D., {et~al.} 2020, Monthly Notices of the Royal Astronomical Society, 497, 3770

\bibitem[{Guillochon {et~al.}(2018)Guillochon, Nicholl, Villar, Mockler, Narayan, Mandel, Berger, \& Williams}]{guillochon2018mosfit}
Guillochon, J., Nicholl, M., Villar, V.~A., {et~al.} 2018, The Astrophysical Journal Supplement Series, 236, 6

\bibitem[{{Guillochon} {et~al.}(2016){Guillochon}, {Parrent}, \& {Margutti}}]{guillochon16}
{Guillochon}, J., {Parrent}, J., \& {Margutti}, R. 2016, ArXiv e-prints.
\newblock \doarXiv{1605.01054}

\bibitem[{Harris {et~al.}(2020)Harris, Millman, Van Der~Walt, Gommers, Virtanen, Cournapeau, Wieser, Taylor, Berg, Smith, {et~al.}}]{harris2020array}
Harris, C.~R., Millman, K.~J., Van Der~Walt, S.~J., {et~al.} 2020, Nature, 585, 357

\bibitem[{Hlo{\v{z}}ek {et~al.}(2020)Hlo{\v{z}}ek, Ponder, Malz, Dai, Narayan, Ishida, Allam~Jr, Bahmanyar, Biswas, Galbany, {et~al.}}]{hlovzek2020results}
Hlo{\v{z}}ek, R., Ponder, K., Malz, A., {et~al.} 2020, arXiv preprint arXiv:2012.12392

\bibitem[{Ho {et~al.}(2023)Ho, Perley, Gal-Yam, Lunnan, Sollerman, Schulze, Das, Dobie, Yao, Fremling, {et~al.}}]{ho2023photometric}
Ho, A.~Y., Perley, D.~A., Gal-Yam, A., {et~al.} 2023, The Astrophysical Journal, 949, 120

\bibitem[{Horesh {et~al.}(2013)Horesh, Kulkarni, Corsi, Frail, Cenko, Ben-Ami, Gal-Yam, Yaron, Arcavi, Kasliwal, {et~al.}}]{horesh2013ptf}
Horesh, A., Kulkarni, S.~R., Corsi, A., {et~al.} 2013, The Astrophysical Journal, 778, 63

\bibitem[{Hosseinzadeh {et~al.}(2019)Hosseinzadeh, McCully, Zabludoff, Arcavi, French, Howell, Berger, \& Hiramatsu}]{hosseinzadeh2019type}
Hosseinzadeh, G., McCully, C., Zabludoff, A.~I., {et~al.} 2019, The Astrophysical Journal Letters, 871, L9

\bibitem[{Hosseinzadeh {et~al.}(2017)Hosseinzadeh, Arcavi, Valenti, McCully, Howell, Johansson, Sollerman, Pastorello, Benetti, Cao, {et~al.}}]{hosseinzadeh2017type}
Hosseinzadeh, G., Arcavi, I., Valenti, S., {et~al.} 2017, The Astrophysical Journal, 836, 158

\bibitem[{Hunter(2007)}]{hunter2007matplotlib}
Hunter, J.~D. 2007, Computing in science \& engineering, 9, 90

\bibitem[{Ivezi{\'c} {et~al.}(2019)Ivezi{\'c}, Kahn, Tyson, Abel, Acosta, Allsman, Alonso, AlSayyad, Anderson, Andrew, {et~al.}}]{ivezic2019lsst}
Ivezi{\'c}, {\v{Z}}., Kahn, S.~M., Tyson, J.~A., {et~al.} 2019, The Astrophysical Journal, 873, 111

\bibitem[{{Karamehmetoglu} {et~al.}(2023){Karamehmetoglu}, {Sollerman}, {Taddia}, {Barbarino}, {Feindt}, {Fremling}, {Gal-Yam}, {Kasliwal}, {Petrushevska}, {Schulze}, {Stritzinger}, \& {Zapartas}}]{Karamehmetoglu2023supernovae}
{Karamehmetoglu}, E., {Sollerman}, J., {Taddia}, F., {et~al.} 2023, \aap, 678, A87, \dodoi{10.1051/0004-6361/202245231}

\bibitem[{{Karthik Yadavalli} {et~al.}(2023){Karthik Yadavalli}, {Villar}, {Izzo}, {Zenati}, {Foley}, {Wheeler}, {Angus}, {B{\'a}nhidi}, {Auchettl}, {Imre B{\'\i}r{\'o}}, {B{\'o}di}, {Bodola}, {de Boer}, {Chambers}, {Chornock}, {Coulter}, {Cs{\'a}nyi}, {Cseh}, {Dandu}, {Davis}, {Braden Dickinson}, {Farias}, {Gall}, {Gao}, {Jacobson-Galan}, {Khetan}, {Kilpatrick}, {K{\"o}nyves-T{\'o}th}, {Kriskovics}, {LeBaron}, {Loertscher}, {Le Saux}, {Margutti}, {Magnier}, {McGill}, {Miao}, {P{\'a}l}, {P{\'a}l}, {Pan}, {Politsch}, {Ransome}, {Ramirez-Ruiz}, {Rest}, {Rest}, {Robinson}, {Sears}, {Scheer}, {S{\'o}dor}, {Swift}, {Sz{\'e}kely}, {Szak{\'a}ts}, {Szalai}, {Taggart}, {Venkatraman}, {Vink{\'o}}, {Yang}, \& {Zhou}}]{2023arXiv230812991K}
{Karthik Yadavalli}, S., {Villar}, V.~A., {Izzo}, L., {et~al.} 2023, arXiv e-prints, arXiv:2308.12991, \dodoi{10.48550/arXiv.2308.12991}

\bibitem[{{Kessler} {et~al.}(2009){Kessler}, {Bernstein}, {Cinabro}, {Dilday}, {Frieman}, {Jha}, {Kuhlmann}, {Miknaitis}, {Sako}, {Taylor}, \& {Vanderplas}}]{2009PASP..121.1028K}
{Kessler}, R., {Bernstein}, J.~P., {Cinabro}, D., {et~al.} 2009, \pasp, 121, 1028, \dodoi{10.1086/605984}

\bibitem[{Kessler {et~al.}(2019)Kessler, Narayan, Avelino, Bachelet, Biswas, Brown, Chernoff, Connolly, Dai, Daniel, {et~al.}}]{kessler2019models}
Kessler, R., Narayan, G., Avelino, A., {et~al.} 2019, Publications of the Astronomical Society of the Pacific, 131, 094501

\bibitem[{{Kessler} {et~al.}(2019){Kessler}, {Narayan}, {Avelino}, {Bachelet}, {Biswas}, {Brown}, {Chernoff}, {Connolly}, {Dai}, {Daniel}, {Di Stefano}, {Drout}, {Galbany}, {Gonz{\'a}lez-Gait{\'a}n}, {Graham}, {Hlo{\v{z}}ek}, {Ishida}, {Guillochon}, {Jha}, {Jones}, {Mandel}, {Muthukrishna}, {O'Grady}, {Peters}, {Pierel}, {Ponder}, {Pr{\v{s}}a}, {Rodney}, {Villar}, {LSST Dark Energy Science Collaboration}, \& {Transient and Variable Stars Science Collaboration}}]{2019PASP..131i4501K}
{Kessler}, R., {Narayan}, G., {Avelino}, A., {et~al.} 2019, \pasp, 131, 094501, \dodoi{10.1088/1538-3873/ab26f1}

\bibitem[{Kilpatrick {et~al.}(2016)Kilpatrick, Foley, Abramson, Pan, Lu, Williams, Treu, Siebert, Fassnacht, \& Max}]{10.1093/mnras/stw3082}
Kilpatrick, C.~D., Foley, R.~J., Abramson, L.~E., {et~al.} 2016, Monthly Notices of the Royal Astronomical Society, 465, 4650, \dodoi{10.1093/mnras/stw3082}

\bibitem[{Krige(1951)}]{Krige51}
Krige, D.~G. 1951, Journal of the Chemical, Metallurgical and Mining Society of South Africa, 52, 119, \dodoi{10.2307/3006914}

\bibitem[{Kulkarni(2013)}]{kulkarni2013intermediate}
Kulkarni, S. 2013, The Astronomer's Telegram, 4807, 1

\bibitem[{{Kumar} {et~al.}(2013){Kumar}, {Pandey}, {Sahu}, {Vinko}, {Moskvitin}, {Anupama}, {Bhatt}, {Ordasi}, {Nagy}, {Sokolov}, {Sokolova}, {Komarova}, {Kumar}, {Bose}, {Roy}, \& {Sagar}}]{2013MNRAS.431..308K}
{Kumar}, B., {Pandey}, S.~B., {Sahu}, D.~K., {et~al.} 2013, \mnras, 431, 308, \dodoi{10.1093/mnras/stt162}

\bibitem[{Kumar {et~al.}(2013)Kumar, Pandey, Sahu, Vinko, Moskvitin, Anupama, Bhatt, Ordasi, Nagy, Sokolov, Sokolova, Komarova, Kumar, Bose, Roy, \& Sagar}]{10.1093/mnras/stt162}
Kumar, B., Pandey, S.~B., Sahu, D.~K., {et~al.} 2013, Monthly Notices of the Royal Astronomical Society, 431, 308, \dodoi{10.1093/mnras/stt162}

\bibitem[{Law {et~al.}(2009)Law, Kulkarni, Dekany, Ofek, Quimby, Nugent, Surace, Grillmair, Bloom, Kasliwal, {et~al.}}]{law2009palomar}
Law, N.~M., Kulkarni, S.~R., Dekany, R.~G., {et~al.} 2009, Publications of the Astronomical Society of the Pacific, 121, 1395

\bibitem[{Li(2007)}]{li2007shock}
Li, L.-X. 2007, Monthly Notices of the Royal Astronomical Society, 375, 240

\bibitem[{{Liu} \& {Modjaz}(2014)}]{liu14}
{Liu}, Y., \& {Modjaz}, M. 2014, ArXiv e-prints.
\newblock \doarXiv{1405.1437}

\bibitem[{{Liu} {et~al.}(2017){Liu}, {Modjaz}, \& {Bianco}}]{liu17}
{Liu}, Y.-Q., {Modjaz}, M., \& {Bianco}, F.~B. 2017, \apj, 845, 85, \dodoi{10.3847/1538-4357/aa7f74}

\bibitem[{Liu {et~al.}(2016)Liu, Modjaz, Bianco, \& Graur}]{liu16}
Liu, Y.-Q., Modjaz, M., Bianco, F.~B., \& Graur, O. 2016, The Astrophysical Journal, 827, 90

\bibitem[{{Lokken} {et~al.}(2023){Lokken}, {Gagliano}, {Narayan}, {Hlo{\v{z}}ek}, {Kessler}, {Crenshaw}, {Salo}, {Alves}, {Chatterjee}, {Vincenzi}, {Malz}, \& {LSST Dark Energy Science Collaboration}}]{2023MNRAS.520.2887L}
{Lokken}, M., {Gagliano}, A., {Narayan}, G., {et~al.} 2023, \mnras, 520, 2887, \dodoi{10.1093/mnras/stad302}

\bibitem[{Lyman {et~al.}(2016)Lyman, Bersier, James, Mazzali, Eldridge, Fraser, \& Pian}]{lyman2016bolometric}
Lyman, J., Bersier, D., James, P., {et~al.} 2016, Monthly Notices of the Royal Astronomical Society, 457, 328

\bibitem[{{Malesani} {et~al.}(2009){Malesani}, {Fynbo}, {Hjorth}, {Leloudas}, {Sollerman}, {Stritzinger}, {Vreeswijk}, {et~al.}}]{malesani09}
{Malesani}, D., {Fynbo}, J.~P.~U., {Hjorth}, J., {et~al.} 2009, \apjl, 692, L84, \dodoi{10.1088/0004-637X/692/2/L84}

\bibitem[{{Malz} {et~al.}(2019){Malz}, {Hlo{\v{z}}ek}, {Allam}, {Bahmanyar}, {Biswas}, {Dai}, {Galbany}, {Ishida}, {Jha}, {Jones}, {Kessler}, {Lochner}, {Mahabal}, {Mandel}, {Mart{\'\i}nez-Galarza}, {McEwen}, {Muthukrishna}, {Narayan}, {Peiris}, {Peters}, {Ponder}, {Setzer}, {(the LSST Dark Energy Science Collaboration}, {LSST Transients}, \& {Variable Stars Science Collaboration}}]{Malz2019Photometric}
{Malz}, A.~I., {Hlo{\v{z}}ek}, R., {Allam}, T., J., {et~al.} 2019, \aj, 158, 171, \dodoi{10.3847/1538-3881/ab3a2f}

\bibitem[{{McAllister} {et~al.}(2017){McAllister}, {Littlefair}, {Dhillon}, {Marsh}, {Ashley}, {Bours}, {Breedt}, {Hardy}, {Hermes}, {Kengkriangkrai}, {Kerry}, {Rattanasoon}, \& {Sahman}}]{McAllister17}
{McAllister}, M.~J., {Littlefair}, S.~P., {Dhillon}, V.~S., {et~al.} 2017, \mnras, 464, 1353, \dodoi{10.1093/mnras/stw2417}

\bibitem[{McKinney {et~al.}(2011)}]{mckinney2011pandas}
McKinney, W., {et~al.} 2011, Python for high performance and scientific computing, 14, 1

\bibitem[{Melandri {et~al.}(2014)Melandri, Pian, D’elia, D’Avanzo, Della~Valle, Mazzali, Tagliaferri, Cano, Levan, M$\Delta$oller, {et~al.}}]{melandri2014diversity}
Melandri, A., Pian, E., D’elia, V., {et~al.} 2014, Astronomy \& Astrophysics, 567, A29

\bibitem[{Mirabal {et~al.}(2006)Mirabal, Halpern, An, Thorstensen, \& Terndrup}]{mirabal2006grb}
Mirabal, N., Halpern, J., An, D., Thorstensen, J., \& Terndrup, D. 2006, The Astrophysical Journal Letters, 643, L99

\bibitem[{Modjaz {et~al.}(2019)Modjaz, Guti{\'e}rrez, \& Arcavi}]{modjaz2019new}
Modjaz, M., Guti{\'e}rrez, C.~P., \& Arcavi, I. 2019, Nature Astronomy, 3, 717

\bibitem[{Modjaz {et~al.}(2016)Modjaz, Liu, Bianco, \& Graur}]{modjaz2016spectral}
Modjaz, M., Liu, Y.~Q., Bianco, F.~B., \& Graur, O. 2016, The Astrophysical Journal, 832, 108

\bibitem[{Modjaz {et~al.}(2009)Modjaz, Li, Butler, Chornock, Perley, Blondin, Bloom, Filippenko, Kirshner, Kocevski, {et~al.}}]{modjaz2009shock}
Modjaz, M., Li, W., Butler, N., {et~al.} 2009, The Astrophysical Journal, 702, 226

\bibitem[{{Modjaz} {et~al.}(2014){Modjaz}, {Blondin}, {Kirshner}, {Matheson}, {Berlind}, {Bianco}, {Calkins}, {Challis}, {Garnavich}, {Hicken}, {Jha}, {Liu}, \& {Marion}}]{modjaz14}
{Modjaz}, M., {Blondin}, S., {Kirshner}, R.~P., {et~al.} 2014, \aj, 147, 99, \dodoi{10.1088/0004-6256/147/5/99}

\bibitem[{Morales-Garoffolo {et~al.}(2014)Morales-Garoffolo, Elias-Rosa, Benetti, Taubenberger, Cappellaro, Pastorello, Klauser, Valenti, Howerton, Ochner, Schramm, Siviero, Tartaglia, \& Tomasella}]{10.1093/mnras/stu1837}
Morales-Garoffolo, A., Elias-Rosa, N., Benetti, S., {et~al.} 2014, Monthly Notices of the Royal Astronomical Society, 445, 1647, \dodoi{10.1093/mnras/stu1837}

\bibitem[{{Morales-Garoffolo} {et~al.}(2014){Morales-Garoffolo}, {Elias-Rosa}, {Benetti}, {Taubenberger}, {Cappellaro}, {Pastorello}, {Klauser}, {Valenti}, {Howerton}, {Ochner}, {Schramm}, {Siviero}, {Tartaglia}, \& {Tomasella}}]{moralesgaroffolo14}
{Morales-Garoffolo}, A., {Elias-Rosa}, N., {Benetti}, S., {et~al.} 2014, \mnras, 445, 1647, \dodoi{10.1093/mnras/stu1837}

\bibitem[{{Morales-Garoffolo} {et~al.}(2015{\natexlab{a}}){Morales-Garoffolo}, {Elias-Rosa}, {Bersten}, {Jerkstrand}, {Taubenberger}, {Benetti}, {Cappellaro}, {Kotak}, {Pastorello}, {Bufano}, {Dom{\'\i}nguez}, {Ergon}, {Fraser}, {Gao}, {Garc{\'\i}a}, {Howell}, {Isern}, {Smartt}, {Tomasella}, \& {Valenti}}]{2015MNRAS.454...95M}
{Morales-Garoffolo}, A., {Elias-Rosa}, N., {Bersten}, M., {et~al.} 2015{\natexlab{a}}, \mnras, 454, 95, \dodoi{10.1093/mnras/stv1972}

\bibitem[{{Morales-Garoffolo} {et~al.}(2015{\natexlab{b}}){Morales-Garoffolo}, {Elias-Rosa}, {Bersten}, {Jerkstrand}, {Taubenberger}, {Benetti}, {Cappellaro}, {Kotak}, {Pastorello}, {Bufano}, {Dom{\'{\i}}nguez}, {Ergon}, {Fraser}, {Gao}, {Garc{\'{\i}}a}, {Howell}, {Isern}, {Smartt}, {Tomasella}, \& {Valenti}}]{moralesgaroffolo15}
---. 2015{\natexlab{b}}, \mnras, 454, 95, \dodoi{10.1093/mnras/stv1972}

\bibitem[{Morrell \& Hamuy(2003)}]{morrell2003supernovae}
Morrell, N., \& Hamuy, M. 2003, International Astronomical Union Circular, 8203, 2

\bibitem[{Najita {et~al.}(2016)Najita, Willman, Finkbeiner, Foley, Hawley, Newman, Rudnick, Simon, Trilling, Street, {et~al.}}]{najita2016maximizing}
Najita, J., Willman, B., Finkbeiner, D.~P., {et~al.} 2016, arXiv preprint arXiv:1610.01661

\bibitem[{{Narayan} \& {ELAsTiCC Team}(2023)}]{2023AAS...24111701N}
{Narayan}, G., \& {ELAsTiCC Team}. 2023, in American Astronomical Society Meeting Abstracts, Vol.~55, American Astronomical Society Meeting Abstracts, 117.01

\bibitem[{Nowogrodzki(2019)}]{nowogrodzki2019support}
Nowogrodzki, A. 2019, Nature, 571, 133

\bibitem[{{Oates} {et~al.}(2012){Oates}, {Bayless}, {Stritzinger}, {Prichard}, {Prieto}, {Immler}, {Brown}, {Breeveld}, {De Pasquale}, {Kuin}, {Hamuy}, {Holland}, {Taddia}, \& {Roming}}]{2012MNRAS.424.1297O}
{Oates}, S.~R., {Bayless}, A.~J., {Stritzinger}, M.~D., {et~al.} 2012, \mnras, 424, 1297, \dodoi{10.1111/j.1365-2966.2012.21311.x}

\bibitem[{{Okyudo} {et~al.}(1993){Okyudo}, {Kato}, {Ishida}, {Tokimasa}, \& {Yamaoka}}]{1993PASJ...45L..63O}
{Okyudo}, M., {Kato}, T., {Ishida}, T., {Tokimasa}, N., \& {Yamaoka}, H. 1993, \pasj, 45, L63

\bibitem[{{Orellana} \& {Bersten}(2022)}]{2022A&A...667A..92O}
{Orellana}, M., \& {Bersten}, M.~C. 2022, \aap, 667, A92, \dodoi{10.1051/0004-6361/202244124}

\bibitem[{{Pastorello} {et~al.}(2015){Pastorello}, {Benetti}, {Brown}, {Tsvetkov}, {Inserra}, {Taubenberger}, {Tomasella}, {Fraser}, {Rich}, {Botticella}, {Bufano}, {Cappellaro}, {Ergon}, {Gorbovskoy}, {Harutyunyan}, {Huang}, {Kotak}, {Lipunov}, {Magill}, {Miluzio}, {Morrell}, {Ochner}, {Smartt}, {Sollerman}, {Spiro}, {Stritzinger}, {Turatto}, {Valenti}, {Wang}, {Wright}, {Yurkov}, {Zampieri}, \& {Zhang}}]{pastorello15}
{Pastorello}, A., {Benetti}, S., {Brown}, P.~J., {et~al.} 2015, \mnras, 449, 1921, \dodoi{10.1093/mnras/stu2745}

\bibitem[{Pellegrino {et~al.}(2023)Pellegrino, Hiramatsu, Arcavi, Howell, Bostroem, Brown, Burke, Elias-Rosa, Itagaki, Kaneda, {et~al.}}]{pellegrino2023sn}
Pellegrino, C., Hiramatsu, D., Arcavi, I., {et~al.} 2023, The Astrophysical Journal, 954, 35

\bibitem[{{Pellegrino} {et~al.}(2023){Pellegrino}, {Hiramatsu}, {Arcavi}, {Howell}, {Bostroem}, {Brown}, {Burke}, {Elias-Rosa}, {Itagaki}, {Kaneda}, {McCully}, {Modjaz}, {Padilla Gonzalez}, {Pritchard}, \& {Yesmin}}]{Pellegrino2023SN2020bio}
{Pellegrino}, C., {Hiramatsu}, D., {Arcavi}, I., {et~al.} 2023, \apj, 954, 35, \dodoi{10.3847/1538-4357/ace595}

\bibitem[{{Perley} {et~al.}(2020){Perley}, {Fremling}, {Sollerman}, {Miller}, {Dahiwale}, {Sharma}, {Bellm}, {Biswas}, {Brink}, {Bruch}, {De}, {Dekany}, {Drake}, {Duev}, {Filippenko}, {Gal-Yam}, {Goobar}, {Graham}, {Graham}, {Ho}, {Irani}, {Kasliwal}, {Kim}, {Kulkarni}, {Mahabal}, {Masci}, {Modak}, {Neill}, {Nordin}, {Riddle}, {Soumagnac}, {Strotjohann}, {Schulze}, {Taggart}, {Tzanidakis}, {Walters}, \& {Yan}}]{2020ApJ...904...35P}
{Perley}, D.~A., {Fremling}, C., {Sollerman}, J., {et~al.} 2020, \apj, 904, 35, \dodoi{10.3847/1538-4357/abbd98}

\bibitem[{Phillips(1993)}]{phillips1993absolute}
Phillips, M.~M. 1993, The Astrophysical Journal, 413, L105

\bibitem[{Pian {et~al.}(2017)Pian, Tomasella, Cappellaro, Benetti, Mazzali, Baltay, Branchesi, Brocato, Campana, Copperwheat, {et~al.}}]{pian2017optical}
Pian, E., Tomasella, L., Cappellaro, E., {et~al.} 2017, Monthly Notices of the Royal Astronomical Society, 466, 1848

\bibitem[{{Pietra}(1955)}]{pietra55}
{Pietra}, S. 1955, \memsai, 26, 185

\bibitem[{{Pignata} {et~al.}(2011){Pignata}, {Stritzinger}, {Soderberg}, {Mazzali}, {Phillips}, {Morrell}, {Anderson}, {Boldt}, {Campillay}, {Contreras}, {Folatelli}, {F{\"o}rster}, {Gonz{\'a}lez}, {Hamuy}, {Krzeminski}, {Maza}, {Roth}, {Salgado}, {Levesque}, {Rest}, {Crain}, {Foster}, {Haislip}, {Ivarsen}, {LaCluyze}, {Nysewander}, \& {Reichart}}]{pignata11}
{Pignata}, G., {Stritzinger}, M., {Soderberg}, A., {et~al.} 2011, \apj, 728, 14, \dodoi{10.1088/0004-637X/728/1/14}

\bibitem[{{Piro}(2015)}]{piro15}
{Piro}, A.~L. 2015, \apjl, 808, L51, \dodoi{10.1088/2041-8205/808/2/L51}

\bibitem[{Piro {et~al.}(2021)Piro, Haynie, \& Yao}]{piro2021shock}
Piro, A.~L., Haynie, A., \& Yao, Y. 2021, The Astrophysical Journal, 909, 209

\bibitem[{{Polin} {et~al.}(2021){Polin}, {Nugent}, \& {Kasen}}]{Polin2021Nebular}
{Polin}, A., {Nugent}, P., \& {Kasen}, D. 2021, \apj, 906, 65, \dodoi{10.3847/1538-4357/abcccc}

\bibitem[{Prentice {et~al.}(2016)Prentice, Mazzali, Pian, Gal-Yam, Kulkarni, Rubin, Corsi, Fremling, Sollerman, Yaron, {et~al.}}]{prentice2016bolometric}
Prentice, S., Mazzali, P., Pian, E., {et~al.} 2016, Monthly Notices of the Royal Astronomical Society, 458, 2973

\bibitem[{{Prentice} {et~al.}(2019){Prentice}, {Ashall}, {James}, {Short}, {Mazzali}, {Bersier}, {Crowther}, {Barbarino}, {Chen}, {Copperwheat}, {Darnley}, {Denneau}, {Elias-Rosa}, {Fraser}, {Galbany}, {Gal-Yam}, {Harmanen}, {Howell}, {Hosseinzadeh}, {Inserra}, {Kankare}, {Karamehmetoglu}, {Lamb}, {Limongi}, {Maguire}, {McCully}, {Olivares E}, {Piascik}, {Pignata}, {Reichart}, {Rest}, {Reynolds}, {Rodr{\'\i}guez}, {Saario}, {Schulze}, {Smartt}, {Smith}, {Sollerman}, {Stalder}, {Sullivan}, {Taddia}, {Valenti}, {Vergani}, {Williams}, \& {Young}}]{Prentice2019Investigating}
{Prentice}, S.~J., {Ashall}, C., {James}, P.~A., {et~al.} 2019, \mnras, 485, 1559, \dodoi{10.1093/mnras/sty3399}

\bibitem[{Press \& Teukolsky(1990)}]{press1990savitzky}
Press, W.~H., \& Teukolsky, S.~A. 1990, Computers in Physics, 4, 669

\bibitem[{{Pritchard} {et~al.}(2014){Pritchard}, {Roming}, {Brown}, {Bayless}, \& {Frey}}]{Pritchard14}
{Pritchard}, T.~A., {Roming}, P.~W.~A., {Brown}, P.~J., {Bayless}, A.~J., \& {Frey}, L.~H. 2014, \apj, 787, 157, \dodoi{10.1088/0004-637X/787/2/157}

\bibitem[{{Pritchard} {et~al.}(2021){Pritchard}, {Bensch}, {Modjaz}, {Williamson}, {Th{\"o}ne}, {Vink{\'o}}, {Bianco}, {Bostroem}, {Burke}, {Garc{\'\i}a-Benito}, {Galbany}, {Hiramatsu}, {Howell}, {Izzo}, {Kann}, {McCully}, {Pellegrino}, {de Ugarte Postigo}, {Valenti}, {Wang}, {Wheeler}, {Xiang}, {S{\'a}rneczky}, {B{\'o}di}, {Cseh}, {Tarczay-Neh{\'e}z}, {Kriskovics}, {Ordasi}, {P{\'a}l}, {Szak{\'a}ts}, \& {Vida}}]{Pritchard2021SN2018gep}
{Pritchard}, T.~A., {Bensch}, K., {Modjaz}, M., {et~al.} 2021, \apj, 915, 121, \dodoi{10.3847/1538-4357/ac00bc}

\bibitem[{Pruzhinskaya {et~al.}(2019)Pruzhinskaya, Malanchev, Kornilov, Ishida, Mondon, Volnova, \& Korolev}]{pruzhinskaya2019anomaly}
Pruzhinskaya, M.~V., Malanchev, K.~L., Kornilov, M.~V., {et~al.} 2019, Monthly Notices of the Royal Astronomical Society, 489, 3591

\bibitem[{Qu \& Sako(2021)}]{qu2021photometric}
Qu, H., \& Sako, M. 2021, arXiv preprint arXiv:2111.05539

\bibitem[{{Qu} \& {Sako}(2022)}]{Qu22}
{Qu}, H., \& {Sako}, M. 2022, \aj, 163, 57, \dodoi{10.3847/1538-3881/ac39a1}

\bibitem[{Reese {et~al.}(2015)Reese, Spencer, \& Ball}]{Reese15}
Reese, C.~S., Spencer, B.~S., \& Ball, E.~L. 2015, Statistical Analysis and Data Mining: The ASA Data Science Journal, 8, 302, \dodoi{10.1002/sam.11295}

\bibitem[{{Rho} {et~al.}(2021){Rho}, {Evans}, {Geballe}, {Banerjee}, {Hoeflich}, {Shahbandeh}, {Valenti}, {Yoon}, {Jin}, {Williamson}, {Modjaz}, {Hiramatsu}, {Howell}, {Pellegrino}, {Vink{\'o}}, {Cartier}, {Burke}, {McCully}, {An}, {Cha}, {Pritchard}, {Wang}, {Andrews}, {Galbany}, {Van Dyk}, {Graham}, {Blinnikov}, {Joshi}, {P{\'a}l}, {Kriskovics}, {Ordasi}, {Szakats}, {Vida}, {Chen}, {Li}, {Zhang}, \& {Yan}}]{2021ApJ...908..232R}
{Rho}, J., {Evans}, A., {Geballe}, T.~R., {et~al.} 2021, \apj, 908, 232, \dodoi{10.3847/1538-4357/abd850}

\bibitem[{{Richmond} {et~al.}(1994){Richmond}, {Treffers}, {Filippenko}, {Paik}, {Leibundgut}, {Schulman}, \& {Cox}}]{1994AJ....107.1022R}
{Richmond}, M.~W., {Treffers}, R.~R., {Filippenko}, A.~V., {et~al.} 1994, \aj, 107, 1022, \dodoi{10.1086/116915}

\bibitem[{{Ryle} \& {Smith}(1948)}]{ryle48}
{Ryle}, M., \& {Smith}, F.~G. 1948, \nat, 162, 462, \dodoi{10.1038/162462a0}

\bibitem[{{Sako} {et~al.}(2014){Sako}, {Bassett}, {Becker}, {Brown}, {Campbell}, {Cane}, {Cinabro}, {D'Andrea}, {Dawson}, {DeJongh}, {Depoy}, {Dilday}, {Doi}, {Filippenko}, {Fischer}, {Foley}, {Frieman}, {Galbany}, {Garnavich}, {Goobar}, {Gupta}, {Hill}, {Hayden}, {Hlozek}, {Holtzman}, {Hopp}, {Jha}, {Kessler}, {Kollatschny}, {Leloudas}, {Marriner}, {Marshall}, {Miquel}, {Morokuma}, {Mosher}, {Nichol}, {Nordin}, {Olmstead}, {Ostman}, {Prieto}, {Richmond}, {Romani}, {Sollerman}, {Stritzinger}, {Schneider}, {Smith}, {Wheeler}, {Yasuda}, \& {Zheng}}]{sako14}
{Sako}, M., {Bassett}, B., {Becker}, A.~C., {et~al.} 2014, ArXiv e-prints.
\newblock \doarXiv{1401.3317}

\bibitem[{Sapir \& Waxman(2017)}]{sapir2017uv}
Sapir, N., \& Waxman, E. 2017, The Astrophysical Journal, 838, 130

\bibitem[{{Schulze} {et~al.}(2014){Schulze}, {Malesani}, {Cucchiara}, {Tanvir}, {Kr{\"u}hler}, {de Ugarte Postigo}, {Leloudas}, {Lyman}, {Bersier}, {Wiersema}, {Perley}, {Schady}, {Gorosabel}, {Anderson}, {Castro-Tirado}, {Cenko}, {De Cia}, {Ellerbroek}, {Fynbo}, {Greiner}, {Hjorth}, {Kann}, {Kaper}, {Klose}, {Levan}, {Mart{\'{\i}}n}, {O'Brien}, {Page}, {Pignata}, {Rapaport}, {S{\'a}nchez-Ram{\'{\i}}rez}, {Sollerman}, {Smith}, {Sparre}, {Th{\"o}ne}, {Watson}, {Xu}, {Bauer}, {Bayliss}, {Bj{\"o}rnsson}, {Bremer}, {Cano}, {Covino}, {D'Elia}, {Frail}, {Geier}, {Goldoni}, {Hartoog}, {Jakobsson}, {Korhonen}, {Lee}, {Milvang-Jensen}, {Nardini}, {Nicuesa Guelbenzu}, {Oguri}, {Pandey}, {Petitpas}, {Rossi}, {Sandberg}, {Schmidl}, {Tagliaferri}, {Tilanus}, {Winters}, {Wright}, \& {Wuyts}}]{schulze14}
{Schulze}, S., {Malesani}, D., {Cucchiara}, A., {et~al.} 2014, \aap, 566, A102, \dodoi{10.1051/0004-6361/201423387}

\bibitem[{Schulze {et~al.}(2021)Schulze, Yaron, Sollerman, Leloudas, Gal, Wright, Lunnan, Gal-Yam, Ofek, Perley, {et~al.}}]{schulze2021palomar}
Schulze, S., Yaron, O., Sollerman, J., {et~al.} 2021, The Astrophysical Journal Supplement Series, 255, 29

\bibitem[{{Shivvers} {et~al.}(2016){Shivvers}, {Modjaz}, {Zheng}, {Filippenko}, {Silverman}, {Liu}, {Matheson}, {Pastorello}, {Graur}, {Foley}, {Chornock}, {Smith}, {Leaman}, \& {Benetti}}]{shivvers16}
{Shivvers}, I., {Modjaz}, M., {Zheng}, W., {et~al.} 2016, ArXiv e-prints.
\newblock \doarXiv{1609.02922}

\bibitem[{Soderberg {et~al.}(2008)Soderberg, Berger, Page, Schady, Parrent, Pooley, Wang, Ofek, Cucchiara, Rau, {et~al.}}]{soderberg2008extremely}
Soderberg, A.~M., Berger, E., Page, K., {et~al.} 2008, Nature, 453, 469

\bibitem[{Sollerman {et~al.}(2021)Sollerman, Yang, Perley, Schulze, Fremling, Kasliwal, Shin, \& Racine}]{sollerman2021maximum}
Sollerman, J., Yang, S., Perley, D., {et~al.} 2021, arXiv preprint arXiv:2109.14339

\bibitem[{Stritzinger {et~al.}(2018{\natexlab{a}})Stritzinger, Anderson, Contreras, Heinrich-Josties, Morrell, Phillips, Anais, Boldt, Busta, Burns, {et~al.}}]{stritzinger2018carnegie1}
Stritzinger, M., Anderson, J., Contreras, C., {et~al.} 2018{\natexlab{a}}, Astronomy \& Astrophysics, 609, A134

\bibitem[{Stritzinger {et~al.}(2018{\natexlab{b}})Stritzinger, Taddia, Burns, Phillips, Bersten, Contreras, Folatelli, Holmbo, Hsiao, Hoeflich, {et~al.}}]{stritzinger2018carnegie2}
Stritzinger, M., Taddia, F., Burns, C., {et~al.} 2018{\natexlab{b}}, Astronomy \& Astrophysics, 609, A135

\bibitem[{Stritzinger {et~al.}(2018{\natexlab{c}})Stritzinger, Taddia, Burns, Phillips, Bersten, Contreras, Folatelli, Holmbo, Hsiao, Hoeflich, {et~al.}}]{stritzinger2018carnegie3}
---. 2018{\natexlab{c}}, Astronomy \& Astrophysics, 609, A135

\bibitem[{{Taddia} {et~al.}(2015){Taddia}, {Sollerman}, {Leloudas}, {Stritzinger}, {Valenti}, {Galbany}, {Kessler}, {Schneider}, \& {Wheeler}}]{taddia15}
{Taddia}, F., {Sollerman}, J., {Leloudas}, G., {et~al.} 2015, \aap, 574, A60, \dodoi{10.1051/0004-6361/201423915}

\bibitem[{Taddia {et~al.}(2016)Taddia, Fremling, Sollerman, Corsi, Gal-Yam, Karamehmetoglu, Lunnan, Bue, Ergon, Kasliwal, {et~al.}}]{taddia2016iptf15dtg}
Taddia, F., Fremling, C., Sollerman, J., {et~al.} 2016, Astronomy \& Astrophysics, 592, A89

\bibitem[{Taddia {et~al.}(2018)Taddia, Stritzinger, Bersten, Baron, Burns, Contreras, Holmbo, Hsiao, Morrell, Phillips, {et~al.}}]{taddia2018carnegie}
Taddia, F., Stritzinger, M., Bersten, M., {et~al.} 2018, Astronomy \& Astrophysics, 609, A136

\bibitem[{{Thomsen} {et~al.}(2004){Thomsen}, {Hjorth}, {Watson}, {Gorosabel}, {Fynbo}, {Jensen}, {Andersen}, {Dall}, {Rasmussen}, {Bruntt}, {Laurikainen}, {Augusteijn}, {Pursimo}, {Germany}, {Jakobsson}, \& {Pedersen}}]{thomsen04}
{Thomsen}, B., {Hjorth}, J., {Watson}, D., {et~al.} 2004, \aap, 419, L21, \dodoi{10.1051/0004-6361:20040133}

\bibitem[{Thornton {et~al.}(2024)Thornton, Villar, Gomez, \& Hosseinzadeh}]{thornton2024extrabol}
Thornton, I., Villar, V.~A., Gomez, S., \& Hosseinzadeh, G. 2024, Research Notes of the AAS, 8, 48

\bibitem[{Vacca \& Leibundgut(1996)}]{vacca96}
Vacca, W.~D., \& Leibundgut, B. 1996, The Astrophysical Journal, 471, L37, \dodoi{10.1086/310323}

\bibitem[{Vacca \& Leibundgut(1997)}]{vacca97}
---. 1997, in Thermonuclear Supernovae (Springer Science + Business Media), 65--75, \dodoi{10.1007/978-94-011-5710-0_4}

\bibitem[{Valenti {et~al.}(2014)Valenti, Yuan, Taubenberger, Maguire, Pastorello, Benetti, Smartt, Cappellaro, Howell, Bildsten, {et~al.}}]{valenti2014pessto}
Valenti, S., Yuan, F., Taubenberger, S., {et~al.} 2014, Monthly Notices of the Royal Astronomical Society, 437, 1519

\bibitem[{Vincenzi {et~al.}(2019)Vincenzi, Sullivan, Firth, Guti{\'e}rrez, Frohmaier, Smith, Angus, \& Nichol}]{vincenzi2019spectrophotometric}
Vincenzi, M., Sullivan, M., Firth, R., {et~al.} 2019, Monthly Notices of the Royal Astronomical Society, 489, 5802

\bibitem[{{Vincenzi} {et~al.}(2019){Vincenzi}, {Sullivan}, {Firth}, {Guti{\'e}rrez}, {Frohmaier}, {Smith}, {Angus}, \& {Nichol}}]{2019MNRAS.489.5802V}
{Vincenzi}, M., {Sullivan}, M., {Firth}, R.~E., {et~al.} 2019, \mnras, 489, 5802, \dodoi{10.1093/mnras/stz2448}

\bibitem[{Walker {et~al.}(2014)Walker, Mazzali, Pian, Hurley, Arcavi, Cenko, Gal-Yam, Horesh, Kasliwal, Poznanski, {et~al.}}]{walker2014optical}
Walker, E., Mazzali, P., Pian, E., {et~al.} 2014, Monthly Notices of the Royal Astronomical Society, 442, 2768

\bibitem[{Waxman \& Katz(2017)}]{Waxman_2017}
Waxman, E., \& Katz, B. 2017, Shock Breakout Theory (Springer International Publishing), 967–1015, \dodoi{10.1007/978-3-319-21846-5_33}

\bibitem[{Xiang {et~al.}(2019)Xiang, Wang, Mo, Wang, Smartt, Fraser, Ehgamberdiev, Mirzaqulov, Zhang, Zhang, {et~al.}}]{xiang2019observations}
Xiang, D., Wang, X., Mo, J., {et~al.} 2019, The Astrophysical Journal, 871, 176

\bibitem[{Yaron \& Gal-Yam(2012)}]{Yaron_2012}
Yaron, O., \& Gal-Yam, A. 2012, Publications of the Astronomical Society of the Pacific, 124, 668, \dodoi{10.1086/666656}

\bibitem[{{Zapartas} {et~al.}(2017){Zapartas}, {de Mink}, {Van Dyk}, {Fox}, {Smith}, {Bostroem}, {de Koter}, {Filippenko}, {Izzard}, {Kelly}, {Neijssel}, {Renzo}, \& {Ryder}}]{2017ApJ...842..125Z}
{Zapartas}, E., {de Mink}, S.~E., {Van Dyk}, S.~D., {et~al.} 2017, \apj, 842, 125, \dodoi{10.3847/1538-4357/aa7467}

\bibitem[{{Zs{\'\i}ros} {et~al.}(2022){Zs{\'\i}ros}, {Nagy}, \& {Szalai}}]{2022MNRAS.509.3235Z}
{Zs{\'\i}ros}, S., {Nagy}, A.~P., \& {Szalai}, T. 2022, \mnras, 509, 3235, \dodoi{10.1093/mnras/stab3075}

\end{thebibliography}
